\documentclass[%
 aps,prx,twocolumn,superscriptaddress,
 amsmath,amssymb,amsfonts,
 floatfix
]{revtex4-2}
\usepackage[T1]{fontenc}
\usepackage{lmodern}
\usepackage{microtype}
\usepackage{graphicx}
\usepackage{bm}
\usepackage{appendix}
\usepackage[dvipsnames]{xcolor}
\usepackage{hyperref}
\hypersetup{
    colorlinks=true,
    citecolor=Green,
    linkcolor=Red,
    urlcolor=NavyBlue
}
\usepackage[capitalise]{cleveref}
\usepackage{soul}
\usepackage{mathtools}
\usepackage{siunitx}

\let\oldhat\hat
\renewcommand{\vec}[1]{\mathbf{#1}}
\renewcommand{\hat}[1]{\mathbf{\oldhat{#1}}}

\DeclarePairedDelimiter\abs{\lvert}{\rvert}
\DeclarePairedDelimiter\expval{\langle}{\rangle}

\begin{abstract}
In a lattice model subject to a perpendicular magnetic field, when the lattice constant is comparable to the magnetic length, one enters the ``Hofstadter regime,'' where continuum Landau levels become fractal magnetic Bloch bands. Strong mixing between bands alters the nature of the resulting quantum phases compared to the continuum limit; lattice potential, magnetic field, and Coulomb interaction must be treated on equal footing. 
Using determinant quantum Monte Carlo (DQMC) and density matrix renormalization group (DMRG) techniques, we study this regime numerically in the context of the Hubbard-Hofstadter model on a triangular lattice. 
In the field-filling phase diagram, we find a broad wedge-shaped region of ferromagnetic ground states for filling factor $\nu \leq 1$, bounded below by filling factor $\nu = 1$ and bounded above by half-filling the lowest Hofstadter subband. We observe signatures of SU(2) quantum Hall ferromagnetism at filling factors $\nu=1$ and $\nu=3$.
The phases near $\nu=1$ are particle-hole asymmetric, and we observe a rapid decrease in ground state spin polarization consistent with the formation of skyrmions only on the electron doped side. At large fields, above the ferromagnetic wedge, we observe a low-spin metallic region with spin correlations peaked at small momenta. We argue that the phenomenology of this region likely results from exchange interaction mixing fractal Hofstadter subbands. The phase diagram derived beyond the continuum limit points to a rich landscape to explore interaction effects in magnetic Bloch bands.
\end{abstract}

\begin{document} 

\title{Particle-hole asymmetric ferromagnetism and spin textures in the triangular Hubbard-Hofstadter model}

\author{Jixun K. Ding}
\email{jxding@stanford.edu}
\affiliation{Stanford Institute for Materials and Energy Sciences,
SLAC National Accelerator Laboratory, 2575 Sand Hill Road, Menlo Park, CA 94025, USA}
\affiliation{Department of Applied Physics, Stanford University, Stanford, CA 94305, USA}

\author{Luhang Yang}
\affiliation{Stanford Institute for Materials and Energy Sciences,
SLAC National Accelerator Laboratory, 2575 Sand Hill Road, Menlo Park, CA 94025, USA}
\affiliation{Department of Physics, Northeastern University, Boston, MA 02115, USA}  

\author{Wen O. Wang}
\affiliation{Stanford Institute for Materials and Energy Sciences,
SLAC National Accelerator Laboratory, 2575 Sand Hill Road, Menlo Park, CA 94025, USA}
\affiliation{Department of Applied Physics, Stanford University, Stanford, CA 94305, USA}

\author{Ziyan Zhu}
\affiliation{Stanford Institute for Materials and Energy Sciences,
SLAC National Accelerator Laboratory, 2575 Sand Hill Road, Menlo Park, CA 94025, USA}

\author{Cheng Peng}
\affiliation{Stanford Institute for Materials and Energy Sciences,
SLAC National Accelerator Laboratory, 2575 Sand Hill Road, Menlo Park, CA 94025, USA}

\author{Peizhi Mai}
\affiliation{Department of Physics and Institute of Condensed Matter Theory, University of Illinois at Urbana-Champaign, Urbana, IL 61801, USA}

\author{Edwin W. Huang}
\affiliation{Department of Physics and Institute of Condensed Matter Theory, University of Illinois at Urbana-Champaign, Urbana, IL 61801, USA}
\affiliation{Department of Physics and Astronomy, University of Notre Dame, Notre Dame, IN 46556, United States}
\affiliation{Stavropoulos Center for Complex Quantum Matter, University of Notre Dame, Notre Dame, IN 46556, United States}

\author{Brian Moritz}
\affiliation{Stanford Institute for Materials and Energy Sciences,
SLAC National Accelerator Laboratory, 2575 Sand Hill Road, Menlo Park, CA 94025, USA}

\author{Philip W. Phillips}
\affiliation{Department of Physics and Institute of Condensed Matter Theory, University of Illinois at Urbana-Champaign, Urbana, IL 61801, USA}

\author{Benjamin E. Feldman}
\affiliation{Stanford Institute for Materials and Energy Sciences, SLAC National Accelerator Laboratory, Menlo Park, CA 94025, USA}
\affiliation{
Department of Physics, Stanford University, Stanford, CA 94305, USA}

\author{Thomas P. Devereaux}
\email{tpd@stanford.edu}
\affiliation{Stanford Institute for Materials and Energy Sciences,
SLAC National Accelerator Laboratory, 2575 Sand Hill Road, Menlo Park, CA 94025, USA}
\affiliation{
Department of Materials Science and Engineering, Stanford University, Stanford, CA 94305, USA}
\date{\today}

\maketitle 

\section{Introduction} 

Landau levels are paradigmatic examples of topological flat bands. They arise from a simple continuum model of a two-dimensional electron gas under the influence of an out-of-plane orbital magnetic field, and have been instrumental in explaining a wide variety of quantum Hall phenomena~\cite{Tong2016}. 
The effectiveness of the Landau level picture relies on two key assumptions. First, lattice effects are neglected: all Landau levels are flat, have uniform quantum geometry, continuous magnetic translation symmetry, and continuous rotational symmetry. Second, the Landau level spacing, $h\omega_c$, is typically assumed to be the largest energy scale in the problem of interest. Any fully filled and empty bands are inert, so the many-body problem of interacting electrons only needs to be treated within one isolated Landau level. The lowest Landau level, in particular, satisfies a number of desirable analytical properties~\cite{Girvin1986,Parameswaran2012}, some of which have one-to-one correspondence in ideal Chern bands~\cite{Roy2014,Claassen2015,Wang2021,Ledwith2022}. 

While the Landau level picture works well for systems with magnetic length much greater than the lattice constant, including GaAs and monolayer graphene, such a description breaks down in moir\'e materials, whose superlattice constants are comparable with magnetic length at moderate magnetic fields~\cite{Hunt2013}. 
Within this ``Hofstadter regime,'' Landau levels become fractal magnetic Bloch bands~\cite{Hofstadter1976,Wannier1978} and acquire nonzero dispersion, non-uniform Berry curvature, and discrete magnetic translation symmetry~\cite{Zak1964,Xiao2010}. 
It is natural to ask at this point if results obtained using the continuum model persist or generalize appropriately in the strong lattice limit. Conversely, the lattice potential may precipitate new phases of matter that cannot be captured in the continuum model. 
Recent works along this line have focused on the lattice fractional quantum Hall effect~\cite{Kapit2010,Bauer2016} and fractional Chern insulators~\cite{Regnault2011,Sheng2011} which may~\cite{Qi2011,Scaffidi2012} or may not~\cite{Moller2015,Andrews2018,Andrews2021,Bauer2022} be adiabatically connected to the continuum limit.  

Another phenomenon of interest is quantum Hall ferromagnetism (QHFM) and associated skyrmions~\cite{Sondhi1993}. The mathematical description of QHFM has been generalized to a large variety of multicomponent topological flat bands, under the moniker of generalized/flavor/isospin QHFM, or flavor polarization of Chern bands. 
In moir\'e materials, particularly magic-angle twisted bilayer graphene, correlated insulating states at integer Chern numbers~\cite{Nuckolls2020,Choi2021,Park2021,Das2021,Wu2021,Saito2021,Stepanov2021,Yu2022} are abundant, and $\nu=\pm2$ gaps have been identified to host skyrmion excitations~\cite{yu2022spin}. Theoretical proposals suggest that skyrmions exist in flat Chern bands, and may even contribute to superconductivity~\cite{Chatterjee2020,Khalaf2021,Kwan2022,Chatterjee2022}.

In the standard picture for SU(2) QHFM~\cite{Sondhi1993,Fertig1997,Abolfath1997}, each Landau level with $N_\phi$ degenerate single-particle orbitals can accommodate $2N_\phi$ electrons with an internal SU(2) spin degree of freedom. Landau level filling factor $\nu=1$ corresponds to $N_\uparrow + N_\downarrow = N_\phi$, i.e. half-filling the lowest Landau level. Coulomb repulsion between electrons induces a ferromagnetic exchange interaction, $E_{\mathrm{X}}$, so the $\nu=1$ ground state is spin polarized with total spin quantum number $S = N_\phi/2$, even in the absence of Zeeman splitting. 
The lowest energy-charged excitations about the $\nu=1$ quantum Hall ferromagnet are not bare particles and holes, but charge-spin texture bound states known as skyrmions. 
The existence of quantum Hall skyrmions manifests experimentally as rapid depolarization:
doping slightly away from $\nu=1$ dramatically reduces the total spin $S$, as each addition and removal of one unit of charge is associated with a large number of spin flips~\cite{Rezayi1991,Barrett1995,Aifer1996}. 
While ferromagnetic insulating states also occur in higher Landau levels at odd integer filling factors $\nu=3,5,7...$, bare particles and holes, rather than skyrmions, are believed to be the lowest charged quasiparticle excitations in higher Landau levels~\cite{Schmeller1995,Wu1995,Rhone2015}. Additionally, Landau level mixing is thought to destabilize $\nu=1$ skyrmions~\cite{Melik1999,Mihalek2000}. The conditions for skyrmion stability in general Chern bands remains an open question~\cite{Wu2020,Khalaf2022,Schindler2022}.

In order to explore possible phases in interacting lattice quantum Hall systems, and explore the question of skyrmion stability, we use two numerically exact and unbiased methods, determinant quantum Monte Carlo (DQMC)~\cite{White1989,Loh1990} and density matrix renormalization group (DMRG)~\cite{White1992,White1993,Ostlund1995,Dukelsky1998}, to study the Hubbard-Hofstadter (HH) model on a triangular lattice. 
By using unbiased numerical methods, we are able to address the analytically intractable regime of non-flat, non-isolated bands. 

The HH model is a minimal lattice Hamiltonian that incorporates the effects of a strong orbital magnetic field and on-site electron repulsion.  
One important application of the HH model is twisted bilayer transition metal dichalcogenides (TMDs)~\cite{wu2018}. At small twist angles ($\theta \lesssim 2^\circ$), moir\'e Wannier orbitals are well localized and the long-range Coulomb interaction can be ignored. The triangular lattice HH model (with a suitable additional Zeeman term) is then directly applicable to small-angle twisted bilayer TMDs under an external magnetic field~\cite{Morales-Duran2022a}. 
Moreover, a time-reversal symmetric variant~\cite{Cocks2012,Sahay2023} of the HH model can be realized with a pseudo-magnetic field induced by strain. For example, in graphene, periodic or elastic strain leads to a pseudo-magnetic field that gives rise to correlated insulators~\cite{Levy2010,Mao2020}. Finally, the HH model on a triangular lattice can also be realized in cold-atom quantum simulators~\cite{Yang2021,Garwood2022,Mongkolkiattichai2022,Xu2023d} in an artificial gauge field~\cite{Mancini2015}.

The HH model has recently been studied in the square and hexagonal lattice geometry via DQMC~\cite{Ding2022,Mai2023} and in the square geometry via DMRG~\cite{Palm2023}.  
Our results for the triangular lattice reveal a large ferromagnetic wedge for $\nu \leq 1$ which evolves into an incompressible quantum Hall ferromagnet at $\nu=1$. We also observe ground state spin polarization signatures of the $\nu=3$ QHFM. Whether the QHFM state at $\nu=1$ extends to high fields ($\Phi/\Phi_0 \rightarrow 0.5$) depends on the Hubbard interaction strength. 
The $\nu< 1$ ferromagnetic wedge has a sharp interaction-independent upper boundary set by half filling the lowest Hofstadter sub-band. For $\Phi/\Phi_0 \leq 1/3$, upon electron doping slightly away from $\nu=1$, the ground state spin polarization drops precipitously, consistent with a region of skyrmion formation. Above the ferromagnetic wedge, we identify a region of parameter space that is low-spin, metallic, and exhibits spin correlations peaked at small momenta. 
In this region, the exchange interaction is comparable to the fractal Hofstadter bandwidths and band gaps, and Landau level mixing effects are strong. We propose the ground state to be a spin-textured metal precipitated by competing tendencies towards singlet and spin-polarized ground states in close proximity. Our work complements recent analytical studies~\cite{Khalaf2022,Schindler2022} and is generally relevant for understanding the behavior of strongly correlated electrons under large magnetic fields.

\section{Methods}
We study the single-band Hubbard-Hofstadter model
\begin{multline}
H = -\sum_{ ij  \sigma} t_{ij}  \left\{\exp\left[\mathrm{i}\varphi_{ij}\right]c_{i \sigma}^\dagger c_{j \sigma} + \mathrm{h.c.}\right\}  
- \mu \sum_{i \sigma} n_{i\sigma} \\
+U\sum_{i}\left(n_{i\uparrow} - 1/2 \right)\left(n_{i\downarrow} - 1/2\right), \label{eq:hamiltonian}
\end{multline}
on a two-dimensional triangular lattice. The hopping integral $t_{ij} = t$ between nearest neighbor sites $\langle i j\rangle$, and $t_{ij} = 0$ otherwise, $\mu$ is the chemical potential, and $U$ is the on-site Coulomb interaction strength. $c_{i\sigma}^{\dagger}$ ($c_{i\sigma}$) is the creation (annihilation) operator for an electron on site $i$ with spin $\sigma\in\{\uparrow,\downarrow\}$, and $n_{i\sigma} =  c_{i\sigma}^\dagger c_{i\sigma}$ measures the number of electrons of spin $\sigma$ on site $i$. 
A spatially uniform and static orbital magnetic field is introduced by the Peierls substitution via the phase
\begin{equation}
\varphi_{ij} = \dfrac{2\pi}{\Phi_0} \int_{\vec{r}_i}^{\vec{r}_j} \vec{A}\cdot d\bm{\ell}, \label{eq:peierls-phase}
\end{equation}
where $\Phi_0 = h/e$ is the magnetic flux quantum, and $\vec{r}_i = (r_{ix}, r_{iy})$ is the position of site $i$, and the path integral is taken along the shortest straight line path between sites $i$ and $j$. The vector potential $\vec{A}$ generates out-of-plane magnetic field $\vec{B} = B \hat{z}$, and has gauge freedom parametrized by $\alpha \in \mathbb{R}$:
\begin{equation}
    \vec{A}(\vec{r}) = B \begin{bmatrix}
        -\alpha r_y \\
        (1-\alpha) r_x
    \end{bmatrix}. \nonumber
\end{equation}
In this work we use $\alpha = 1/2$, which corresponds to the symmetric gauge $\mathbf{A}(\vec{r})= B(-r_y\hat{x} + r_x\hat{y})/2$. 
We have verified that the results reported in this work do not depend on the choice of gauge. Zeeman coupling is neglected. 
While the Hamiltonian breaks time-reversal, parity, and particle-hole symmetries, it preserves SU(2) spin symmetry. This choice is consistent with the energy hierarchy $E_{\mathrm{Z}} \ll E_{\mathrm{X}}$ (where $E_{\mathrm{Z}} = g\mu_B B$ is Zeeman splitting) required to observe quantum Hall skyrmions. 

The Hamiltonian in~\cref{eq:hamiltonian} is simulated on a finite cluster with lattice constant $a$, and $N_1$ and $N_2$ sites respectively, along the directions 
$\vec{a}_1 = a(1/2 , \sqrt{3}/2)$,
$\vec{a}_2 = a(-1/2, \sqrt{3}/2)$,
which are the primitive lattice vectors of the triangular lattice.
$N = N_1 N_2$ denotes the total number of sites. Modified periodic boundary conditions are implemented, consistent with magnetic translation symmetry~\cite{Assaad2002,Xiao2010}, as described in detail in~\cref{sec:bdy_cond}.
Requiring that the wave function be single-valued on the torus gives the flux quantization condition 
$\Phi/\Phi_0 = N_\phi/N$, where $\Phi=\sqrt{3}Ba^2/2$ is the magnetic flux threading each unit cell and $N_\phi$ is an integer. Denoting the number of electrons in the system as $N_e = N_\uparrow + N_\downarrow$, filling factor $\nu=1$ corresponds to $N_e = N_\phi$. Electron number density $n = N_e/N \in [0,2]$. 

Throughout this text, magnetic field strengths are denoted in units of $\Phi/\Phi_0$ and the lattice constant is set to $a=1$. Following Ref.~\cite{Parameswaran2013}, the terms ``magnetic Bloch band,'' ``Hofstadter band,'' and ``Chern band'' are used interchangeably. We use $n = C(\Phi/\Phi_0) + s$ to parametrize non-interacting gaps and correlated states in $(n,\Phi/\Phi_0)$ parameter space. Matching the notation used in existing literature, when $s=0$, we denote $C$ by $\nu$.

DQMC simulations of~\cref{eq:hamiltonian} are performed on clusters with linear size $N_1 = N_2 = 6, 8, 9, 12,$ and 15. As shown in~\cref{sec:finite_size}, finite-size effects in DQMC data at accessible temperatures are minimal, so we may safely consider DQMC results representative of the thermodynamic limit. Error bars in DQMC results, when shown, denote $\pm 1$ standard error of the mean, estimated by jackknife resampling. Detailed DQMC simulation parameters are listed in~\cref{sec:simulation_parameters}, and fermion sign for a typical set of parameters is shown in~\cref{sec:fermion_sign}. 

DMRG simulations are performed using the ITensor library~\cite{Fishman2022,Fishman2022_2}. Results in the main text are obtained on a $N_1 \times N_2 = 3\times 25$ cluster with cylindrical boundary conditions which is periodic in the short direction and open in the long direction. By keeping the bond dimension up to $m=5000$, we ensure the truncation error is of order $10^{-5}$ or less. In \cref{sec:supplemental-data} we present DMRG calculations on the torus geometry with modified periodic boundary conditions in both directions for a $3\times 6$ cluster to directly compare with DQMC.

\section{Summary of Key Findings}

\begin{figure}[htpb]
    \centering
    \includegraphics[width=\linewidth]{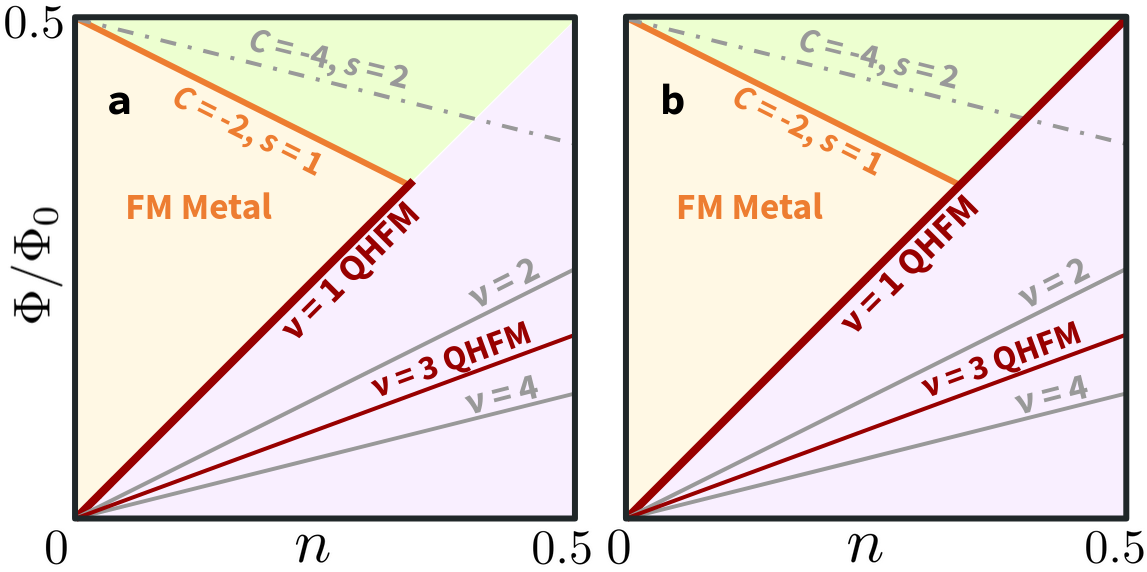}
    \caption{Phase diagram sketches of Hubbard-Hofstadter model on a triangular lattice for \textbf{a} $U/t\lesssim4$ and \textbf{b} $U/t \gtrsim 8$ in the $n \in [0,0.5]$, $\Phi/\Phi_0 \in[0,0.5]$ quadrant of $(n,\Phi/\Phi_0)$ parameter space. 
    Dot-dashed grey lines denote the Chern band gap satisfying the Diophantine equation $n=C (\Phi/\Phi_0) + s$ ($C = -4, s=2$)~\cite{Wannier1978}. Solid orange lines denote $n=-2 (\Phi/\Phi_0) + 1$, the observed upper boundary of FM wedges. Solid grey lines denote even integer Chern band gaps as annotated (the intercept $s=0$ implicitly, so that $C=\nu$). Pale purple shadings indicate metallic states with short-range antiferromagnetic correlations. Pale green shading indicates anomalous metal with small-$q$ spin correlations.}
    \label{fig:phase-diagram}
\end{figure}

In this work we primarily investigate the $n \in [0,0.5]$, $\Phi/\Phi_0 \in[0,0.5]$ corner of parameter space, which encompasses the lowest several Chern bands, and approaches the continuum limit as $n\rightarrow 0$, $\Phi/\Phi_0 \rightarrow 0$. Our main results can be summarized in a schematic phase diagram shown in ~\cref{fig:phase-diagram}. We first highlight the primary findings and then elaborate on these observations.

In a triangular-shaped wedge of parameter space bounded below by $\nu = 1$ and above by $n = -2(\Phi/\Phi_0) + 1$, the ground state is ferromagnetic and metallic, only becoming incompressible at $\nu = 1$. We call this $\nu \leq 1$ region the ``FM wedge,'' and identify the incompressible state at $\nu=1$ as the lattice realization of $\nu=1$ QHFM. 
For $U/t = 4$ in~\cref{fig:phase-diagram}\textbf{a}, QHFM terminates near $(n,\Phi/\Phi_0) = (1/3,1/3)$, while 
for $U/t = 8$ in~\cref{fig:phase-diagram}\textbf{b}, QHFM extends toward the highest field strength $\Phi/\Phi_0=0.5$. Near $\nu=1$, when $\Phi/\Phi_0 < 1/3$, the ground state spin polarization remains maximal upon removal of electrons, but drops rapidly upon addition of electrons, consistent with the formation of skyrmions for only $\nu\rightarrow 1^+$. Outside the FM wedge, for $\nu>1$, except for incompressible states at even filling factors and the partially polarized $\nu=3$ QHFM, the ground state is metallic with short-range antiferromagnetic fluctuations,
exhibiting behavior largely consistent with weakly interacting bands. 

Nontrivial spin features are observed above the FM wedge, for $n \geq -2(\Phi/\Phi_0) + 1$ and $\nu<1$. In this region, the ground state is metallic and low-spin. At finite temperatures, the equal-time spin correlations are peaked at small momenta, different from both nearby ferromagnetic order and the non-interacting Lindhard susceptibility. We suggest that this region hosts a novel spin-textured metallic phase due to the competition between the exchange interaction and Hofstadter subband bandgaps.

\section{Results and Discussion}

\begin{figure}[htpb]
    \includegraphics[width=\linewidth]{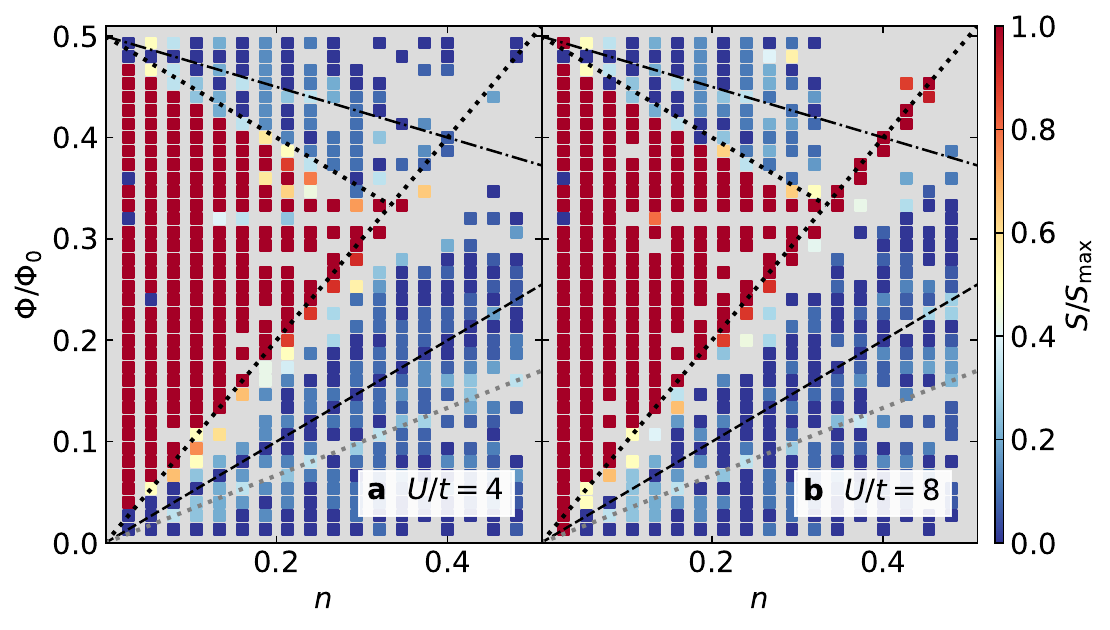}
    \caption{DMRG ground state spin degeneracy for \textbf{a} $U/t=4$ and \textbf{b} $U/t=8$, plotted as $S/S_{\mathrm{max}}$, where $S_{\mathrm{max}} = N_e/2$. States differing in energy by less than $10^{-3} t$ 
    are treated as degenerate. Red color indicates that the ground state is maximally spin polarized, while blue color indicates that the ground state is a singlet configuration. Black dotted lines denote $\nu=1$ and $n=-2(\Phi/\Phi_0) + 1$, black dashed lines denote $\nu=2$, grey dotted lines denote $\nu=3$, and black dot-dashed lines denote $n=-4 (\Phi/\Phi_0) + 2$. DMRG data are obtained on a $3\times25$ cluster with cylindrical boundary conditions.}
    \label{fig:GS-DMRG}
\end{figure}

Zero-temperature DMRG results indicating the ground state spin degeneracy, shown in \cref{fig:GS-DMRG}, illustrate the ``FM wedge,'' interaction dependent $\nu=1$ QHFM, and partially polarized $\nu=3$ QHFM features in the schematic ground state phase diagram in \cref{fig:phase-diagram}. 

These spin-sector resolved DMRG calculations extend the results of previous work~\cite{Palm2023} and produce a more finely-resolved ground-state phase diagram. In particular, we note that the spin polarization of the $\nu=3$ QHFM ($S/S_{\mathrm{max}} = 1/3$) was not observed in previous numerical studies~\cite{Ding2022,Palm2023,Mai2023}. We also note that the upper boundary of the FM wedge appears associated with half-filling the lowest fractal Hofstadter sub-band. This sub-band only exists for $\Phi/\Phi_0 > 1/3$ (see ~\cref{fig:nonint-dos-wannier}). The spin polarization change as one crosses the $n=-2(\Phi/\Phi_0) + 1$ line matches expectations from experimentally observed Hofstadter sub-band ferromagnetism~\cite{Saito2021}.

\begin{figure*}[htpb]
    \includegraphics[width=\linewidth]{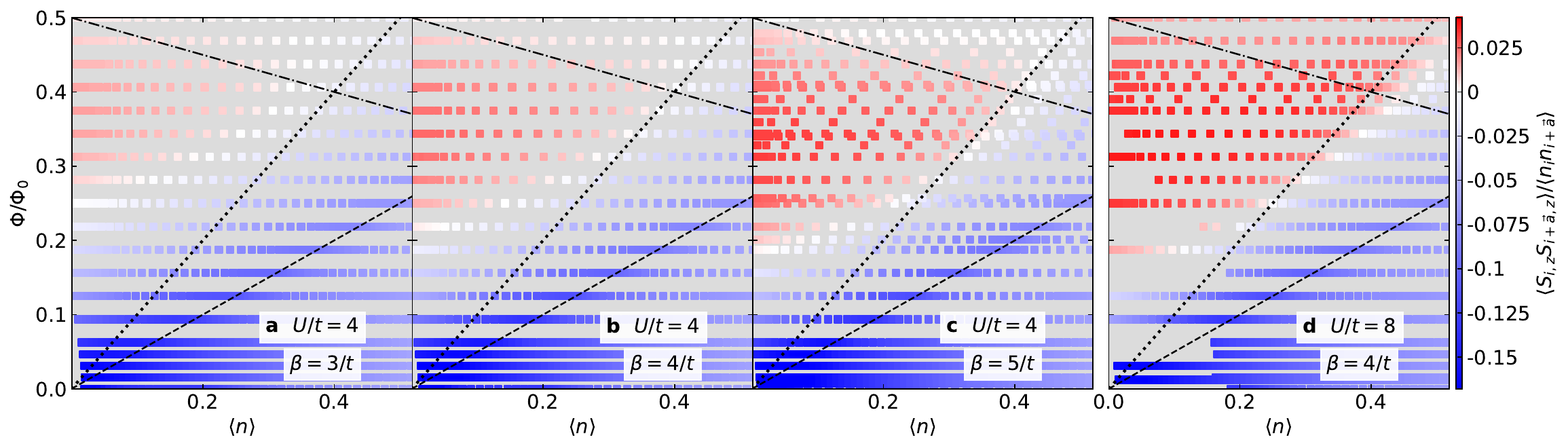} 
    \caption{Equal-time nearest-neighbor spin correlation normalized by nearest-neighbor charge correlation, $\langle S_{i,z} S_{i+\vec{a},z}\rangle/\langle n_i n_{i+\vec{a}}\rangle$, obtained by DQMC. This normalization convention was chosen to ensure that we are not simply observing higher spin correlation values due to higher particle density. \textbf{a}-\textbf{c} show temperature dependence at fixed $U/t=4$; \textbf{b} and \textbf{d} show Hubbard $U$ dependence at fixed inverse temperature $\beta=4/t$. Dotted lines denote $\nu=1$, dashed lines denote $\nu=2$, and dot-dashed lines denote $n=-4 (\Phi/\Phi_0) + 2$. The colorbar is saturated in the minimum direction. DQMC data combine results obtained on clusters of size $8\times 8$, $9\times 9$ and $12\times 12$, an appropriate procedure since finite-size effects are small for this local observable, as shown in \cref{sec:finite_size}. }
    \label{fig:wedge-U-dependence}
\end{figure*}

We compare the ground-state DMRG results with DQMC results to examine the manifestation of ground state features at elevated temperatures. \cref{fig:wedge-U-dependence} shows DQMC data for nearest-neighbor spin correlations $\expval{S_{i,z} S_{i+\vec{a},z}}$, normalized by nearest-neighbor particle density correlations $\langle n_{i} n_{i+\vec{a}}\rangle$. The results reveal a triangular wedge of ferromagnetic correlations at $\nu\lesssim 1$, and antiferromagnetic correlations for $\nu\gtrsim 1$.
\cref{fig:wedge-U-dependence}\textbf{a}-\textbf{c} show for fixed $U/t=4$, local ferromagnetic correlations in the $\nu\lesssim 1$ region become stronger as temperature decreases. Comparing \cref{fig:wedge-U-dependence}\textbf{a} and \textbf{d} shows that at the same temperature, larger Hubbard interaction strength induces stronger local ferromagnetic correlations.
For $U/t=4$, the region of ferromagnetic correlations is roughly bounded above by the straight line $n=-4 (\Phi/\Phi_0) + 2$, down to the lowest accessible temperatures (see~\cref{fig:extend-T} for extended temperature dependence). 
This is different from the $U/t=8$ case shown in \cref{fig:wedge-U-dependence}\textbf{d}, where ferromagnetic correlations persist to largest field $\Phi/\Phi_0=0.5$ (see~\cref{fig:extend-U} for extended interaction dependence). 
The $n=-4 (\Phi/\Phi_0) + 2$ line corresponds to a small Hofstadter sub-band band gap (see dot-dashed line in \cref{fig:nonint-dos-wannier}\textbf{b}) and intersects with the $\nu=1$ line at $(n,\Phi/\Phi_0) = (0.4,0.4)$. In the absence of interactions, states on this line represent singlet ground states. It is therefore a natural boundary for the termination of ferromagnetic correlations. The qualitatively different behavior of $U/t=4$ and $U/t=8$ suggests that the latter corresponds to sufficiently high interaction strength that the induced exchange splitting is larger than the non-interacting Chern band gap, allowing local ferromagnetic to persist to $\Phi/\Phi_0 = 0.5$. The $\nu \lesssim 1$ region of local ferromagnetic correlations in DQMC results overlap with the FM wedge of spin polarized DMRG ground states, except in the limit $\Phi/\Phi_0 \rightarrow 0$, $n \rightarrow 0$, when energy scales are compressed and become smaller than temperatures accessible via DQMC. The short-range antiferromagnetic correlations found by DQMC for $\nu>1$ are also consistent with low-spin ground states found by DMRG. Additionally, DQMC results for charge compressibility exhibit interaction induced minima at $\nu=1$ (\cref{fig:compress-T-dep,fig:corr-butterfly,fig:wannier}) as also observed in previous work~\cite{Mai2023}, which matches the picture of $\nu=1$ QHFM.
Agreement between DQMC and DMRG results ensures that quasi-one-dimensional DMRG results are valid representations of the physics of a two-dimensional system.

Given the evidence for $\nu=1$ QHFM in the HH model, it is natural to seek evidence for skyrmionic ground states near $\nu=1$, as predicted in the continuum limit. \cref{fig:GS-DMRG} shows that for $\Phi/\Phi_0 \leq 1/3$, the ground state upon removing one electron from $\nu=1$ ($N_e = N_\phi-1$) has $S = N_e/2$. These spin numbers are consistent with bare holes being the favored way to remove charge from the ferromagnet~\cite{Barrett1995}. On the other hand, the ground state spin polarization upon adding electrons to $\nu=1$ ($N_e > N_\phi$) drops rapidly to $S\approx 0$, much faster than if bare electrons are added without modifying the ferromagnetic spin background.
These results are consistent with Ref.~\cite{Palm2023}, which suggests that slightly electron doping the $\nu=1$ QHFM may produce a narrow strip of skyrmion ground states. The DMRG simulations are hard to converge in the $\nu\rightarrow 1^+$ region, indicating closely competing energy scales. 

The particle-hole asymmetric nature of charge carriers in the vicinity of $\nu=1$ for $\Phi/\Phi_0 < 1/3$ may be heuristically explained by the extreme short-range nature of the Hubbard interaction. When $\nu\leq 1$, the Hubbard interaction only acts indirectly by favoring spin-polarized ground states. In this case, when electrons of the same spin are added to the system, they can always choose to occupy a vacant Wannier orbital in the lowest Hofstadter band, and avoid incurring the large Hubbard potential energy cost. At $\nu=1$, the electrons form a fully spin-polarized Slater determinant. When we try to add an electron to this state, while keeping the ferromagnetic spin background undisturbed, we end up either placing an electron in a higher Hofstadter band (and thus paying the cost of a band gap), or placing a $\uparrow$ and $\downarrow$ electron in the same Wannier orbital in the lowest Hofstadter band (and thus incurring the Hubbard interaction penalty). Without any Zeeman splitting to explicitly favor a net direction of magnetization, it is natural that the addition of an electron to the $\nu=1$ state should significantly modify the spin background and induce a large number of spin flips. Whether the resulting low-spin state truly consists of large skyrmions however, requires further examination of the Pontryagin density~\cite{Sondhi1993}. 

We emphasize that there is no fundamental reason that a quantum Hall ferromagnet must be accompanied by skyrmions, as evidenced by the $\nu=3,5,7...$ QHFM states which have bare particles and holes as the lowest charged excitations~\cite{Schmeller1995,Wu1995,Rhone2015}. As the question of skyrmion stability is a purely energetic one,
it is possible that band geometry or dispersion reduces skyrmion stability, although how this occurs and to what extent is not clear \emph{a priori}.
While a sufficiently strong, purely local interaction ($\delta$-function potential in the continuum model, Hubbard interaction in this lattice model) appears sufficient to generate ferromagnetic exchange, and thereby induce a quantum Hall ferromagnet, longer-range repulsive interactions may be required for skyrmions to be the lowest energy quasiparticle on both sides of the $\nu=1$ QHFM. This was discussed in the Landau level picture in Refs.~\cite{MacDonald1996,Wojs2002}, but requires further study in the lattice context~\cite{Zhang2019,Repellin2020}. 

\begin{figure}[htpb]
    \includegraphics[width=\linewidth]{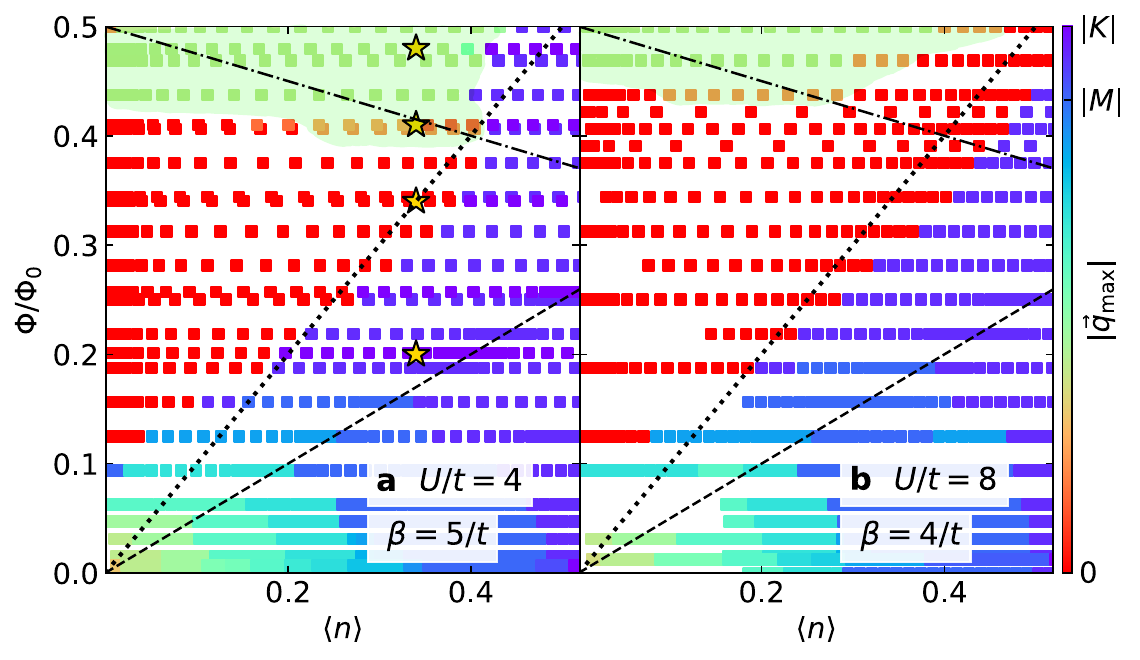} 
    \caption{Distance of peak position of $\chi_s(\vec{q},\tau=0)$ (\cref{eq:suscept-q}) to $\Gamma$, denoted by $|q_{\mathrm{max}}|$. Angular information is suppressed and only distance to $\Gamma$ is shown. \textbf{a} inverse temperature $\beta=5/t$, Hubbard interaction $U/t=4$ and \textbf{b} inverse temperature $\beta=4/t$, Hubbard interaction $U/t=8$. Red color indicates ferromagnet correlations, while blue-purple colors indicate antiferromagnetic correlations peaked at large momentum. 
    In the $\nu<1$ region above the ferromagnetic wedge, spin correlations are peaked at small momentum $\abs{\vec{q}_{\mathrm{max}}} \leq \abs{\vec{K}}/2$.  Yellow stars mark four representative points analyzed in~\cref{fig:suscept-sample}. DQMC data combine results obtained on clusters of size $8\times 8$, $9\times 9$, and $12\times 12$.}
    \label{fig:qmax-Udep}
\end{figure}

We next discuss in more detail the high-field region indicated by pale green shading in \cref{fig:phase-diagram}. 
DQMC results for the peak position of spin correlations in momentum space are shown in \cref{fig:qmax-Udep} (see \cref{fig:extend-qmax-U} for extended interaction dependence). For both $U/t = 4$ and $U/t = 8$, the $\nu>1$ region is dominated by short-range antiferromagnetic correlations peaked at momentum $\vec{K}$~(purple) or $\vec{M}$~(blue), which are high symmetry points in the triangular lattice Brillouin zone. In a roughly triangular shaped region with $\nu \lesssim 1$, spin correlations are peaked at $\Gamma$. This region is smaller for $U/t=4$ than for $U/t=8$, which is also consistent with our findings based on \cref{fig:GS-DMRG} and \cref{fig:wedge-U-dependence}. 
Above the FM wedge, the peak at $\Gamma$ gives way to peaks at small momenta with $\abs{\vec{q}} \leq \abs{\vec{K}}/2$ (green, top of \cref{fig:qmax-Udep}). 
DMRG calculations (\cref{fig:GS-DMRG}) indicate this region has low-spin ground states, and DQMC calculations (\cref{fig:wannier}) indicate this region is charge compressible. For $U/t=4$ in \cref{fig:qmax-Udep}\textbf{a}, as particle density increases to $\langle n\rangle \sim 0.4$, the small-$q$ spin correlations abruptly transition to correlations peaked at $\vec{K}$. For $U/t=8$ in \cref{fig:qmax-Udep}\textbf{b}, as particle density increases, the small-$q$ spin correlations first transition to ferromagnetic correlations near $\nu=1$ then to antiferromagnetic correlations peaked at $\vec{K}$ for $\nu>1$. By comparison, the non-interacting Lindhard susceptibility is peaked near $\vec{K}$ in the entire $n \in [0,0.5]$, $\Phi/\Phi_0 \in[0,0.5]$ parameter space quadrant, as shown in \cref{fig:nonint-suscept}.

\begin{figure}[htpb]
    \includegraphics[width=\linewidth]{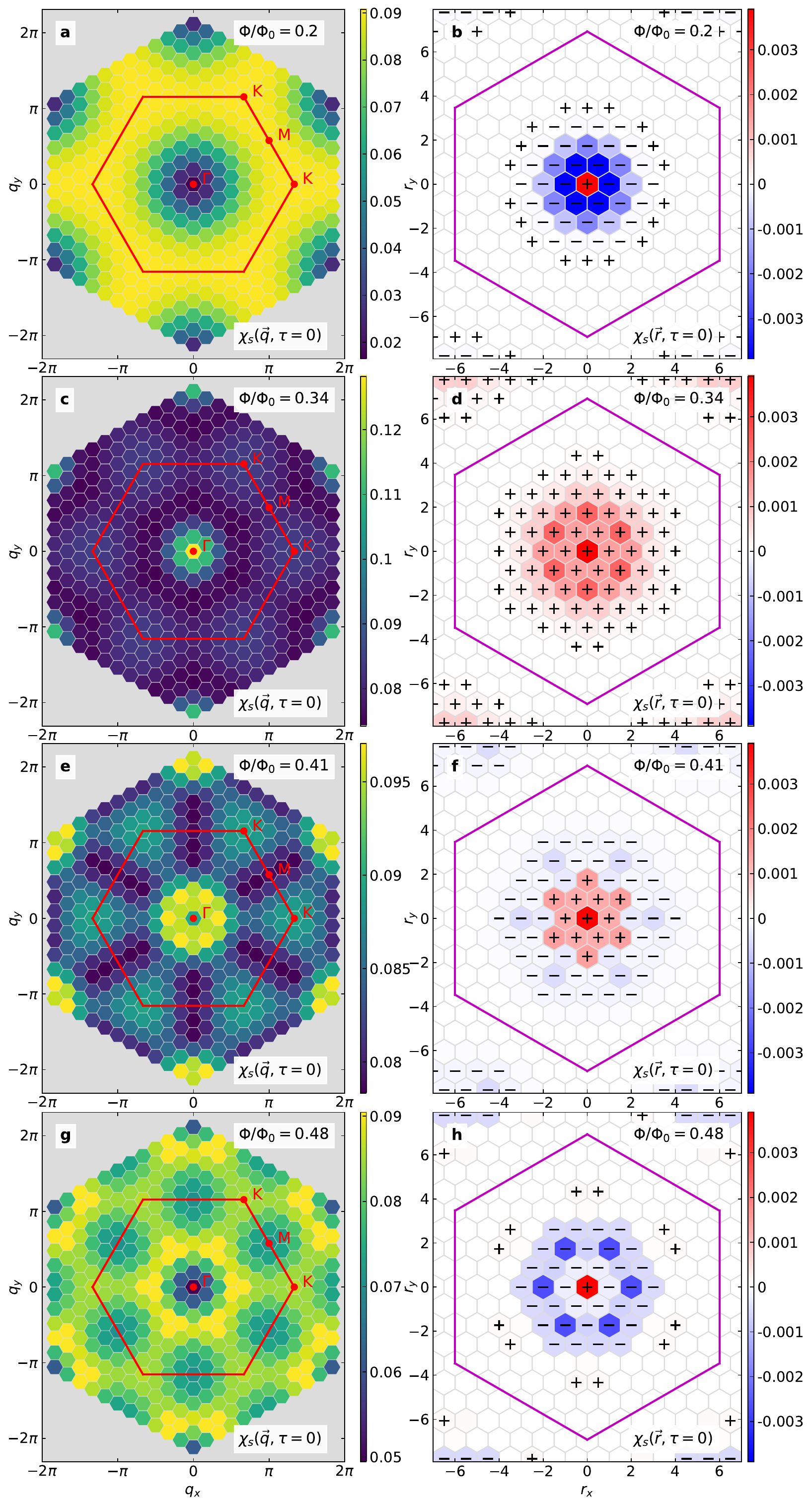} 
    \caption{Equal-time spin correlation for four representative points marked by stars in~\cref{fig:qmax-Udep}. Each row represents a different field strength. All plots have fixed inverse temperature $\beta=6/t$, Hubbard interaction $U/t=4$, and particle density $\langle n\rangle = 0.34$. Each row shows a different field strength, and field strength increases as we move down the plot. The first column (\textbf{a},\textbf{c},\textbf{e},\textbf{g}) shows momentum-space spin correlations $\chi_s(\vec{q},\tau=0)$(\cref{eq:suscept-q}). The second column (\textbf{b},\textbf{d},\textbf{f},\textbf{h}) shows the corresponding real-space spin correlations $\chi_s(\vec{r},\tau=0)$(\cref{eq:suscept-r}). In real-space plots, data points annotated with a $+$ or $-$ sign are nonzero by at least two standard errors, and color bars are saturated in both directions in order to improve contrast. DQMC data are obtained on $12\times 12$ cluster. For finite-size analysis of \textbf{e}-\textbf{h}, see \cref{fig:suscept-15x15}.}
    \label{fig:suscept-sample}
\end{figure}

To further examine the anomalous spin correlations above the FM wedge, we choose four representative points in parameter space (marked by stars in \cref{fig:qmax-Udep}\textbf{a}) which have identical particle density and increasing magnetic field strength. Their equal-time spin correlations in both real and momentum space, shown in \cref{fig:suscept-sample}, demonstrate the dramatic evolution of spin correlations in different phases as field strength increases. For the sake of comparison, the non-interacting Lindhard responses at identical $(n,\Phi/\Phi_0)$ points are shown in \cref{fig:nonint-suscept}.
The point $(n,\Phi/\Phi_0) = (0.34,0.2)$ (shown in \cref{fig:suscept-sample}\textbf{a}-\textbf{b}) lies in the $\nu>1$ low-spin metallic region. It exhibits short-range antiferromagnetic correlations with broad peaks centered at $\vec{K}$. The profile of spin correlations in \cref{fig:suscept-sample}\textbf{a} matches the non-interacting Lindhard susceptibility in \cref{fig:nonint-suscept}\textbf{a} closely, indicating that the $\nu>1$ area is weakly interacting. 
The point $(n,\Phi/\Phi_0) = (0.34,0.34)$ (shown in \cref{fig:suscept-sample}\textbf{c}-\textbf{d}) lies on the $\nu=1$ QHFM line. It exhibits ferromagnetic correlations peaked at $\Gamma$, 
which is consistent with a ferromagnetic ground state. The spin correlations are (unsurprisingly) different from the non-interacting case, \cref{fig:nonint-suscept}\textbf{c}, indicating the strong influence of interactions in the FM wedge. 

The point $(n,\Phi/\Phi_0) = (0.34,0.41)$ (shown in \cref{fig:suscept-sample}\textbf{e}-\textbf{f}) lies above the ground state FM wedge in the anomalous metal region. It exhibits weak spin correlations peaked at $\vec{q} \sim \vec{K}/4$, and the local ferromagnetic correlations give way to antiferromagnetic correlations at larger distances. These spin correlations are reminiscent of the triple-$q$ structures commonly reported in magnetic skyrmion literature~\cite{Okubo2012}, but because this region does not hug the $\nu=1$ QHFM, these spin correlation features should not be interpreted as signatures of standard quantum Hall skyrmions. 
Increasing field strength further, $(n,\Phi/\Phi_0) = (0.34,0.48)$ (shown in \cref{fig:suscept-sample}\textbf{g}-\textbf{h}) exhibits stronger momentum space spin correlations peaked at $\vec{q} \sim \vec{K}/3$ and oscillatory real space correlations. These features of the high-field metal do not depend on system size, as shown in \cref{fig:qmax-fs,fig:suscept-15x15}. 
While it is unclear if the regions shown in \cref{fig:suscept-sample}\textbf{e}-\textbf{f} and \cref{fig:suscept-sample}\textbf{g}-\textbf{h} belong to distinct phases, the results are different from both the neighboring FM wedge and the Lindhard susceptibility, shown in \cref{fig:nonint-suscept}\textbf{e}-\textbf{h}. As the spin susceptibility peaks shown in \cref{fig:suscept-sample}\textbf{e} and \textbf{g} stay at the same position over a broad range of particle density (green shading near the top of \cref{fig:qmax-Udep}), an description in terms of ``fluctuating stripes'' is also inappropriate.

In the non-interacting Hofstadter model, when $\Phi/\Phi_0 > 1/3$, the lowest Hofstadter band splits into a fractal sequence of subbands, with subband gaps lying on straight lines with slope $C = -4,-8,-12,...$, as shown in \cref{fig:nonint-dos-wannier}. Intuition from the Landau level picture tells us to anticipate incompressible spin polarized states on straight lines with slope $C = -2, -6, -10, ...$, and incompressible singlet states on straight lines with slope $C = -4,-8,-12,...$ in $(n,\Phi/\Phi_0)$ phase space. However, only the $C=-2, s =1$ feature is observed in DMRG simulations. 

To see why states above the $C=-2,s=1$ boundary are largely featureless,
we note the Hofstadter subband gaps are much smaller ($\lesssim 0.2t$) than the $C=1$ gap between main Hofstadter bands ($\sim 3t$). Let's suppose that these subband gaps are comparable to or smaller than the Coulomb exchange interaction, so that Landau level mixing effects between subbands are strong and non-perturbative. While the Coulomb exchange interaction favors splitting up each individual subband (and forming a QHFM when each subband is half-filled), a correlation induced gap can't move states belonging to one subband too far before they ``collide'' with an adjacent subband, which is itself experiencing the effects of Coulomb exchange. The most likely net result of such strong inter-subband mixing is all the sub-bands smear together to form a metal. In our case, it appears to be an anomalous (interaction-induced) metal in which competing tendencies toward singlet and spin-polarized ground states compromise to form small-$q$ spin textures. 
Since Hofstadter subbands as $\Phi/\Phi_0 \rightarrow 0.5$ are not equivalent to Landau levels~\cite{Bauer2022}, the $n\geq -2 (\Phi/\Phi_0) + 1$, $\nu<1$ region is a rich breeding ground for quantum fluids with no counterpart in the continuum~\cite{Barkeshli2012}. Our results thus highlight a parameter region that warrants further scrutiny. 

\section{Conclusions}

In this work, we numerically study the Hubbard-Hofstadter model on a triangular lattice using DQMC and DMRG. We observe a large ferromagnetic wedge for $\nu \leq 1$, quantum Hall ferromagnets at $\nu=1$ and $\nu=3$, 
and low-spin ground states elsewhere. 
We find that whether $\nu=1$ QHFM is observed up to $\Phi/\Phi_0 = 0.5$ depends on the Hubbard interaction strength. The $\nu=1$ QHFM exhibits particle-hole asymmetry upon doping, with ground state spin polarization consistent with skyrmions on the electron-doped side.
Finally, we identify a region of parameter space above the ferromagnetic wedge that hosts small-$q$ spin fluctuations. Interactions in this high magnetic field limit may drive a reorganization of finely spaced Chern bands to heretofore unstudied metallic states with spin texture but no magnetic order~\cite{Dong2022a}. 

Our study shows that unbiased numerical methods applied to interacting lattice Chern bands may lead to results that are dramatically different from expectations based on the continuum limit and uncover fundamentally new physics. 
We expect the confluence of lattice potential, magnetic field, and interactions to produce novel states at a wide range of particle densities, particularly as half-filling is approached and strong correlations dominate the complex insulating phases nearby~\cite{Zhu2022}. 

\section{Data and Code Availability}
Aggregated numerical data and analysis routines required to reproduce the figures can be found at \url{10.5281/zenodo.8339843}. Raw simulation data that support the findings of this study are stored on the Sherlock cluster at Stanford University and are available from the corresponding author upon reasonable request.

The most up-to-date version of our DQMC simulation code can be accessed at \url{https://github.com/katherineding/dqmc-dev}.

\section{Acknowledgements}
We are especially thankful for insightful comments and invaluable suggestions by Steve Kivelson, Shivaji Sondhi, Andrei Bernevig, Hongchen Jiang, Junkai Dong, Vladimir Calvera, and Nishchhal Verma. We are also indebted to helpful discussions with Sankar Das Sarma, Arno Kampf, Xiaoliang Qi, Donna Sheng, Emily Zhang, Jiachen Yu, Kyung-Su Kim, Paul Neves, Tomohiro Soejima, Patrick Ledwith, Daniel Parker, Vedant Dhruv, and Chaitanya Murthy. 

This work was supported by the Center for Quantum Sensing and Quantum Materials, a DOE Energy Frontier Research Center, grant DE-SC0021238 (JKD, PM, PWP, BEF, and TPD). Portions of this work (LY, WOW, BM) were supported by U.S. Department of Energy (DOE), Office of Basic Energy Sciences, Division of Materials Sciences and Engineering.
ZZ was supported by a Stanford Science fellowship. 
CP acknowledges the support of the U.S. Department of Energy, Office of Science, Basic Energy Sciences under Award No. DE-SC0022216. 
EWH was supported by the Gordon and Betty Moore Foundation's EPiQS Initiative through grants GBMF 4305 and GBMF 8691. 
Computational work was performed on the Sherlock cluster at Stanford University and on resources of the National Energy Research Scientific Computing Center, supported by the U.S. DOE, Office of Science, under Contract no. DE-AC02-05CH11231.
\bibliography{main}

\begin{thebibliography}{83}%
\makeatletter
\providecommand \@ifxundefined [1]{%
 \@ifx{#1\undefined}
}%
\providecommand \@ifnum [1]{%
 \ifnum #1\expandafter \@firstoftwo
 \else \expandafter \@secondoftwo
 \fi
}%
\providecommand \@ifx [1]{%
 \ifx #1\expandafter \@firstoftwo
 \else \expandafter \@secondoftwo
 \fi
}%
\providecommand \natexlab [1]{#1}%
\providecommand \enquote  [1]{``#1''}%
\providecommand \bibnamefont  [1]{#1}%
\providecommand \bibfnamefont [1]{#1}%
\providecommand \citenamefont [1]{#1}%
\providecommand \href@noop [0]{\@secondoftwo}%
\providecommand \href [0]{\begingroup \@sanitize@url \@href}%
\providecommand \@href[1]{\@@startlink{#1}\@@href}%
\providecommand \@@href[1]{\endgroup#1\@@endlink}%
\providecommand \@sanitize@url [0]{\catcode `\\12\catcode `\$12\catcode
  `\&12\catcode `\#12\catcode `\^12\catcode `\_12\catcode `\%12\relax}%
\providecommand \@@startlink[1]{}%
\providecommand \@@endlink[0]{}%
\providecommand \url  [0]{\begingroup\@sanitize@url \@url }%
\providecommand \@url [1]{\endgroup\@href {#1}{\urlprefix }}%
\providecommand \urlprefix  [0]{URL }%
\providecommand \Eprint [0]{\href }%
\providecommand \doibase [0]{https://doi.org/}%
\providecommand \selectlanguage [0]{\@gobble}%
\providecommand \bibinfo  [0]{\@secondoftwo}%
\providecommand \bibfield  [0]{\@secondoftwo}%
\providecommand \translation [1]{[#1]}%
\providecommand \BibitemOpen [0]{}%
\providecommand \bibitemStop [0]{}%
\providecommand \bibitemNoStop [0]{.\EOS\space}%
\providecommand \EOS [0]{\spacefactor3000\relax}%
\providecommand \BibitemShut  [1]{\csname bibitem#1\endcsname}%
\let\auto@bib@innerbib\@empty
\bibitem [{\citenamefont {Tong}(2016)}]{Tong2016}%
  \BibitemOpen
  \bibfield  {author} {\bibinfo {author} {\bibfnamefont {D.}~\bibnamefont
  {Tong}},\ }\href@noop {} {\bibinfo {title} {Lectures on the quantum hall
  effect}} (\bibinfo {year} {2016}),\ \Eprint
  {https://arxiv.org/abs/1606.06687} {arXiv:1606.06687 [hep-th]} \BibitemShut
  {NoStop}%
\bibitem [{\citenamefont {Girvin}\ \emph {et~al.}(1986)\citenamefont {Girvin},
  \citenamefont {MacDonald},\ and\ \citenamefont {Platzman}}]{Girvin1986}%
  \BibitemOpen
  \bibfield  {author} {\bibinfo {author} {\bibfnamefont {S.~M.}\ \bibnamefont
  {Girvin}}, \bibinfo {author} {\bibfnamefont {A.~H.}\ \bibnamefont
  {MacDonald}},\ and\ \bibinfo {author} {\bibfnamefont {P.~M.}\ \bibnamefont
  {Platzman}},\ }\bibfield  {title} {\bibinfo {title} {Magneto-roton theory of
  collective excitations in the fractional quantum {{Hall}} effect},\ }\href
  {https://doi.org/10.1103/PhysRevB.33.2481} {\bibfield  {journal} {\bibinfo
  {journal} {Physical Review B}\ }\textbf {\bibinfo {volume} {33}},\ \bibinfo
  {pages} {2481} (\bibinfo {year} {1986})}\BibitemShut {NoStop}%
\bibitem [{\citenamefont {Parameswaran}\ \emph {et~al.}(2012)\citenamefont
  {Parameswaran}, \citenamefont {Roy},\ and\ \citenamefont
  {Sondhi}}]{Parameswaran2012}%
  \BibitemOpen
  \bibfield  {author} {\bibinfo {author} {\bibfnamefont {S.~A.}\ \bibnamefont
  {Parameswaran}}, \bibinfo {author} {\bibfnamefont {R.}~\bibnamefont {Roy}},\
  and\ \bibinfo {author} {\bibfnamefont {S.~L.}\ \bibnamefont {Sondhi}},\
  }\bibfield  {title} {\bibinfo {title} {Fractional {{Chern}} insulators and
  the $\ensuremath{W}_\infty$ algebra},\ }\href
  {https://doi.org/10.1103/PhysRevB.85.241308} {\bibfield  {journal} {\bibinfo
  {journal} {Physical Review B}\ }\textbf {\bibinfo {volume} {85}},\ \bibinfo
  {pages} {241308(R)} (\bibinfo {year} {2012})}\BibitemShut {NoStop}%
\bibitem [{\citenamefont {Roy}(2014)}]{Roy2014}%
  \BibitemOpen
  \bibfield  {author} {\bibinfo {author} {\bibfnamefont {R.}~\bibnamefont
  {Roy}},\ }\bibfield  {title} {\bibinfo {title} {Band geometry of fractional
  topological insulators},\ }\href {https://doi.org/10.1103/PhysRevB.90.165139}
  {\bibfield  {journal} {\bibinfo  {journal} {Phys. Rev. B}\ }\textbf {\bibinfo
  {volume} {90}},\ \bibinfo {pages} {165139} (\bibinfo {year}
  {2014})}\BibitemShut {NoStop}%
\bibitem [{\citenamefont {Claassen}\ \emph {et~al.}(2015)\citenamefont
  {Claassen}, \citenamefont {Lee}, \citenamefont {Thomale}, \citenamefont
  {Qi},\ and\ \citenamefont {Devereaux}}]{Claassen2015}%
  \BibitemOpen
  \bibfield  {author} {\bibinfo {author} {\bibfnamefont {M.}~\bibnamefont
  {Claassen}}, \bibinfo {author} {\bibfnamefont {C.~H.}\ \bibnamefont {Lee}},
  \bibinfo {author} {\bibfnamefont {R.}~\bibnamefont {Thomale}}, \bibinfo
  {author} {\bibfnamefont {X.-L.}\ \bibnamefont {Qi}},\ and\ \bibinfo {author}
  {\bibfnamefont {T.~P.}\ \bibnamefont {Devereaux}},\ }\bibfield  {title}
  {\bibinfo {title} {Position-momentum duality and fractional quantum {Hall}
  effect in {Chern} insulators},\ }\href
  {https://doi.org/10.1103/PhysRevLett.114.236802} {\bibfield  {journal}
  {\bibinfo  {journal} {Phys. Rev. Lett.}\ }\textbf {\bibinfo {volume} {114}},\
  \bibinfo {pages} {236802} (\bibinfo {year} {2015})}\BibitemShut {NoStop}%
\bibitem [{\citenamefont {Wang}\ \emph {et~al.}(2021)\citenamefont {Wang},
  \citenamefont {Cano}, \citenamefont {Millis}, \citenamefont {Liu},\ and\
  \citenamefont {Yang}}]{Wang2021}%
  \BibitemOpen
  \bibfield  {author} {\bibinfo {author} {\bibfnamefont {J.}~\bibnamefont
  {Wang}}, \bibinfo {author} {\bibfnamefont {J.}~\bibnamefont {Cano}}, \bibinfo
  {author} {\bibfnamefont {A.~J.}\ \bibnamefont {Millis}}, \bibinfo {author}
  {\bibfnamefont {Z.}~\bibnamefont {Liu}},\ and\ \bibinfo {author}
  {\bibfnamefont {B.}~\bibnamefont {Yang}},\ }\bibfield  {title} {\bibinfo
  {title} {Exact {Landau} level description of geometry and interaction in a
  flatband},\ }\href {https://doi.org/10.1103/PhysRevLett.127.246403}
  {\bibfield  {journal} {\bibinfo  {journal} {Physical Review Letters}\
  }\textbf {\bibinfo {volume} {127}},\ \bibinfo {pages} {246403} (\bibinfo
  {year} {2021})}\BibitemShut {NoStop}%
\bibitem [{\citenamefont {Ledwith}\ \emph {et~al.}(2022)\citenamefont
  {Ledwith}, \citenamefont {Vishwanath},\ and\ \citenamefont
  {Parker}}]{Ledwith2022}%
  \BibitemOpen
  \bibfield  {author} {\bibinfo {author} {\bibfnamefont {P.~J.}\ \bibnamefont
  {Ledwith}}, \bibinfo {author} {\bibfnamefont {A.}~\bibnamefont
  {Vishwanath}},\ and\ \bibinfo {author} {\bibfnamefont {D.~E.}\ \bibnamefont
  {Parker}},\ }\href@noop {} {\bibinfo {title} {Vortexability: A unifying
  criterion for ideal fractional {Chern} insulators}} (\bibinfo {year}
  {2022}),\ \Eprint {https://arxiv.org/abs/2209.15023} {arxiv:2209.15023}
  \BibitemShut {NoStop}%
\bibitem [{\citenamefont {Hunt}\ \emph {et~al.}(2013)\citenamefont {Hunt},
  \citenamefont {{Sanchez-Yamagishi}}, \citenamefont {Young}, \citenamefont
  {Yankowitz}, \citenamefont {LeRoy}, \citenamefont {Watanabe}, \citenamefont
  {Taniguchi}, \citenamefont {Moon}, \citenamefont {Koshino}, \citenamefont
  {{Jarillo-Herrero}},\ and\ \citenamefont {Ashoori}}]{Hunt2013}%
  \BibitemOpen
  \bibfield  {author} {\bibinfo {author} {\bibfnamefont {B.}~\bibnamefont
  {Hunt}}, \bibinfo {author} {\bibfnamefont {J.~D.}\ \bibnamefont
  {{Sanchez-Yamagishi}}}, \bibinfo {author} {\bibfnamefont {A.~F.}\
  \bibnamefont {Young}}, \bibinfo {author} {\bibfnamefont {M.}~\bibnamefont
  {Yankowitz}}, \bibinfo {author} {\bibfnamefont {B.~J.}\ \bibnamefont
  {LeRoy}}, \bibinfo {author} {\bibfnamefont {K.}~\bibnamefont {Watanabe}},
  \bibinfo {author} {\bibfnamefont {T.}~\bibnamefont {Taniguchi}}, \bibinfo
  {author} {\bibfnamefont {P.}~\bibnamefont {Moon}}, \bibinfo {author}
  {\bibfnamefont {M.}~\bibnamefont {Koshino}}, \bibinfo {author} {\bibfnamefont
  {P.}~\bibnamefont {{Jarillo-Herrero}}},\ and\ \bibinfo {author}
  {\bibfnamefont {R.~C.}\ \bibnamefont {Ashoori}},\ }\bibfield  {title}
  {\bibinfo {title} {Massive {{Dirac}} fermions and {{Hofstadter}} butterfly in
  a van {{Der Waals}} heterostructure},\ }\href
  {https://doi.org/10.1126/science.1237240} {\bibfield  {journal} {\bibinfo
  {journal} {Science}\ }\textbf {\bibinfo {volume} {340}},\ \bibinfo {pages}
  {1427} (\bibinfo {year} {2013})}\BibitemShut {NoStop}%
\bibitem [{\citenamefont {Hofstadter}(1976)}]{Hofstadter1976}%
  \BibitemOpen
  \bibfield  {author} {\bibinfo {author} {\bibfnamefont {D.~R.}\ \bibnamefont
  {Hofstadter}},\ }\bibfield  {title} {\bibinfo {title} {Energy levels and wave
  functions of {{Bloch}} electrons in rational and irrational magnetic
  fields},\ }\href {https://doi.org/10.1103/PhysRevB.14.2239} {\bibfield
  {journal} {\bibinfo  {journal} {Physical Review B}\ }\textbf {\bibinfo
  {volume} {14}},\ \bibinfo {pages} {2239} (\bibinfo {year}
  {1976})}\BibitemShut {NoStop}%
\bibitem [{\citenamefont {Wannier}(1978)}]{Wannier1978}%
  \BibitemOpen
  \bibfield  {author} {\bibinfo {author} {\bibfnamefont {G.~H.}\ \bibnamefont
  {Wannier}},\ }\bibfield  {title} {\bibinfo {title} {A result not dependent on
  rationality for {Bloch} electrons in a magnetic field},\ }\href
  {https://doi.org/10.1002/pssb.2220880243} {\bibfield  {journal} {\bibinfo
  {journal} {Physica Status Solidi (b)}\ }\textbf {\bibinfo {volume} {88}},\
  \bibinfo {pages} {757} (\bibinfo {year} {1978})}\BibitemShut {NoStop}%
\bibitem [{\citenamefont {Zak}(1964)}]{Zak1964}%
  \BibitemOpen
  \bibfield  {author} {\bibinfo {author} {\bibfnamefont {J.}~\bibnamefont
  {Zak}},\ }\bibfield  {title} {\bibinfo {title} {Magnetic translation group},\
  }\href {https://doi.org/10.1103/PhysRev.134.A1602} {\bibfield  {journal}
  {\bibinfo  {journal} {Physical Review}\ }\textbf {\bibinfo {volume} {134}},\
  \bibinfo {pages} {A1602} (\bibinfo {year} {1964})}\BibitemShut {NoStop}%
\bibitem [{\citenamefont {Xiao}\ \emph {et~al.}(2010)\citenamefont {Xiao},
  \citenamefont {Chang},\ and\ \citenamefont {Niu}}]{Xiao2010}%
  \BibitemOpen
  \bibfield  {author} {\bibinfo {author} {\bibfnamefont {D.}~\bibnamefont
  {Xiao}}, \bibinfo {author} {\bibfnamefont {M.-C.}\ \bibnamefont {Chang}},\
  and\ \bibinfo {author} {\bibfnamefont {Q.}~\bibnamefont {Niu}},\ }\bibfield
  {title} {\bibinfo {title} {Berry phase effects on electronic properties},\
  }\href {https://doi.org/10.1103/RevModPhys.82.1959} {\bibfield  {journal}
  {\bibinfo  {journal} {Rev. Mod. Phys.}\ }\textbf {\bibinfo {volume} {82}},\
  \bibinfo {pages} {1959} (\bibinfo {year} {2010})}\BibitemShut {NoStop}%
\bibitem [{\citenamefont {Kapit}\ and\ \citenamefont
  {Mueller}(2010)}]{Kapit2010}%
  \BibitemOpen
  \bibfield  {author} {\bibinfo {author} {\bibfnamefont {E.}~\bibnamefont
  {Kapit}}\ and\ \bibinfo {author} {\bibfnamefont {E.}~\bibnamefont
  {Mueller}},\ }\bibfield  {title} {\bibinfo {title} {Exact parent hamiltonian
  for the quantum {Hall} states in a lattice},\ }\href
  {https://doi.org/10.1103/PhysRevLett.105.215303} {\bibfield  {journal}
  {\bibinfo  {journal} {Physical Review Letters}\ }\textbf {\bibinfo {volume}
  {105}},\ \bibinfo {pages} {215303} (\bibinfo {year} {2010})}\BibitemShut
  {NoStop}%
\bibitem [{\citenamefont {Bauer}\ \emph {et~al.}(2016)\citenamefont {Bauer},
  \citenamefont {Jackson},\ and\ \citenamefont {Roy}}]{Bauer2016}%
  \BibitemOpen
  \bibfield  {author} {\bibinfo {author} {\bibfnamefont {D.}~\bibnamefont
  {Bauer}}, \bibinfo {author} {\bibfnamefont {T.~S.}\ \bibnamefont {Jackson}},\
  and\ \bibinfo {author} {\bibfnamefont {R.}~\bibnamefont {Roy}},\ }\bibfield
  {title} {\bibinfo {title} {Quantum geometry and stability of the fractional
  quantum {Hall} effect in the {Hofstadter} model},\ }\href
  {https://doi.org/10.1103/PhysRevB.93.235133} {\bibfield  {journal} {\bibinfo
  {journal} {Physical Review B}\ }\textbf {\bibinfo {volume} {93}},\ \bibinfo
  {pages} {235133} (\bibinfo {year} {2016})}\BibitemShut {NoStop}%
\bibitem [{\citenamefont {Regnault}\ and\ \citenamefont
  {Bernevig}(2011)}]{Regnault2011}%
  \BibitemOpen
  \bibfield  {author} {\bibinfo {author} {\bibfnamefont {N.}~\bibnamefont
  {Regnault}}\ and\ \bibinfo {author} {\bibfnamefont {B.~A.}\ \bibnamefont
  {Bernevig}},\ }\bibfield  {title} {\bibinfo {title} {Fractional {Chern}
  insulator},\ }\href {https://doi.org/10.1103/PhysRevX.1.021014} {\bibfield
  {journal} {\bibinfo  {journal} {Physical Review X}\ }\textbf {\bibinfo
  {volume} {1}},\ \bibinfo {pages} {021014} (\bibinfo {year}
  {2011})}\BibitemShut {NoStop}%
\bibitem [{\citenamefont {Sheng}\ \emph {et~al.}(2011)\citenamefont {Sheng},
  \citenamefont {Gu}, \citenamefont {Sun},\ and\ \citenamefont
  {Sheng}}]{Sheng2011}%
  \BibitemOpen
  \bibfield  {author} {\bibinfo {author} {\bibfnamefont {D.~N.}\ \bibnamefont
  {Sheng}}, \bibinfo {author} {\bibfnamefont {Z.-C.}\ \bibnamefont {Gu}},
  \bibinfo {author} {\bibfnamefont {K.}~\bibnamefont {Sun}},\ and\ \bibinfo
  {author} {\bibfnamefont {L.}~\bibnamefont {Sheng}},\ }\bibfield  {title}
  {\bibinfo {title} {Fractional quantum {{Hall}} effect in the absence of
  {{Landau}} levels},\ }\href {https://doi.org/10.1038/ncomms1380} {\bibfield
  {journal} {\bibinfo  {journal} {Nature Communications}\ }\textbf {\bibinfo
  {volume} {2}},\ \bibinfo {pages} {389} (\bibinfo {year} {2011})}\BibitemShut
  {NoStop}%
\bibitem [{\citenamefont {Qi}(2011)}]{Qi2011}%
  \BibitemOpen
  \bibfield  {author} {\bibinfo {author} {\bibfnamefont {X.-L.}\ \bibnamefont
  {Qi}},\ }\bibfield  {title} {\bibinfo {title} {Generic wave-function
  description of fractional quantum anomalous {Hall} states and fractional
  topological insulators},\ }\href
  {https://doi.org/10.1103/PhysRevLett.107.126803} {\bibfield  {journal}
  {\bibinfo  {journal} {Physical Review Letters}\ }\textbf {\bibinfo {volume}
  {107}},\ \bibinfo {pages} {126803} (\bibinfo {year} {2011})}\BibitemShut
  {NoStop}%
\bibitem [{\citenamefont {Scaffidi}\ and\ \citenamefont
  {M\"oller}(2012)}]{Scaffidi2012}%
  \BibitemOpen
  \bibfield  {author} {\bibinfo {author} {\bibfnamefont {T.}~\bibnamefont
  {Scaffidi}}\ and\ \bibinfo {author} {\bibfnamefont {G.}~\bibnamefont
  {M\"oller}},\ }\bibfield  {title} {\bibinfo {title} {Adiabatic continuation
  of fractional {Chern} insulators to fractional quantum {Hall} states},\
  }\href {https://doi.org/10.1103/PhysRevLett.109.246805} {\bibfield  {journal}
  {\bibinfo  {journal} {Phys. Rev. Lett.}\ }\textbf {\bibinfo {volume} {109}},\
  \bibinfo {pages} {246805} (\bibinfo {year} {2012})}\BibitemShut {NoStop}%
\bibitem [{\citenamefont {M{\"o}ller}\ and\ \citenamefont
  {Cooper}(2015)}]{Moller2015}%
  \BibitemOpen
  \bibfield  {author} {\bibinfo {author} {\bibfnamefont {G.}~\bibnamefont
  {M{\"o}ller}}\ and\ \bibinfo {author} {\bibfnamefont {N.~R.}\ \bibnamefont
  {Cooper}},\ }\bibfield  {title} {\bibinfo {title} {Fractional {Chern}
  insulators in {{Harper-Hofstadter}} bands with higher {Chern} number},\
  }\href {https://doi.org/10.1103/PhysRevLett.115.126401} {\bibfield  {journal}
  {\bibinfo  {journal} {Physical Review Letters}\ }\textbf {\bibinfo {volume}
  {115}},\ \bibinfo {pages} {126401} (\bibinfo {year} {2015})}\BibitemShut
  {NoStop}%
\bibitem [{\citenamefont {Andrews}\ and\ \citenamefont
  {M{\"o}ller}(2018)}]{Andrews2018}%
  \BibitemOpen
  \bibfield  {author} {\bibinfo {author} {\bibfnamefont {B.}~\bibnamefont
  {Andrews}}\ and\ \bibinfo {author} {\bibfnamefont {G.}~\bibnamefont
  {M{\"o}ller}},\ }\bibfield  {title} {\bibinfo {title} {Stability of
  fractional {Chern} insulators in the effective continuum limit of
  {{Harper-Hofstadter}} bands with {Chern} number $|{C}| >1$},\ }\href
  {https://doi.org/10.1103/PhysRevB.97.035159} {\bibfield  {journal} {\bibinfo
  {journal} {Physical Review B}\ }\textbf {\bibinfo {volume} {97}},\ \bibinfo
  {pages} {035159} (\bibinfo {year} {2018})}\BibitemShut {NoStop}%
\bibitem [{\citenamefont {Andrews}\ \emph {et~al.}(2021)\citenamefont
  {Andrews}, \citenamefont {Neupert},\ and\ \citenamefont
  {M{\"o}ller}}]{Andrews2021}%
  \BibitemOpen
  \bibfield  {author} {\bibinfo {author} {\bibfnamefont {B.}~\bibnamefont
  {Andrews}}, \bibinfo {author} {\bibfnamefont {T.}~\bibnamefont {Neupert}},\
  and\ \bibinfo {author} {\bibfnamefont {G.}~\bibnamefont {M{\"o}ller}},\
  }\bibfield  {title} {\bibinfo {title} {Stability, phase transitions, and
  numerical breakdown of fractional {Chern} insulators in higher {Chern} bands
  of the {Hofstadter} model},\ }\href
  {https://doi.org/10.1103/PhysRevB.104.125107} {\bibfield  {journal} {\bibinfo
   {journal} {Physical Review B}\ }\textbf {\bibinfo {volume} {104}},\ \bibinfo
  {pages} {125107} (\bibinfo {year} {2021})}\BibitemShut {NoStop}%
\bibitem [{\citenamefont {Bauer}\ \emph {et~al.}(2022)\citenamefont {Bauer},
  \citenamefont {Talkington}, \citenamefont {Harper}, \citenamefont {Andrews},\
  and\ \citenamefont {Roy}}]{Bauer2022}%
  \BibitemOpen
  \bibfield  {author} {\bibinfo {author} {\bibfnamefont {D.}~\bibnamefont
  {Bauer}}, \bibinfo {author} {\bibfnamefont {S.}~\bibnamefont {Talkington}},
  \bibinfo {author} {\bibfnamefont {F.}~\bibnamefont {Harper}}, \bibinfo
  {author} {\bibfnamefont {B.}~\bibnamefont {Andrews}},\ and\ \bibinfo {author}
  {\bibfnamefont {R.}~\bibnamefont {Roy}},\ }\bibfield  {title} {\bibinfo
  {title} {Fractional {Chern} insulators with a non-{Landau} level continuum
  limit},\ }\href {https://doi.org/10.1103/PhysRevB.105.045144} {\bibfield
  {journal} {\bibinfo  {journal} {Physical Review B}\ }\textbf {\bibinfo
  {volume} {105}},\ \bibinfo {pages} {045144} (\bibinfo {year}
  {2022})}\BibitemShut {NoStop}%
\bibitem [{\citenamefont {Sondhi}\ \emph {et~al.}(1993)\citenamefont {Sondhi},
  \citenamefont {Karlhede}, \citenamefont {Kivelson},\ and\ \citenamefont
  {Rezayi}}]{Sondhi1993}%
  \BibitemOpen
  \bibfield  {author} {\bibinfo {author} {\bibfnamefont {S.~L.}\ \bibnamefont
  {Sondhi}}, \bibinfo {author} {\bibfnamefont {A.}~\bibnamefont {Karlhede}},
  \bibinfo {author} {\bibfnamefont {S.~A.}\ \bibnamefont {Kivelson}},\ and\
  \bibinfo {author} {\bibfnamefont {E.~H.}\ \bibnamefont {Rezayi}},\ }\bibfield
   {title} {\bibinfo {title} {Skyrmions and the crossover from the integer to
  fractional quantum {Hall} effect at small {Zeeman} energies},\ }\href
  {https://doi.org/10.1103/PhysRevB.47.16419} {\bibfield  {journal} {\bibinfo
  {journal} {Phys. Rev. B}\ }\textbf {\bibinfo {volume} {47}},\ \bibinfo
  {pages} {16419} (\bibinfo {year} {1993})}\BibitemShut {NoStop}%
\bibitem [{\citenamefont {Nuckolls}\ \emph {et~al.}(2020)\citenamefont
  {Nuckolls}, \citenamefont {Oh}, \citenamefont {Wong}, \citenamefont {Lian},
  \citenamefont {Watanabe}, \citenamefont {Taniguchi}, \citenamefont
  {Bernevig},\ and\ \citenamefont {Yazdani}}]{Nuckolls2020}%
  \BibitemOpen
  \bibfield  {author} {\bibinfo {author} {\bibfnamefont {K.~P.}\ \bibnamefont
  {Nuckolls}}, \bibinfo {author} {\bibfnamefont {M.}~\bibnamefont {Oh}},
  \bibinfo {author} {\bibfnamefont {D.}~\bibnamefont {Wong}}, \bibinfo {author}
  {\bibfnamefont {B.}~\bibnamefont {Lian}}, \bibinfo {author} {\bibfnamefont
  {K.}~\bibnamefont {Watanabe}}, \bibinfo {author} {\bibfnamefont
  {T.}~\bibnamefont {Taniguchi}}, \bibinfo {author} {\bibfnamefont {B.~A.}\
  \bibnamefont {Bernevig}},\ and\ \bibinfo {author} {\bibfnamefont
  {A.}~\bibnamefont {Yazdani}},\ }\bibfield  {title} {\bibinfo {title}
  {Strongly correlated {Chern} insulators in magic-angle twisted bilayer
  graphene},\ }\href {https://doi.org/10.1038/s41586-020-3028-8} {\bibfield
  {journal} {\bibinfo  {journal} {Nature}\ }\textbf {\bibinfo {volume} {588}},\
  \bibinfo {pages} {610} (\bibinfo {year} {2020})}\BibitemShut {NoStop}%
\bibitem [{\citenamefont {Choi}\ \emph {et~al.}(2021)\citenamefont {Choi},
  \citenamefont {Kim}, \citenamefont {Peng}, \citenamefont {Thomson},
  \citenamefont {Lewandowski}, \citenamefont {Polski}, \citenamefont {Zhang},
  \citenamefont {Arora}, \citenamefont {Watanabe}, \citenamefont {Taniguchi},
  \citenamefont {Alicea},\ and\ \citenamefont {{Nadj-Perge}}}]{Choi2021}%
  \BibitemOpen
  \bibfield  {author} {\bibinfo {author} {\bibfnamefont {Y.}~\bibnamefont
  {Choi}}, \bibinfo {author} {\bibfnamefont {H.}~\bibnamefont {Kim}}, \bibinfo
  {author} {\bibfnamefont {Y.}~\bibnamefont {Peng}}, \bibinfo {author}
  {\bibfnamefont {A.}~\bibnamefont {Thomson}}, \bibinfo {author} {\bibfnamefont
  {C.}~\bibnamefont {Lewandowski}}, \bibinfo {author} {\bibfnamefont
  {R.}~\bibnamefont {Polski}}, \bibinfo {author} {\bibfnamefont
  {Y.}~\bibnamefont {Zhang}}, \bibinfo {author} {\bibfnamefont {H.~S.}\
  \bibnamefont {Arora}}, \bibinfo {author} {\bibfnamefont {K.}~\bibnamefont
  {Watanabe}}, \bibinfo {author} {\bibfnamefont {T.}~\bibnamefont {Taniguchi}},
  \bibinfo {author} {\bibfnamefont {J.}~\bibnamefont {Alicea}},\ and\ \bibinfo
  {author} {\bibfnamefont {S.}~\bibnamefont {{Nadj-Perge}}},\ }\bibfield
  {title} {\bibinfo {title} {Correlation-driven topological phases in
  magic-angle twisted bilayer graphene},\ }\href
  {https://doi.org/10.1038/s41586-020-03159-7} {\bibfield  {journal} {\bibinfo
  {journal} {Nature}\ }\textbf {\bibinfo {volume} {589}},\ \bibinfo {pages}
  {536} (\bibinfo {year} {2021})}\BibitemShut {NoStop}%
\bibitem [{\citenamefont {Park}\ \emph {et~al.}(2021)\citenamefont {Park},
  \citenamefont {Cao}, \citenamefont {Watanabe}, \citenamefont {Taniguchi},\
  and\ \citenamefont {{Jarillo-Herrero}}}]{Park2021}%
  \BibitemOpen
  \bibfield  {author} {\bibinfo {author} {\bibfnamefont {J.~M.}\ \bibnamefont
  {Park}}, \bibinfo {author} {\bibfnamefont {Y.}~\bibnamefont {Cao}}, \bibinfo
  {author} {\bibfnamefont {K.}~\bibnamefont {Watanabe}}, \bibinfo {author}
  {\bibfnamefont {T.}~\bibnamefont {Taniguchi}},\ and\ \bibinfo {author}
  {\bibfnamefont {P.}~\bibnamefont {{Jarillo-Herrero}}},\ }\bibfield  {title}
  {\bibinfo {title} {Flavour {Hund}'s coupling, {Chern} gaps and charge
  diffusivity in moir\'e graphene},\ }\href
  {https://doi.org/10.1038/s41586-021-03366-w} {\bibfield  {journal} {\bibinfo
  {journal} {Nature}\ }\textbf {\bibinfo {volume} {592}},\ \bibinfo {pages}
  {43} (\bibinfo {year} {2021})}\BibitemShut {NoStop}%
\bibitem [{\citenamefont {Das}\ \emph {et~al.}(2021)\citenamefont {Das},
  \citenamefont {Lu}, \citenamefont {{Herzog-Arbeitman}}, \citenamefont {Song},
  \citenamefont {Watanabe}, \citenamefont {Taniguchi}, \citenamefont
  {Bernevig},\ and\ \citenamefont {Efetov}}]{Das2021}%
  \BibitemOpen
  \bibfield  {author} {\bibinfo {author} {\bibfnamefont {I.}~\bibnamefont
  {Das}}, \bibinfo {author} {\bibfnamefont {X.}~\bibnamefont {Lu}}, \bibinfo
  {author} {\bibfnamefont {J.}~\bibnamefont {{Herzog-Arbeitman}}}, \bibinfo
  {author} {\bibfnamefont {Z.-D.}\ \bibnamefont {Song}}, \bibinfo {author}
  {\bibfnamefont {K.}~\bibnamefont {Watanabe}}, \bibinfo {author}
  {\bibfnamefont {T.}~\bibnamefont {Taniguchi}}, \bibinfo {author}
  {\bibfnamefont {B.~A.}\ \bibnamefont {Bernevig}},\ and\ \bibinfo {author}
  {\bibfnamefont {D.~K.}\ \bibnamefont {Efetov}},\ }\bibfield  {title}
  {\bibinfo {title} {Symmetry-broken {{Chern}} insulators and {{Rashba-like
  Landau-level}} crossings in magic-angle bilayer graphene},\ }\href
  {https://doi.org/10.1038/s41567-021-01186-3} {\bibfield  {journal} {\bibinfo
  {journal} {Nature Physics}\ }\textbf {\bibinfo {volume} {17}},\ \bibinfo
  {pages} {710} (\bibinfo {year} {2021})}\BibitemShut {NoStop}%
\bibitem [{\citenamefont {Wu}\ \emph {et~al.}(2021)\citenamefont {Wu},
  \citenamefont {Zhang}, \citenamefont {Watanabe}, \citenamefont {Taniguchi},\
  and\ \citenamefont {Andrei}}]{Wu2021}%
  \BibitemOpen
  \bibfield  {author} {\bibinfo {author} {\bibfnamefont {S.}~\bibnamefont
  {Wu}}, \bibinfo {author} {\bibfnamefont {Z.}~\bibnamefont {Zhang}}, \bibinfo
  {author} {\bibfnamefont {K.}~\bibnamefont {Watanabe}}, \bibinfo {author}
  {\bibfnamefont {T.}~\bibnamefont {Taniguchi}},\ and\ \bibinfo {author}
  {\bibfnamefont {E.~Y.}\ \bibnamefont {Andrei}},\ }\bibfield  {title}
  {\bibinfo {title} {Chern insulators, van {Hove} singularities and topological
  flat bands in magic-angle twisted bilayer graphene},\ }\href
  {https://doi.org/10.1038/s41563-020-00911-2} {\bibfield  {journal} {\bibinfo
  {journal} {Nature Materials}\ }\textbf {\bibinfo {volume} {20}},\ \bibinfo
  {pages} {488} (\bibinfo {year} {2021})}\BibitemShut {NoStop}%
\bibitem [{\citenamefont {Saito}\ \emph {et~al.}(2021)\citenamefont {Saito},
  \citenamefont {Ge}, \citenamefont {Rademaker}, \citenamefont {Watanabe},
  \citenamefont {Taniguchi}, \citenamefont {Abanin},\ and\ \citenamefont
  {Young}}]{Saito2021}%
  \BibitemOpen
  \bibfield  {author} {\bibinfo {author} {\bibfnamefont {Y.}~\bibnamefont
  {Saito}}, \bibinfo {author} {\bibfnamefont {J.}~\bibnamefont {Ge}}, \bibinfo
  {author} {\bibfnamefont {L.}~\bibnamefont {Rademaker}}, \bibinfo {author}
  {\bibfnamefont {K.}~\bibnamefont {Watanabe}}, \bibinfo {author}
  {\bibfnamefont {T.}~\bibnamefont {Taniguchi}}, \bibinfo {author}
  {\bibfnamefont {D.~A.}\ \bibnamefont {Abanin}},\ and\ \bibinfo {author}
  {\bibfnamefont {A.~F.}\ \bibnamefont {Young}},\ }\bibfield  {title} {\bibinfo
  {title} {Hofstadter subband ferromagnetism and symmetry-broken {{Chern}}
  insulators in twisted bilayer graphene},\ }\href
  {https://doi.org/10.1038/s41567-020-01129-4} {\bibfield  {journal} {\bibinfo
  {journal} {Nature Physics}\ }\textbf {\bibinfo {volume} {17}},\ \bibinfo
  {pages} {478} (\bibinfo {year} {2021})}\BibitemShut {NoStop}%
\bibitem [{\citenamefont {Stepanov}\ \emph {et~al.}(2021)\citenamefont
  {Stepanov}, \citenamefont {Xie}, \citenamefont {Taniguchi}, \citenamefont
  {Watanabe}, \citenamefont {Lu}, \citenamefont {MacDonald}, \citenamefont
  {Bernevig},\ and\ \citenamefont {Efetov}}]{Stepanov2021}%
  \BibitemOpen
  \bibfield  {author} {\bibinfo {author} {\bibfnamefont {P.}~\bibnamefont
  {Stepanov}}, \bibinfo {author} {\bibfnamefont {M.}~\bibnamefont {Xie}},
  \bibinfo {author} {\bibfnamefont {T.}~\bibnamefont {Taniguchi}}, \bibinfo
  {author} {\bibfnamefont {K.}~\bibnamefont {Watanabe}}, \bibinfo {author}
  {\bibfnamefont {X.}~\bibnamefont {Lu}}, \bibinfo {author} {\bibfnamefont
  {A.~H.}\ \bibnamefont {MacDonald}}, \bibinfo {author} {\bibfnamefont {B.~A.}\
  \bibnamefont {Bernevig}},\ and\ \bibinfo {author} {\bibfnamefont {D.~K.}\
  \bibnamefont {Efetov}},\ }\bibfield  {title} {\bibinfo {title} {Competing
  zero-field {Chern} insulators in superconducting twisted bilayer graphene},\
  }\href {https://doi.org/10.1103/PhysRevLett.127.197701} {\bibfield  {journal}
  {\bibinfo  {journal} {Physical Review Letters}\ }\textbf {\bibinfo {volume}
  {127}},\ \bibinfo {pages} {197701} (\bibinfo {year} {2021})}\BibitemShut
  {NoStop}%
\bibitem [{\citenamefont {Yu}\ \emph {et~al.}(2022)\citenamefont {Yu},
  \citenamefont {Foutty}, \citenamefont {Han}, \citenamefont {Barber},
  \citenamefont {Schattner}, \citenamefont {Watanabe}, \citenamefont
  {Taniguchi}, \citenamefont {Phillips}, \citenamefont {Shen}, \citenamefont
  {Kivelson},\ and\ \citenamefont {Feldman}}]{Yu2022}%
  \BibitemOpen
  \bibfield  {author} {\bibinfo {author} {\bibfnamefont {J.}~\bibnamefont
  {Yu}}, \bibinfo {author} {\bibfnamefont {B.~A.}\ \bibnamefont {Foutty}},
  \bibinfo {author} {\bibfnamefont {Z.}~\bibnamefont {Han}}, \bibinfo {author}
  {\bibfnamefont {M.~E.}\ \bibnamefont {Barber}}, \bibinfo {author}
  {\bibfnamefont {Y.}~\bibnamefont {Schattner}}, \bibinfo {author}
  {\bibfnamefont {K.}~\bibnamefont {Watanabe}}, \bibinfo {author}
  {\bibfnamefont {T.}~\bibnamefont {Taniguchi}}, \bibinfo {author}
  {\bibfnamefont {P.}~\bibnamefont {Phillips}}, \bibinfo {author}
  {\bibfnamefont {Z.-X.}\ \bibnamefont {Shen}}, \bibinfo {author}
  {\bibfnamefont {S.~A.}\ \bibnamefont {Kivelson}},\ and\ \bibinfo {author}
  {\bibfnamefont {B.~E.}\ \bibnamefont {Feldman}},\ }\bibfield  {title}
  {\bibinfo {title} {Correlated {Hofstadter} spectrum and flavour phase diagram
  in magic-angle twisted bilayer graphene},\ }\href
  {https://doi.org/10.1038/s41567-022-01589-w} {\bibfield  {journal} {\bibinfo
  {journal} {Nature Physics}\ }\textbf {\bibinfo {volume} {18}},\ \bibinfo
  {pages} {825} (\bibinfo {year} {2022})}\BibitemShut {NoStop}%
\bibitem [{\citenamefont {Yu}\ \emph {et~al.}(2023)\citenamefont {Yu},
  \citenamefont {Foutty}, \citenamefont {Kwan}, \citenamefont {Barber},
  \citenamefont {Watanabe}, \citenamefont {Taniguchi}, \citenamefont {Shen},
  \citenamefont {Parameswaran},\ and\ \citenamefont {Feldman}}]{yu2022spin}%
  \BibitemOpen
  \bibfield  {author} {\bibinfo {author} {\bibfnamefont {J.}~\bibnamefont
  {Yu}}, \bibinfo {author} {\bibfnamefont {B.~A.}\ \bibnamefont {Foutty}},
  \bibinfo {author} {\bibfnamefont {Y.~H.}\ \bibnamefont {Kwan}}, \bibinfo
  {author} {\bibfnamefont {M.~E.}\ \bibnamefont {Barber}}, \bibinfo {author}
  {\bibfnamefont {K.}~\bibnamefont {Watanabe}}, \bibinfo {author}
  {\bibfnamefont {T.}~\bibnamefont {Taniguchi}}, \bibinfo {author}
  {\bibfnamefont {Z.-X.}\ \bibnamefont {Shen}}, \bibinfo {author}
  {\bibfnamefont {S.~A.}\ \bibnamefont {Parameswaran}},\ and\ \bibinfo {author}
  {\bibfnamefont {B.~E.}\ \bibnamefont {Feldman}},\ }\bibfield  {title}
  {\bibinfo {title} {Spin skyrmion gaps as signatures of strong-coupling
  insulators in magic-angle twisted bilayer graphene},\ }\href
  {https://doi.org/10.1038/s41467-023-42275-6} {\bibfield  {journal} {\bibinfo
  {journal} {Nature Communications}\ }\textbf {\bibinfo {volume} {14}},\
  \bibinfo {pages} {6679} (\bibinfo {year} {2023})}\BibitemShut {NoStop}%
\bibitem [{\citenamefont {Chatterjee}\ \emph {et~al.}(2020)\citenamefont
  {Chatterjee}, \citenamefont {Bultinck},\ and\ \citenamefont
  {Zaletel}}]{Chatterjee2020}%
  \BibitemOpen
  \bibfield  {author} {\bibinfo {author} {\bibfnamefont {S.}~\bibnamefont
  {Chatterjee}}, \bibinfo {author} {\bibfnamefont {N.}~\bibnamefont
  {Bultinck}},\ and\ \bibinfo {author} {\bibfnamefont {M.~P.}\ \bibnamefont
  {Zaletel}},\ }\bibfield  {title} {\bibinfo {title} {Symmetry breaking and
  skyrmionic transport in twisted bilayer graphene},\ }\href
  {https://doi.org/10.1103/PhysRevB.101.165141} {\bibfield  {journal} {\bibinfo
   {journal} {Physical Review B}\ }\textbf {\bibinfo {volume} {101}},\ \bibinfo
  {pages} {165141} (\bibinfo {year} {2020})}\BibitemShut {NoStop}%
\bibitem [{\citenamefont {Khalaf}\ \emph {et~al.}(2021)\citenamefont {Khalaf},
  \citenamefont {Chatterjee}, \citenamefont {Bultinck}, \citenamefont
  {Zaletel},\ and\ \citenamefont {Vishwanath}}]{Khalaf2021}%
  \BibitemOpen
  \bibfield  {author} {\bibinfo {author} {\bibfnamefont {E.}~\bibnamefont
  {Khalaf}}, \bibinfo {author} {\bibfnamefont {S.}~\bibnamefont {Chatterjee}},
  \bibinfo {author} {\bibfnamefont {N.}~\bibnamefont {Bultinck}}, \bibinfo
  {author} {\bibfnamefont {M.~P.}\ \bibnamefont {Zaletel}},\ and\ \bibinfo
  {author} {\bibfnamefont {A.}~\bibnamefont {Vishwanath}},\ }\bibfield  {title}
  {\bibinfo {title} {Charged skyrmions and topological origin of
  superconductivity in magic-angle graphene},\ }\href
  {https://doi.org/10.1126/sciadv.abf5299} {\bibfield  {journal} {\bibinfo
  {journal} {Science Advances}\ }\textbf {\bibinfo {volume} {7}},\ \bibinfo
  {pages} {eabf5299} (\bibinfo {year} {2021})}\BibitemShut {NoStop}%
\bibitem [{\citenamefont {Kwan}\ \emph {et~al.}(2022)\citenamefont {Kwan},
  \citenamefont {Wagner}, \citenamefont {Bultinck}, \citenamefont {Simon},\
  and\ \citenamefont {Parameswaran}}]{Kwan2022}%
  \BibitemOpen
  \bibfield  {author} {\bibinfo {author} {\bibfnamefont {Y.~H.}\ \bibnamefont
  {Kwan}}, \bibinfo {author} {\bibfnamefont {G.}~\bibnamefont {Wagner}},
  \bibinfo {author} {\bibfnamefont {N.}~\bibnamefont {Bultinck}}, \bibinfo
  {author} {\bibfnamefont {S.~H.}\ \bibnamefont {Simon}},\ and\ \bibinfo
  {author} {\bibfnamefont {S.~A.}\ \bibnamefont {Parameswaran}},\ }\bibfield
  {title} {\bibinfo {title} {Skyrmions in twisted bilayer graphene: Stability,
  pairing, and crystallization},\ }\href
  {https://doi.org/10.1103/PhysRevX.12.031020} {\bibfield  {journal} {\bibinfo
  {journal} {Physical Review X}\ }\textbf {\bibinfo {volume} {12}},\ \bibinfo
  {pages} {031020} (\bibinfo {year} {2022})}\BibitemShut {NoStop}%
\bibitem [{\citenamefont {Chatterjee}\ \emph {et~al.}(2022)\citenamefont
  {Chatterjee}, \citenamefont {Ippoliti},\ and\ \citenamefont
  {Zaletel}}]{Chatterjee2022}%
  \BibitemOpen
  \bibfield  {author} {\bibinfo {author} {\bibfnamefont {S.}~\bibnamefont
  {Chatterjee}}, \bibinfo {author} {\bibfnamefont {M.}~\bibnamefont
  {Ippoliti}},\ and\ \bibinfo {author} {\bibfnamefont {M.~P.}\ \bibnamefont
  {Zaletel}},\ }\bibfield  {title} {\bibinfo {title} {Skyrmion
  superconductivity: {{DMRG}} evidence for a topological route to
  superconductivity},\ }\href {https://doi.org/10.1103/PhysRevB.106.035421}
  {\bibfield  {journal} {\bibinfo  {journal} {Physical Review B}\ }\textbf
  {\bibinfo {volume} {106}},\ \bibinfo {pages} {035421} (\bibinfo {year}
  {2022})}\BibitemShut {NoStop}%
\bibitem [{\citenamefont {Fertig}\ \emph {et~al.}(1997)\citenamefont {Fertig},
  \citenamefont {Brey}, \citenamefont {C{\^o}t{\'e}}, \citenamefont
  {MacDonald}, \citenamefont {Karlhede},\ and\ \citenamefont
  {Sondhi}}]{Fertig1997}%
  \BibitemOpen
  \bibfield  {author} {\bibinfo {author} {\bibfnamefont {H.~A.}\ \bibnamefont
  {Fertig}}, \bibinfo {author} {\bibfnamefont {L.}~\bibnamefont {Brey}},
  \bibinfo {author} {\bibfnamefont {R.}~\bibnamefont {C{\^o}t{\'e}}}, \bibinfo
  {author} {\bibfnamefont {A.~H.}\ \bibnamefont {MacDonald}}, \bibinfo {author}
  {\bibfnamefont {A.}~\bibnamefont {Karlhede}},\ and\ \bibinfo {author}
  {\bibfnamefont {S.~L.}\ \bibnamefont {Sondhi}},\ }\bibfield  {title}
  {\bibinfo {title} {Hartree-{Fock} theory of skyrmions in quantum {{Hall}}
  ferromagnets},\ }\href {https://doi.org/10.1103/PhysRevB.55.10671} {\bibfield
   {journal} {\bibinfo  {journal} {Physical Review B}\ }\textbf {\bibinfo
  {volume} {55}},\ \bibinfo {pages} {10671} (\bibinfo {year}
  {1997})}\BibitemShut {NoStop}%
\bibitem [{\citenamefont {Abolfath}\ \emph {et~al.}(1997)\citenamefont
  {Abolfath}, \citenamefont {Palacios}, \citenamefont {Fertig}, \citenamefont
  {Girvin},\ and\ \citenamefont {MacDonald}}]{Abolfath1997}%
  \BibitemOpen
  \bibfield  {author} {\bibinfo {author} {\bibfnamefont {M.}~\bibnamefont
  {Abolfath}}, \bibinfo {author} {\bibfnamefont {J.~J.}\ \bibnamefont
  {Palacios}}, \bibinfo {author} {\bibfnamefont {H.~A.}\ \bibnamefont
  {Fertig}}, \bibinfo {author} {\bibfnamefont {S.~M.}\ \bibnamefont {Girvin}},\
  and\ \bibinfo {author} {\bibfnamefont {A.~H.}\ \bibnamefont {MacDonald}},\
  }\bibfield  {title} {\bibinfo {title} {Critical comparison of classical field
  theory and microscopic wave functions for skyrmions in quantum {Hall}
  ferromagnets},\ }\href {https://doi.org/10.1103/PhysRevB.56.6795} {\bibfield
  {journal} {\bibinfo  {journal} {Phys. Rev. B}\ }\textbf {\bibinfo {volume}
  {56}},\ \bibinfo {pages} {6795} (\bibinfo {year} {1997})}\BibitemShut
  {NoStop}%
\bibitem [{\citenamefont {Rezayi}(1991)}]{Rezayi1991}%
  \BibitemOpen
  \bibfield  {author} {\bibinfo {author} {\bibfnamefont {E.~H.}\ \bibnamefont
  {Rezayi}},\ }\bibfield  {title} {\bibinfo {title} {Wave functions and other
  properties of spin-reversed quasiparticles at $\ensuremath\{\nu\}=1/m$
  {{Landau-level}} occupation},\ }\href
  {https://doi.org/10.1103/PhysRevB.43.5944} {\bibfield  {journal} {\bibinfo
  {journal} {Physical Review B}\ }\textbf {\bibinfo {volume} {43}},\ \bibinfo
  {pages} {5944} (\bibinfo {year} {1991})}\BibitemShut {NoStop}%
\bibitem [{\citenamefont {Barrett}\ \emph {et~al.}(1995)\citenamefont
  {Barrett}, \citenamefont {Dabbagh}, \citenamefont {Pfeiffer}, \citenamefont
  {West},\ and\ \citenamefont {Tycko}}]{Barrett1995}%
  \BibitemOpen
  \bibfield  {author} {\bibinfo {author} {\bibfnamefont {S.~E.}\ \bibnamefont
  {Barrett}}, \bibinfo {author} {\bibfnamefont {G.}~\bibnamefont {Dabbagh}},
  \bibinfo {author} {\bibfnamefont {L.~N.}\ \bibnamefont {Pfeiffer}}, \bibinfo
  {author} {\bibfnamefont {K.~W.}\ \bibnamefont {West}},\ and\ \bibinfo
  {author} {\bibfnamefont {R.}~\bibnamefont {Tycko}},\ }\bibfield  {title}
  {\bibinfo {title} {Optically pumped {NMR} evidence for finite-size skyrmions
  in {GaAs} quantum wells near landau level filling
  $\mathit{\ensuremath{\nu}}=1$},\ }\href
  {https://doi.org/10.1103/PhysRevLett.74.5112} {\bibfield  {journal} {\bibinfo
   {journal} {Phys. Rev. Lett.}\ }\textbf {\bibinfo {volume} {74}},\ \bibinfo
  {pages} {5112} (\bibinfo {year} {1995})}\BibitemShut {NoStop}%
\bibitem [{\citenamefont {Aifer}\ \emph {et~al.}(1996)\citenamefont {Aifer},
  \citenamefont {Goldberg},\ and\ \citenamefont {Broido}}]{Aifer1996}%
  \BibitemOpen
  \bibfield  {author} {\bibinfo {author} {\bibfnamefont {E.~H.}\ \bibnamefont
  {Aifer}}, \bibinfo {author} {\bibfnamefont {B.~B.}\ \bibnamefont
  {Goldberg}},\ and\ \bibinfo {author} {\bibfnamefont {D.~A.}\ \bibnamefont
  {Broido}},\ }\bibfield  {title} {\bibinfo {title} {Evidence of skyrmion
  excitations about
  $\mathit{\ensuremath{\nu}}\phantom{\rule{0ex}{0ex}}=\phantom{\rule{0ex}{0ex}}1$
  in $\mathit{n}$-modulation-doped single quantum wells by interband optical
  transmission},\ }\href {https://doi.org/10.1103/PhysRevLett.76.680}
  {\bibfield  {journal} {\bibinfo  {journal} {Phys. Rev. Lett.}\ }\textbf
  {\bibinfo {volume} {76}},\ \bibinfo {pages} {680} (\bibinfo {year}
  {1996})}\BibitemShut {NoStop}%
\bibitem [{\citenamefont {Schmeller}\ \emph {et~al.}(1995)\citenamefont
  {Schmeller}, \citenamefont {Eisenstein}, \citenamefont {Pfeiffer},\ and\
  \citenamefont {West}}]{Schmeller1995}%
  \BibitemOpen
  \bibfield  {author} {\bibinfo {author} {\bibfnamefont {A.}~\bibnamefont
  {Schmeller}}, \bibinfo {author} {\bibfnamefont {J.~P.}\ \bibnamefont
  {Eisenstein}}, \bibinfo {author} {\bibfnamefont {L.~N.}\ \bibnamefont
  {Pfeiffer}},\ and\ \bibinfo {author} {\bibfnamefont {K.~W.}\ \bibnamefont
  {West}},\ }\bibfield  {title} {\bibinfo {title} {Evidence for skyrmions and
  single spin flips in the integer quantized {Hall} effect},\ }\href
  {https://doi.org/10.1103/PhysRevLett.75.4290} {\bibfield  {journal} {\bibinfo
   {journal} {Phys. Rev. Lett.}\ }\textbf {\bibinfo {volume} {75}},\ \bibinfo
  {pages} {4290} (\bibinfo {year} {1995})}\BibitemShut {NoStop}%
\bibitem [{\citenamefont {Wu}\ and\ \citenamefont {Sondhi}(1995)}]{Wu1995}%
  \BibitemOpen
  \bibfield  {author} {\bibinfo {author} {\bibfnamefont {X.-G.}\ \bibnamefont
  {Wu}}\ and\ \bibinfo {author} {\bibfnamefont {S.~L.}\ \bibnamefont
  {Sondhi}},\ }\bibfield  {title} {\bibinfo {title} {Skyrmions in higher
  {{Landau}} levels},\ }\href {https://doi.org/10.1103/PhysRevB.51.14725}
  {\bibfield  {journal} {\bibinfo  {journal} {Physical Review B}\ }\textbf
  {\bibinfo {volume} {51}},\ \bibinfo {pages} {14725} (\bibinfo {year}
  {1995})}\BibitemShut {NoStop}%
\bibitem [{\citenamefont {Rhone}\ \emph {et~al.}(2015)\citenamefont {Rhone},
  \citenamefont {Tiemann},\ and\ \citenamefont {Muraki}}]{Rhone2015}%
  \BibitemOpen
  \bibfield  {author} {\bibinfo {author} {\bibfnamefont {T.~D.}\ \bibnamefont
  {Rhone}}, \bibinfo {author} {\bibfnamefont {L.}~\bibnamefont {Tiemann}},\
  and\ \bibinfo {author} {\bibfnamefont {K.}~\bibnamefont {Muraki}},\
  }\bibfield  {title} {\bibinfo {title} {{NMR} probing of spin and charge order
  near odd-integer filling in the second {Landau} level},\ }\href
  {https://doi.org/10.1103/PhysRevB.92.041301} {\bibfield  {journal} {\bibinfo
  {journal} {Phys. Rev. B}\ }\textbf {\bibinfo {volume} {92}},\ \bibinfo
  {pages} {041301(R)} (\bibinfo {year} {2015})}\BibitemShut {NoStop}%
\bibitem [{\citenamefont {{Melik-Alaverdian}}\ \emph
  {et~al.}(1999)\citenamefont {{Melik-Alaverdian}}, \citenamefont {Bonesteel},\
  and\ \citenamefont {Ortiz}}]{Melik1999}%
  \BibitemOpen
  \bibfield  {author} {\bibinfo {author} {\bibfnamefont {V.}~\bibnamefont
  {{Melik-Alaverdian}}}, \bibinfo {author} {\bibfnamefont {N.~E.}\ \bibnamefont
  {Bonesteel}},\ and\ \bibinfo {author} {\bibfnamefont {G.}~\bibnamefont
  {Ortiz}},\ }\bibfield  {title} {\bibinfo {title} {Skyrmion physics beyond the
  lowest {{Landau-level}} approximation},\ }\href
  {https://doi.org/10.1103/PhysRevB.60.R8501} {\bibfield  {journal} {\bibinfo
  {journal} {Physical Review B}\ }\textbf {\bibinfo {volume} {60}},\ \bibinfo
  {pages} {R8501} (\bibinfo {year} {1999})}\BibitemShut {NoStop}%
\bibitem [{\citenamefont {Mihalek}\ and\ \citenamefont
  {Fertig}(2000)}]{Mihalek2000}%
  \BibitemOpen
  \bibfield  {author} {\bibinfo {author} {\bibfnamefont {I.}~\bibnamefont
  {Mihalek}}\ and\ \bibinfo {author} {\bibfnamefont {H.~A.}\ \bibnamefont
  {Fertig}},\ }\bibfield  {title} {\bibinfo {title} {Landau-level mixing and
  skyrmion stability in quantum {{Hall}} ferromagnets},\ }\href
  {https://doi.org/10.1103/PhysRevB.62.13573} {\bibfield  {journal} {\bibinfo
  {journal} {Physical Review B}\ }\textbf {\bibinfo {volume} {62}},\ \bibinfo
  {pages} {13573} (\bibinfo {year} {2000})}\BibitemShut {NoStop}%
\bibitem [{\citenamefont {Wu}\ and\ \citenamefont {Das~Sarma}(2020)}]{Wu2020}%
  \BibitemOpen
  \bibfield  {author} {\bibinfo {author} {\bibfnamefont {F.}~\bibnamefont
  {Wu}}\ and\ \bibinfo {author} {\bibfnamefont {S.}~\bibnamefont {Das~Sarma}},\
  }\bibfield  {title} {\bibinfo {title} {Quantum geometry and stability of
  moir\'e flatband ferromagnetism},\ }\href
  {https://doi.org/10.1103/PhysRevB.102.165118} {\bibfield  {journal} {\bibinfo
   {journal} {Phys. Rev. B}\ }\textbf {\bibinfo {volume} {102}},\ \bibinfo
  {pages} {165118} (\bibinfo {year} {2020})}\BibitemShut {NoStop}%
\bibitem [{\citenamefont {Khalaf}\ and\ \citenamefont
  {Vishwanath}(2022)}]{Khalaf2022}%
  \BibitemOpen
  \bibfield  {author} {\bibinfo {author} {\bibfnamefont {E.}~\bibnamefont
  {Khalaf}}\ and\ \bibinfo {author} {\bibfnamefont {A.}~\bibnamefont
  {Vishwanath}},\ }\bibfield  {title} {\bibinfo {title} {Baby skyrmions in
  {Chern} ferromagnets and topological mechanism for spin-polaron formation in
  twisted bilayer graphene},\ }\href
  {https://doi.org/10.1038/s41467-022-33673-3} {\bibfield  {journal} {\bibinfo
  {journal} {Nature Communications}\ }\textbf {\bibinfo {volume} {13}},\
  \bibinfo {pages} {6245} (\bibinfo {year} {2022})}\BibitemShut {NoStop}%
\bibitem [{\citenamefont {Schindler}\ \emph {et~al.}(2022)\citenamefont
  {Schindler}, \citenamefont {Vafek},\ and\ \citenamefont
  {Bernevig}}]{Schindler2022}%
  \BibitemOpen
  \bibfield  {author} {\bibinfo {author} {\bibfnamefont {F.}~\bibnamefont
  {Schindler}}, \bibinfo {author} {\bibfnamefont {O.}~\bibnamefont {Vafek}},\
  and\ \bibinfo {author} {\bibfnamefont {B.~A.}\ \bibnamefont {Bernevig}},\
  }\bibfield  {title} {\bibinfo {title} {Trions in twisted bilayer graphene},\
  }\href {https://doi.org/10.1103/PhysRevB.105.155135} {\bibfield  {journal}
  {\bibinfo  {journal} {Physical Review B}\ }\textbf {\bibinfo {volume}
  {105}},\ \bibinfo {pages} {155135} (\bibinfo {year} {2022})}\BibitemShut
  {NoStop}%
\bibitem [{\citenamefont {White}\ \emph {et~al.}(1989)\citenamefont {White},
  \citenamefont {Scalapino}, \citenamefont {Sugar}, \citenamefont {Loh},
  \citenamefont {Gubernatis},\ and\ \citenamefont {Scalettar}}]{White1989}%
  \BibitemOpen
  \bibfield  {author} {\bibinfo {author} {\bibfnamefont {S.~R.}\ \bibnamefont
  {White}}, \bibinfo {author} {\bibfnamefont {D.~J.}\ \bibnamefont
  {Scalapino}}, \bibinfo {author} {\bibfnamefont {R.~L.}\ \bibnamefont
  {Sugar}}, \bibinfo {author} {\bibfnamefont {E.~Y.}\ \bibnamefont {Loh}},
  \bibinfo {author} {\bibfnamefont {J.~E.}\ \bibnamefont {Gubernatis}},\ and\
  \bibinfo {author} {\bibfnamefont {R.~T.}\ \bibnamefont {Scalettar}},\
  }\bibfield  {title} {\bibinfo {title} {Numerical study of the two-dimensional
  {Hubbard} model},\ }\href {https://doi.org/10.1103/PhysRevB.40.506}
  {\bibfield  {journal} {\bibinfo  {journal} {Physical Review B}\ }\textbf
  {\bibinfo {volume} {40}},\ \bibinfo {pages} {506} (\bibinfo {year}
  {1989})}\BibitemShut {NoStop}%
\bibitem [{\citenamefont {Loh}\ \emph {et~al.}(1990)\citenamefont {Loh},
  \citenamefont {Gubernatis}, \citenamefont {Scalettar}, \citenamefont {White},
  \citenamefont {Scalapino},\ and\ \citenamefont {Sugar}}]{Loh1990}%
  \BibitemOpen
  \bibfield  {author} {\bibinfo {author} {\bibfnamefont {E.~Y.}\ \bibnamefont
  {Loh}}, \bibinfo {author} {\bibfnamefont {J.~E.}\ \bibnamefont {Gubernatis}},
  \bibinfo {author} {\bibfnamefont {R.~T.}\ \bibnamefont {Scalettar}}, \bibinfo
  {author} {\bibfnamefont {S.~R.}\ \bibnamefont {White}}, \bibinfo {author}
  {\bibfnamefont {D.~J.}\ \bibnamefont {Scalapino}},\ and\ \bibinfo {author}
  {\bibfnamefont {R.~L.}\ \bibnamefont {Sugar}},\ }\bibfield  {title} {\bibinfo
  {title} {Sign problem in the numerical simulation of many-electron systems},\
  }\href {https://doi.org/10.1103/PhysRevB.41.9301} {\bibfield  {journal}
  {\bibinfo  {journal} {Physical Review B}\ }\textbf {\bibinfo {volume} {41}},\
  \bibinfo {pages} {9301} (\bibinfo {year} {1990})}\BibitemShut {NoStop}%
\bibitem [{\citenamefont {White}(1992)}]{White1992}%
  \BibitemOpen
  \bibfield  {author} {\bibinfo {author} {\bibfnamefont {S.~R.}\ \bibnamefont
  {White}},\ }\bibfield  {title} {\bibinfo {title} {Density matrix formulation
  for quantum renormalization groups},\ }\href
  {https://doi.org/10.1103/PhysRevLett.69.2863} {\bibfield  {journal} {\bibinfo
   {journal} {Phys. Rev. Lett.}\ }\textbf {\bibinfo {volume} {69}},\ \bibinfo
  {pages} {2863} (\bibinfo {year} {1992})}\BibitemShut {NoStop}%
\bibitem [{\citenamefont {White}(1993)}]{White1993}%
  \BibitemOpen
  \bibfield  {author} {\bibinfo {author} {\bibfnamefont {S.~R.}\ \bibnamefont
  {White}},\ }\bibfield  {title} {\bibinfo {title} {Density-matrix algorithms
  for quantum renormalization groups},\ }\href
  {https://doi.org/10.1103/PhysRevB.48.10345} {\bibfield  {journal} {\bibinfo
  {journal} {Phys. Rev. B}\ }\textbf {\bibinfo {volume} {48}},\ \bibinfo
  {pages} {10345} (\bibinfo {year} {1993})}\BibitemShut {NoStop}%
\bibitem [{\citenamefont {\"Ostlund}\ and\ \citenamefont
  {Rommer}(1995)}]{Ostlund1995}%
  \BibitemOpen
  \bibfield  {author} {\bibinfo {author} {\bibfnamefont {S.}~\bibnamefont
  {\"Ostlund}}\ and\ \bibinfo {author} {\bibfnamefont {S.}~\bibnamefont
  {Rommer}},\ }\bibfield  {title} {\bibinfo {title} {Thermodynamic limit of
  density matrix renormalization},\ }\href
  {https://doi.org/10.1103/PhysRevLett.75.3537} {\bibfield  {journal} {\bibinfo
   {journal} {Phys. Rev. Lett.}\ }\textbf {\bibinfo {volume} {75}},\ \bibinfo
  {pages} {3537} (\bibinfo {year} {1995})}\BibitemShut {NoStop}%
\bibitem [{\citenamefont {Dukelsky}\ \emph {et~al.}(1998)\citenamefont
  {Dukelsky}, \citenamefont {Mart\'{\i}n-Delgado}, \citenamefont {Nishino},\
  and\ \citenamefont {Sierra}}]{Dukelsky1998}%
  \BibitemOpen
  \bibfield  {author} {\bibinfo {author} {\bibfnamefont {J.}~\bibnamefont
  {Dukelsky}}, \bibinfo {author} {\bibfnamefont {M.~A.}\ \bibnamefont
  {Mart\'{\i}n-Delgado}}, \bibinfo {author} {\bibfnamefont {T.}~\bibnamefont
  {Nishino}},\ and\ \bibinfo {author} {\bibfnamefont {G.}~\bibnamefont
  {Sierra}},\ }\bibfield  {title} {\bibinfo {title} {Equivalence of the
  variational matrix product method and the density matrix renormalization
  group applied to spin chains},\ }\href
  {https://doi.org/10.1209/epl/i1998-00381-x} {\bibfield  {journal} {\bibinfo
  {journal} {Europhys. Lett.}\ }\textbf {\bibinfo {volume} {43}},\ \bibinfo
  {pages} {457} (\bibinfo {year} {1998})}\BibitemShut {NoStop}%
\bibitem [{\citenamefont {Wu}\ \emph {et~al.}(2018)\citenamefont {Wu},
  \citenamefont {Lovorn}, \citenamefont {Tutuc},\ and\ \citenamefont
  {MacDonald}}]{wu2018}%
  \BibitemOpen
  \bibfield  {author} {\bibinfo {author} {\bibfnamefont {F.}~\bibnamefont
  {Wu}}, \bibinfo {author} {\bibfnamefont {T.}~\bibnamefont {Lovorn}}, \bibinfo
  {author} {\bibfnamefont {E.}~\bibnamefont {Tutuc}},\ and\ \bibinfo {author}
  {\bibfnamefont {A.~H.}\ \bibnamefont {MacDonald}},\ }\bibfield  {title}
  {\bibinfo {title} {Hubbard model physics in transition metal dichalcogenide
  moir\'e bands},\ }\href {https://doi.org/10.1103/PhysRevLett.121.026402}
  {\bibfield  {journal} {\bibinfo  {journal} {Phys. Rev. Lett.}\ }\textbf
  {\bibinfo {volume} {121}},\ \bibinfo {pages} {026402} (\bibinfo {year}
  {2018})}\BibitemShut {NoStop}%
\bibitem [{\citenamefont {{Morales-Dur{\'a}n}}\ \emph
  {et~al.}(2022)\citenamefont {{Morales-Dur{\'a}n}}, \citenamefont {Hu},
  \citenamefont {Potasz},\ and\ \citenamefont
  {MacDonald}}]{Morales-Duran2022a}%
  \BibitemOpen
  \bibfield  {author} {\bibinfo {author} {\bibfnamefont {N.}~\bibnamefont
  {{Morales-Dur{\'a}n}}}, \bibinfo {author} {\bibfnamefont {N.~C.}\
  \bibnamefont {Hu}}, \bibinfo {author} {\bibfnamefont {P.}~\bibnamefont
  {Potasz}},\ and\ \bibinfo {author} {\bibfnamefont {A.~H.}\ \bibnamefont
  {MacDonald}},\ }\bibfield  {title} {\bibinfo {title} {Nonlocal interactions
  in {{moir{\'e} Hubbard}} systems},\ }\href
  {https://doi.org/10.1103/PhysRevLett.128.217202} {\bibfield  {journal}
  {\bibinfo  {journal} {Physical Review Letters}\ }\textbf {\bibinfo {volume}
  {128}},\ \bibinfo {pages} {217202} (\bibinfo {year} {2022})}\BibitemShut
  {NoStop}%
\bibitem [{\citenamefont {Cocks}\ \emph {et~al.}(2012)\citenamefont {Cocks},
  \citenamefont {Orth}, \citenamefont {Rachel}, \citenamefont {Buchhold},
  \citenamefont {Le~Hur},\ and\ \citenamefont {Hofstetter}}]{Cocks2012}%
  \BibitemOpen
  \bibfield  {author} {\bibinfo {author} {\bibfnamefont {D.}~\bibnamefont
  {Cocks}}, \bibinfo {author} {\bibfnamefont {P.~P.}\ \bibnamefont {Orth}},
  \bibinfo {author} {\bibfnamefont {S.}~\bibnamefont {Rachel}}, \bibinfo
  {author} {\bibfnamefont {M.}~\bibnamefont {Buchhold}}, \bibinfo {author}
  {\bibfnamefont {K.}~\bibnamefont {Le~Hur}},\ and\ \bibinfo {author}
  {\bibfnamefont {W.}~\bibnamefont {Hofstetter}},\ }\bibfield  {title}
  {\bibinfo {title} {Time-reversal-invariant {{Hofstadter-Hubbard}} model with
  ultracold fermions},\ }\href {https://doi.org/10.1103/PhysRevLett.109.205303}
  {\bibfield  {journal} {\bibinfo  {journal} {Physical Review Letters}\
  }\textbf {\bibinfo {volume} {109}},\ \bibinfo {pages} {205303} (\bibinfo
  {year} {2012})}\BibitemShut {NoStop}%
\bibitem [{\citenamefont {Sahay}\ \emph {et~al.}(2023)\citenamefont {Sahay},
  \citenamefont {Divic}, \citenamefont {Parker}, \citenamefont {Soejima},
  \citenamefont {Anand}, \citenamefont {Hauschild}, \citenamefont
  {Aidelsburger}, \citenamefont {Vishwanath}, \citenamefont {Chatterjee},
  \citenamefont {Yao},\ and\ \citenamefont {Zaletel}}]{Sahay2023}%
  \BibitemOpen
  \bibfield  {author} {\bibinfo {author} {\bibfnamefont {R.}~\bibnamefont
  {Sahay}}, \bibinfo {author} {\bibfnamefont {S.}~\bibnamefont {Divic}},
  \bibinfo {author} {\bibfnamefont {D.~E.}\ \bibnamefont {Parker}}, \bibinfo
  {author} {\bibfnamefont {T.}~\bibnamefont {Soejima}}, \bibinfo {author}
  {\bibfnamefont {S.}~\bibnamefont {Anand}}, \bibinfo {author} {\bibfnamefont
  {J.}~\bibnamefont {Hauschild}}, \bibinfo {author} {\bibfnamefont
  {M.}~\bibnamefont {Aidelsburger}}, \bibinfo {author} {\bibfnamefont
  {A.}~\bibnamefont {Vishwanath}}, \bibinfo {author} {\bibfnamefont
  {S.}~\bibnamefont {Chatterjee}}, \bibinfo {author} {\bibfnamefont {N.~Y.}\
  \bibnamefont {Yao}},\ and\ \bibinfo {author} {\bibfnamefont {M.~P.}\
  \bibnamefont {Zaletel}},\ }\href {https://doi.org/10.48550/arXiv.2308.10935}
  {\bibinfo {title} {Superconductivity in a topological lattice model with
  strong repulsion}} (\bibinfo {year} {2023}),\ \Eprint
  {https://arxiv.org/abs/2308.10935} {arxiv:2308.10935} \BibitemShut {NoStop}%
\bibitem [{\citenamefont {Levy}\ \emph {et~al.}(2010)\citenamefont {Levy},
  \citenamefont {Burke}, \citenamefont {Meaker}, \citenamefont {Panlasigui},
  \citenamefont {Zettl}, \citenamefont {Guinea}, \citenamefont {Neto},\ and\
  \citenamefont {Crommie}}]{Levy2010}%
  \BibitemOpen
  \bibfield  {author} {\bibinfo {author} {\bibfnamefont {N.}~\bibnamefont
  {Levy}}, \bibinfo {author} {\bibfnamefont {S.~A.}\ \bibnamefont {Burke}},
  \bibinfo {author} {\bibfnamefont {K.~L.}\ \bibnamefont {Meaker}}, \bibinfo
  {author} {\bibfnamefont {M.}~\bibnamefont {Panlasigui}}, \bibinfo {author}
  {\bibfnamefont {A.}~\bibnamefont {Zettl}}, \bibinfo {author} {\bibfnamefont
  {F.}~\bibnamefont {Guinea}}, \bibinfo {author} {\bibfnamefont {A.~H.~C.}\
  \bibnamefont {Neto}},\ and\ \bibinfo {author} {\bibfnamefont {M.~F.}\
  \bibnamefont {Crommie}},\ }\bibfield  {title} {\bibinfo {title}
  {Strain-induced pseudo-magnetic fields greater than 300 {{Tesla}} in graphene
  nanobubbles},\ }\href {https://doi.org/10.1126/science.1191700} {\bibfield
  {journal} {\bibinfo  {journal} {Science}\ }\textbf {\bibinfo {volume}
  {329}},\ \bibinfo {pages} {544} (\bibinfo {year} {2010})}\BibitemShut
  {NoStop}%
\bibitem [{\citenamefont {Mao}\ \emph {et~al.}(2020)\citenamefont {Mao},
  \citenamefont {Milovanovi{\'c}}, \citenamefont {An{\dj}elkovi{\'c}},
  \citenamefont {Lai}, \citenamefont {Cao}, \citenamefont {Watanabe},
  \citenamefont {Taniguchi}, \citenamefont {Covaci}, \citenamefont {Peeters},
  \citenamefont {Geim}, \citenamefont {Jiang},\ and\ \citenamefont
  {Andrei}}]{Mao2020}%
  \BibitemOpen
  \bibfield  {author} {\bibinfo {author} {\bibfnamefont {J.}~\bibnamefont
  {Mao}}, \bibinfo {author} {\bibfnamefont {S.~P.}\ \bibnamefont
  {Milovanovi{\'c}}}, \bibinfo {author} {\bibfnamefont {M.}~\bibnamefont
  {An{\dj}elkovi{\'c}}}, \bibinfo {author} {\bibfnamefont {X.}~\bibnamefont
  {Lai}}, \bibinfo {author} {\bibfnamefont {Y.}~\bibnamefont {Cao}}, \bibinfo
  {author} {\bibfnamefont {K.}~\bibnamefont {Watanabe}}, \bibinfo {author}
  {\bibfnamefont {T.}~\bibnamefont {Taniguchi}}, \bibinfo {author}
  {\bibfnamefont {L.}~\bibnamefont {Covaci}}, \bibinfo {author} {\bibfnamefont
  {F.~M.}\ \bibnamefont {Peeters}}, \bibinfo {author} {\bibfnamefont {A.~K.}\
  \bibnamefont {Geim}}, \bibinfo {author} {\bibfnamefont {Y.}~\bibnamefont
  {Jiang}},\ and\ \bibinfo {author} {\bibfnamefont {E.~Y.}\ \bibnamefont
  {Andrei}},\ }\bibfield  {title} {\bibinfo {title} {Evidence of flat bands and
  correlated states in buckled graphene superlattices},\ }\href
  {https://doi.org/10.1038/s41586-020-2567-3} {\bibfield  {journal} {\bibinfo
  {journal} {Nature}\ }\textbf {\bibinfo {volume} {584}},\ \bibinfo {pages}
  {215} (\bibinfo {year} {2020})}\BibitemShut {NoStop}%
\bibitem [{\citenamefont {Yang}\ \emph {et~al.}(2021)\citenamefont {Yang},
  \citenamefont {Liu}, \citenamefont {Mongkolkiattichai},\ and\ \citenamefont
  {Schauss}}]{Yang2021}%
  \BibitemOpen
  \bibfield  {author} {\bibinfo {author} {\bibfnamefont {J.}~\bibnamefont
  {Yang}}, \bibinfo {author} {\bibfnamefont {L.}~\bibnamefont {Liu}}, \bibinfo
  {author} {\bibfnamefont {J.}~\bibnamefont {Mongkolkiattichai}},\ and\
  \bibinfo {author} {\bibfnamefont {P.}~\bibnamefont {Schauss}},\ }\bibfield
  {title} {\bibinfo {title} {Site-resolved imaging of ultracold fermions in a
  triangular-lattice quantum gas microscope},\ }\href
  {https://doi.org/10.1103/PRXQuantum.2.020344} {\bibfield  {journal} {\bibinfo
   {journal} {PRX Quantum}\ }\textbf {\bibinfo {volume} {2}},\ \bibinfo {pages}
  {020344} (\bibinfo {year} {2021})}\BibitemShut {NoStop}%
\bibitem [{\citenamefont {Garwood}\ \emph {et~al.}(2022)\citenamefont
  {Garwood}, \citenamefont {Mongkolkiattichai}, \citenamefont {Liu},
  \citenamefont {Yang},\ and\ \citenamefont {Schauss}}]{Garwood2022}%
  \BibitemOpen
  \bibfield  {author} {\bibinfo {author} {\bibfnamefont {D.}~\bibnamefont
  {Garwood}}, \bibinfo {author} {\bibfnamefont {J.}~\bibnamefont
  {Mongkolkiattichai}}, \bibinfo {author} {\bibfnamefont {L.}~\bibnamefont
  {Liu}}, \bibinfo {author} {\bibfnamefont {J.}~\bibnamefont {Yang}},\ and\
  \bibinfo {author} {\bibfnamefont {P.}~\bibnamefont {Schauss}},\ }\bibfield
  {title} {\bibinfo {title} {Site-resolved observables in the doped
  spin-imbalanced triangular {Hubbard} model},\ }\href
  {https://doi.org/10.1103/PhysRevA.106.013310} {\bibfield  {journal} {\bibinfo
   {journal} {Phys. Rev. A}\ }\textbf {\bibinfo {volume} {106}},\ \bibinfo
  {pages} {013310} (\bibinfo {year} {2022})}\BibitemShut {NoStop}%
\bibitem [{\citenamefont {Mongkolkiattichai}\ \emph {et~al.}(2023)\citenamefont
  {Mongkolkiattichai}, \citenamefont {Liu}, \citenamefont {Garwood},
  \citenamefont {Yang},\ and\ \citenamefont {Schauss}}]{Mongkolkiattichai2022}%
  \BibitemOpen
  \bibfield  {author} {\bibinfo {author} {\bibfnamefont {J.}~\bibnamefont
  {Mongkolkiattichai}}, \bibinfo {author} {\bibfnamefont {L.}~\bibnamefont
  {Liu}}, \bibinfo {author} {\bibfnamefont {D.}~\bibnamefont {Garwood}},
  \bibinfo {author} {\bibfnamefont {J.}~\bibnamefont {Yang}},\ and\ \bibinfo
  {author} {\bibfnamefont {P.}~\bibnamefont {Schauss}},\ }\bibfield  {title}
  {\bibinfo {title} {Quantum gas microscopy of fermionic triangular-lattice
  {Mott} insulators},\ }\href {https://doi.org/10.1103/PhysRevA.108.L061301}
  {\bibfield  {journal} {\bibinfo  {journal} {Phys. Rev. A}\ }\textbf {\bibinfo
  {volume} {108}},\ \bibinfo {pages} {L061301} (\bibinfo {year}
  {2023})}\BibitemShut {NoStop}%
\bibitem [{\citenamefont {Xu}\ \emph {et~al.}(2023)\citenamefont {Xu},
  \citenamefont {Kendrick}, \citenamefont {Kale}, \citenamefont {Gang},
  \citenamefont {Ji}, \citenamefont {Scalettar}, \citenamefont {Lebrat},\ and\
  \citenamefont {Greiner}}]{Xu2023d}%
  \BibitemOpen
  \bibfield  {author} {\bibinfo {author} {\bibfnamefont {M.}~\bibnamefont
  {Xu}}, \bibinfo {author} {\bibfnamefont {L.~H.}\ \bibnamefont {Kendrick}},
  \bibinfo {author} {\bibfnamefont {A.}~\bibnamefont {Kale}}, \bibinfo {author}
  {\bibfnamefont {Y.}~\bibnamefont {Gang}}, \bibinfo {author} {\bibfnamefont
  {G.}~\bibnamefont {Ji}}, \bibinfo {author} {\bibfnamefont {R.~T.}\
  \bibnamefont {Scalettar}}, \bibinfo {author} {\bibfnamefont {M.}~\bibnamefont
  {Lebrat}},\ and\ \bibinfo {author} {\bibfnamefont {M.}~\bibnamefont
  {Greiner}},\ }\bibfield  {title} {\bibinfo {title} {Frustration- and
  doping-induced magnetism in a {Fermi}-{Hubbard} simulator},\ }\href
  {https://doi.org/10.1038/s41586-023-06280-5} {\bibfield  {journal} {\bibinfo
  {journal} {Nature}\ }\textbf {\bibinfo {volume} {620}},\ \bibinfo {pages}
  {971} (\bibinfo {year} {2023})}\BibitemShut {NoStop}%
\bibitem [{\citenamefont {Mancini}\ \emph {et~al.}(2015)\citenamefont
  {Mancini}, \citenamefont {Pagano}, \citenamefont {Cappellini}, \citenamefont
  {Livi}, \citenamefont {Rider}, \citenamefont {Catani}, \citenamefont {Sias},
  \citenamefont {Zoller}, \citenamefont {Inguscio}, \citenamefont {Dalmonte},\
  and\ \citenamefont {Fallani}}]{Mancini2015}%
  \BibitemOpen
  \bibfield  {author} {\bibinfo {author} {\bibfnamefont {M.}~\bibnamefont
  {Mancini}}, \bibinfo {author} {\bibfnamefont {G.}~\bibnamefont {Pagano}},
  \bibinfo {author} {\bibfnamefont {G.}~\bibnamefont {Cappellini}}, \bibinfo
  {author} {\bibfnamefont {L.}~\bibnamefont {Livi}}, \bibinfo {author}
  {\bibfnamefont {M.}~\bibnamefont {Rider}}, \bibinfo {author} {\bibfnamefont
  {J.}~\bibnamefont {Catani}}, \bibinfo {author} {\bibfnamefont
  {C.}~\bibnamefont {Sias}}, \bibinfo {author} {\bibfnamefont {P.}~\bibnamefont
  {Zoller}}, \bibinfo {author} {\bibfnamefont {M.}~\bibnamefont {Inguscio}},
  \bibinfo {author} {\bibfnamefont {M.}~\bibnamefont {Dalmonte}},\ and\
  \bibinfo {author} {\bibfnamefont {L.}~\bibnamefont {Fallani}},\ }\bibfield
  {title} {\bibinfo {title} {Observation of chiral edge states with neutral
  fermions in synthetic {Hall} ribbons},\ }\href
  {https://doi.org/10.1126/science.aaa8736} {\bibfield  {journal} {\bibinfo
  {journal} {Science}\ }\textbf {\bibinfo {volume} {349}},\ \bibinfo {pages}
  {1510} (\bibinfo {year} {2015})}\BibitemShut {NoStop}%
\bibitem [{\citenamefont {Ding}\ \emph {et~al.}(2022)\citenamefont {Ding},
  \citenamefont {Wang}, \citenamefont {Moritz}, \citenamefont {Schattner},
  \citenamefont {Huang},\ and\ \citenamefont {Devereaux}}]{Ding2022}%
  \BibitemOpen
  \bibfield  {author} {\bibinfo {author} {\bibfnamefont {J.~K.}\ \bibnamefont
  {Ding}}, \bibinfo {author} {\bibfnamefont {W.~O.}\ \bibnamefont {Wang}},
  \bibinfo {author} {\bibfnamefont {B.}~\bibnamefont {Moritz}}, \bibinfo
  {author} {\bibfnamefont {Y.}~\bibnamefont {Schattner}}, \bibinfo {author}
  {\bibfnamefont {E.~W.}\ \bibnamefont {Huang}},\ and\ \bibinfo {author}
  {\bibfnamefont {T.~P.}\ \bibnamefont {Devereaux}},\ }\bibfield  {title}
  {\bibinfo {title} {Thermodynamics of correlated electrons in a magnetic
  field},\ }\href {https://doi.org/10.1038/s42005-022-00968-2} {\bibfield
  {journal} {\bibinfo  {journal} {Communications Physics}\ }\textbf {\bibinfo
  {volume} {5}},\ \bibinfo {pages} {204} (\bibinfo {year} {2022})}\BibitemShut
  {NoStop}%
\bibitem [{\citenamefont {Mai}\ \emph {et~al.}(2023)\citenamefont {Mai},
  \citenamefont {Huang}, \citenamefont {Yu}, \citenamefont {Feldman},\ and\
  \citenamefont {Phillips}}]{Mai2023}%
  \BibitemOpen
  \bibfield  {author} {\bibinfo {author} {\bibfnamefont {P.}~\bibnamefont
  {Mai}}, \bibinfo {author} {\bibfnamefont {E.~W.}\ \bibnamefont {Huang}},
  \bibinfo {author} {\bibfnamefont {J.}~\bibnamefont {Yu}}, \bibinfo {author}
  {\bibfnamefont {B.~E.}\ \bibnamefont {Feldman}},\ and\ \bibinfo {author}
  {\bibfnamefont {P.~W.}\ \bibnamefont {Phillips}},\ }\bibfield  {title}
  {\bibinfo {title} {Interaction-driven spontaneous ferromagnetic insulating
  states with odd {Chern} numbers},\ }\href
  {https://doi.org/10.1038/s41535-023-00544-z} {\bibfield  {journal} {\bibinfo
  {journal} {npj Quantum Materials}\ }\textbf {\bibinfo {volume} {8}},\
  \bibinfo {pages} {1} (\bibinfo {year} {2023})}\BibitemShut {NoStop}%
\bibitem [{\citenamefont {Palm}\ \emph {et~al.}(2023)\citenamefont {Palm},
  \citenamefont {Kurttutan}, \citenamefont {Bohrdt}, \citenamefont
  {Schollw{\"o}ck},\ and\ \citenamefont {Grusdt}}]{Palm2023}%
  \BibitemOpen
  \bibfield  {author} {\bibinfo {author} {\bibfnamefont {F.~A.}\ \bibnamefont
  {Palm}}, \bibinfo {author} {\bibfnamefont {M.}~\bibnamefont {Kurttutan}},
  \bibinfo {author} {\bibfnamefont {A.}~\bibnamefont {Bohrdt}}, \bibinfo
  {author} {\bibfnamefont {U.}~\bibnamefont {Schollw{\"o}ck}},\ and\ \bibinfo
  {author} {\bibfnamefont {F.}~\bibnamefont {Grusdt}},\ }\bibfield  {title}
  {\bibinfo {title} {Ferromagnetism and skyrmions in the
  {Hofstadter}-{Fermi}-{Hubbard} model},\ }\href
  {https://doi.org/10.1088/1367-2630/acb963} {\bibfield  {journal} {\bibinfo
  {journal} {New Journal of Physics}\ }\textbf {\bibinfo {volume} {25}},\
  \bibinfo {pages} {023021} (\bibinfo {year} {2023})}\BibitemShut {NoStop}%
\bibitem [{\citenamefont {Assaad}(2002)}]{Assaad2002}%
  \BibitemOpen
  \bibfield  {author} {\bibinfo {author} {\bibfnamefont {F.~F.}\ \bibnamefont
  {Assaad}},\ }\bibfield  {title} {\bibinfo {title} {Depleted {Kondo} lattices:
  Quantum {Monte} {Carlo} and mean-field calculations},\ }\href
  {https://doi.org/10.1103/PhysRevB.65.115104} {\bibfield  {journal} {\bibinfo
  {journal} {Phys. Rev. B}\ }\textbf {\bibinfo {volume} {65}},\ \bibinfo
  {pages} {115104} (\bibinfo {year} {2002})}\BibitemShut {NoStop}%
\bibitem [{\citenamefont {Parameswaran}\ \emph {et~al.}(2013)\citenamefont
  {Parameswaran}, \citenamefont {Roy},\ and\ \citenamefont
  {Sondhi}}]{Parameswaran2013}%
  \BibitemOpen
  \bibfield  {author} {\bibinfo {author} {\bibfnamefont {S.~A.}\ \bibnamefont
  {Parameswaran}}, \bibinfo {author} {\bibfnamefont {R.}~\bibnamefont {Roy}},\
  and\ \bibinfo {author} {\bibfnamefont {S.~L.}\ \bibnamefont {Sondhi}},\
  }\bibfield  {title} {\bibinfo {title} {Fractional quantum {{Hall}} physics in
  topological flat bands},\ }\href {https://doi.org/10.1016/j.crhy.2013.04.003}
  {\bibfield  {journal} {\bibinfo  {journal} {Comptes Rendus Physique}\
  }\bibinfo {series} {Topological Insulators / {{Isolants}} Topologiques},\
  \textbf {\bibinfo {volume} {14}},\ \bibinfo {pages} {816} (\bibinfo {year}
  {2013})}\BibitemShut {NoStop}%
\bibitem [{\citenamefont {Fishman}\ \emph
  {et~al.}(2022{\natexlab{a}})\citenamefont {Fishman}, \citenamefont {White},\
  and\ \citenamefont {Stoudenmire}}]{Fishman2022}%
  \BibitemOpen
  \bibfield  {author} {\bibinfo {author} {\bibfnamefont {M.}~\bibnamefont
  {Fishman}}, \bibinfo {author} {\bibfnamefont {S.~R.}\ \bibnamefont {White}},\
  and\ \bibinfo {author} {\bibfnamefont {E.~M.}\ \bibnamefont {Stoudenmire}},\
  }\bibfield  {title} {\bibinfo {title} {The {ITensor} software library for
  tensor network calculations},\ }\href
  {https://doi.org/10.21468/SciPostPhysCodeb.4} {\bibfield  {journal} {\bibinfo
   {journal} {SciPost Phys. Codebases}\ ,\ \bibinfo {pages} {4}} (\bibinfo
  {year} {2022}{\natexlab{a}})}\BibitemShut {NoStop}%
\bibitem [{\citenamefont {Fishman}\ \emph
  {et~al.}(2022{\natexlab{b}})\citenamefont {Fishman}, \citenamefont {White},\
  and\ \citenamefont {Stoudenmire}}]{Fishman2022_2}%
  \BibitemOpen
  \bibfield  {author} {\bibinfo {author} {\bibfnamefont {M.}~\bibnamefont
  {Fishman}}, \bibinfo {author} {\bibfnamefont {S.~R.}\ \bibnamefont {White}},\
  and\ \bibinfo {author} {\bibfnamefont {E.~M.}\ \bibnamefont {Stoudenmire}},\
  }\bibfield  {title} {\bibinfo {title} {Codebase release 0.3 for {ITensor}},\
  }\href {https://doi.org/10.21468/SciPostPhysCodeb.4-r0.3} {\bibfield
  {journal} {\bibinfo  {journal} {SciPost Phys. Codebases}\ ,\ \bibinfo {pages}
  {4}} (\bibinfo {year} {2022}{\natexlab{b}})}\BibitemShut {NoStop}%
\bibitem [{\citenamefont {MacDonald}\ \emph {et~al.}(1996)\citenamefont
  {MacDonald}, \citenamefont {Fertig},\ and\ \citenamefont
  {Brey}}]{MacDonald1996}%
  \BibitemOpen
  \bibfield  {author} {\bibinfo {author} {\bibfnamefont {A.~H.}\ \bibnamefont
  {MacDonald}}, \bibinfo {author} {\bibfnamefont {H.~A.}\ \bibnamefont
  {Fertig}},\ and\ \bibinfo {author} {\bibfnamefont {L.}~\bibnamefont {Brey}},\
  }\bibfield  {title} {\bibinfo {title} {Skyrmions without sigma models in
  quantum {Hall} ferromagnets},\ }\href
  {https://doi.org/10.1103/PhysRevLett.76.2153} {\bibfield  {journal} {\bibinfo
   {journal} {Phys. Rev. Lett.}\ }\textbf {\bibinfo {volume} {76}},\ \bibinfo
  {pages} {2153} (\bibinfo {year} {1996})}\BibitemShut {NoStop}%
\bibitem [{\citenamefont {W{\'o}js}\ and\ \citenamefont
  {Quinn}(2002)}]{Wojs2002}%
  \BibitemOpen
  \bibfield  {author} {\bibinfo {author} {\bibfnamefont {A.}~\bibnamefont
  {W{\'o}js}}\ and\ \bibinfo {author} {\bibfnamefont {J.~J.}\ \bibnamefont
  {Quinn}},\ }\bibfield  {title} {\bibinfo {title} {Spin excitation spectra of
  integral and fractional quantum {{Hall}} systems},\ }\href
  {https://doi.org/10.1103/PhysRevB.66.045323} {\bibfield  {journal} {\bibinfo
  {journal} {Physical Review B}\ }\textbf {\bibinfo {volume} {66}},\ \bibinfo
  {pages} {045323} (\bibinfo {year} {2002})}\BibitemShut {NoStop}%
\bibitem [{\citenamefont {Zhang}\ \emph {et~al.}(2019)\citenamefont {Zhang},
  \citenamefont {Mao}, \citenamefont {Cao}, \citenamefont {{Jarillo-Herrero}},\
  and\ \citenamefont {Senthil}}]{Zhang2019}%
  \BibitemOpen
  \bibfield  {author} {\bibinfo {author} {\bibfnamefont {Y.-H.}\ \bibnamefont
  {Zhang}}, \bibinfo {author} {\bibfnamefont {D.}~\bibnamefont {Mao}}, \bibinfo
  {author} {\bibfnamefont {Y.}~\bibnamefont {Cao}}, \bibinfo {author}
  {\bibfnamefont {P.}~\bibnamefont {{Jarillo-Herrero}}},\ and\ \bibinfo
  {author} {\bibfnamefont {T.}~\bibnamefont {Senthil}},\ }\bibfield  {title}
  {\bibinfo {title} {Nearly flat {Chern} bands in moir\'e superlattices},\
  }\href {https://doi.org/10.1103/PhysRevB.99.075127} {\bibfield  {journal}
  {\bibinfo  {journal} {Physical Review B}\ }\textbf {\bibinfo {volume} {99}},\
  \bibinfo {pages} {075127} (\bibinfo {year} {2019})}\BibitemShut {NoStop}%
\bibitem [{\citenamefont {Repellin}\ and\ \citenamefont
  {Senthil}(2020)}]{Repellin2020}%
  \BibitemOpen
  \bibfield  {author} {\bibinfo {author} {\bibfnamefont {C.}~\bibnamefont
  {Repellin}}\ and\ \bibinfo {author} {\bibfnamefont {T.}~\bibnamefont
  {Senthil}},\ }\bibfield  {title} {\bibinfo {title} {Chern bands of twisted
  bilayer graphene: Fractional {Chern} insulators and spin phase transition},\
  }\href {https://doi.org/10.1103/PhysRevResearch.2.023238} {\bibfield
  {journal} {\bibinfo  {journal} {Phys. Rev. Res.}\ }\textbf {\bibinfo {volume}
  {2}},\ \bibinfo {pages} {023238} (\bibinfo {year} {2020})}\BibitemShut
  {NoStop}%
\bibitem [{\citenamefont {Okubo}\ \emph {et~al.}(2012)\citenamefont {Okubo},
  \citenamefont {Chung},\ and\ \citenamefont {Kawamura}}]{Okubo2012}%
  \BibitemOpen
  \bibfield  {author} {\bibinfo {author} {\bibfnamefont {T.}~\bibnamefont
  {Okubo}}, \bibinfo {author} {\bibfnamefont {S.}~\bibnamefont {Chung}},\ and\
  \bibinfo {author} {\bibfnamefont {H.}~\bibnamefont {Kawamura}},\ }\bibfield
  {title} {\bibinfo {title} {Multiple-$q$ states and the skyrmion lattice of
  the triangular-lattice {Heisenberg} antiferromagnet under magnetic fields},\
  }\href {https://doi.org/10.1103/PhysRevLett.108.017206} {\bibfield  {journal}
  {\bibinfo  {journal} {Phys. Rev. Lett.}\ }\textbf {\bibinfo {volume} {108}},\
  \bibinfo {pages} {017206} (\bibinfo {year} {2012})}\BibitemShut {NoStop}%
\bibitem [{\citenamefont {Barkeshli}\ and\ \citenamefont
  {McGreevy}(2012)}]{Barkeshli2012}%
  \BibitemOpen
  \bibfield  {author} {\bibinfo {author} {\bibfnamefont {M.}~\bibnamefont
  {Barkeshli}}\ and\ \bibinfo {author} {\bibfnamefont {J.}~\bibnamefont
  {McGreevy}},\ }\bibfield  {title} {\bibinfo {title} {Continuous transitions
  between composite {{Fermi}} liquid and {{Landau Fermi}} liquid: {{A}} route
  to fractionalized {{Mott}} insulators},\ }\href
  {https://doi.org/10.1103/PhysRevB.86.075136} {\bibfield  {journal} {\bibinfo
  {journal} {Physical Review B}\ }\textbf {\bibinfo {volume} {86}},\ \bibinfo
  {pages} {075136} (\bibinfo {year} {2012})}\BibitemShut {NoStop}%
\bibitem [{\citenamefont {Dong}\ and\ \citenamefont
  {Levitov}(2022)}]{Dong2022a}%
  \BibitemOpen
  \bibfield  {author} {\bibinfo {author} {\bibfnamefont {Z.}~\bibnamefont
  {Dong}}\ and\ \bibinfo {author} {\bibfnamefont {L.}~\bibnamefont {Levitov}},\
  }\href {https://doi.org/10.48550/arXiv.2208.02051} {\bibinfo {title} {Chiral
  {{Stoner}} magnetism in {{Dirac}} bands}} (\bibinfo {year} {2022}),\ \Eprint
  {https://arxiv.org/abs/2208.02051} {arxiv:2208.02051 [cond-mat]} \BibitemShut
  {NoStop}%
\bibitem [{\citenamefont {Zhu}\ \emph {et~al.}(2022)\citenamefont {Zhu},
  \citenamefont {Sheng},\ and\ \citenamefont {Vishwanath}}]{Zhu2022}%
  \BibitemOpen
  \bibfield  {author} {\bibinfo {author} {\bibfnamefont {Z.}~\bibnamefont
  {Zhu}}, \bibinfo {author} {\bibfnamefont {D.~N.}\ \bibnamefont {Sheng}},\
  and\ \bibinfo {author} {\bibfnamefont {A.}~\bibnamefont {Vishwanath}},\
  }\bibfield  {title} {\bibinfo {title} {Doped {{Mott}} insulators in the
  triangular-lattice {{Hubbard}} model},\ }\href
  {https://doi.org/10.1103/PhysRevB.105.205110} {\bibfield  {journal} {\bibinfo
   {journal} {Physical Review B}\ }\textbf {\bibinfo {volume} {105}},\ \bibinfo
  {pages} {205110} (\bibinfo {year} {2022})}\BibitemShut {NoStop}%
\bibitem [{\citenamefont {Iglovikov}\ \emph {et~al.}(2015)\citenamefont
  {Iglovikov}, \citenamefont {Khatami},\ and\ \citenamefont
  {Scalettar}}]{Iglovikov2015}%
  \BibitemOpen
  \bibfield  {author} {\bibinfo {author} {\bibfnamefont {V.~I.}\ \bibnamefont
  {Iglovikov}}, \bibinfo {author} {\bibfnamefont {E.}~\bibnamefont {Khatami}},\
  and\ \bibinfo {author} {\bibfnamefont {R.~T.}\ \bibnamefont {Scalettar}},\
  }\bibfield  {title} {\bibinfo {title} {Geometry dependence of the sign
  problem in quantum {Monte} {Carlo} simulations},\ }\href
  {https://doi.org/10.1103/PhysRevB.92.045110} {\bibfield  {journal} {\bibinfo
  {journal} {Phys. Rev. B}\ }\textbf {\bibinfo {volume} {92}},\ \bibinfo
  {pages} {045110} (\bibinfo {year} {2015})}\BibitemShut {NoStop}%
\bibitem [{\citenamefont {Schattner}\ \emph {et~al.}(2016)\citenamefont
  {Schattner}, \citenamefont {Gerlach}, \citenamefont {Trebst},\ and\
  \citenamefont {Berg}}]{Schattner2016}%
  \BibitemOpen
  \bibfield  {author} {\bibinfo {author} {\bibfnamefont {Y.}~\bibnamefont
  {Schattner}}, \bibinfo {author} {\bibfnamefont {M.~H.}\ \bibnamefont
  {Gerlach}}, \bibinfo {author} {\bibfnamefont {S.}~\bibnamefont {Trebst}},\
  and\ \bibinfo {author} {\bibfnamefont {E.}~\bibnamefont {Berg}},\ }\bibfield
  {title} {\bibinfo {title} {Competing orders in a nearly antiferromagnetic
  metal},\ }\href {https://doi.org/10.1103/PhysRevLett.117.097002} {\bibfield
  {journal} {\bibinfo  {journal} {Phys. Rev. Lett.}\ }\textbf {\bibinfo
  {volume} {117}},\ \bibinfo {pages} {097002} (\bibinfo {year}
  {2016})}\BibitemShut {NoStop}%
\end{thebibliography}%

\clearpage

\appendix

\setcounter{figure}{0}
\renewcommand{\thefigure}{S\arabic{figure}}
\renewcommand{\theHfigure}{S\arabic{figure}}

\section{Boundary conditions} 
\label{sec:bdy_cond}

As mentioned in the main text, the real space primitive lattice vectors of the triangular lattice is 
\begin{equation}
\vec{a}_1 = \frac{a}{2}\begin{bmatrix}
    1 \\
    \sqrt{3}
\end{bmatrix}, \quad \vec{a}_2 = \frac{a}{2}\begin{bmatrix}
    -1\\
    \sqrt{3}
\end{bmatrix}. \nonumber
\end{equation}

Each lattice site has 6 nearest neighbors located at $\pm \vec{a}_{1,2,3}$, where $\vec{a}_3 = (a,0)$.

Lattice sites indexed by $i = (i_1, i_2)$ and $j = (j_1, j_2)$ have spatial positions
\begin{align}
    \vec{r}_i &= i_1 \vec{a}_1 + i_2 \vec{a}_2 = 
    \frac{i_1}{2} \begin{bmatrix}
        1\\
        \sqrt{3}
    \end{bmatrix} + \frac{i_2}{2} \begin{bmatrix}
        -1\\
        \sqrt{3}
    \end{bmatrix} \nonumber\\
    &= \begin{bmatrix}
        \dfrac{1}{2}(i_1 - i_2)\\
        \dfrac{\sqrt{3}}{2}(i_1 + i_2)
    \end{bmatrix} = \begin{bmatrix}
        r_{ix}\\
        r_{iy}
    \end{bmatrix}, \\
    \vec{r}_j &= j_1 \vec{a}_1 + j_2 \vec{a}_2 \nonumber \\
    &= 
    \begin{bmatrix}
        \dfrac{1}{2}(j_1 - j_2)\\
        \dfrac{\sqrt{3}}{2}(j_1 + j_2)
    \end{bmatrix} = \begin{bmatrix}
        r_{jx}\\
        r_{jy}
    \end{bmatrix}.
\end{align}
The midpoint between site $i$ and site $j$ has position
\begin{multline}
    \vec{r}_m(i,j) = \frac{1}{2}(\vec{r}_i + \vec{r}_j) = \frac{1}{2}\begin{bmatrix}
        r_{ix} + r_{jx}\\
        r_{iy} + r_{jy}
    \end{bmatrix} \\
    =  \frac{1}{4} \begin{bmatrix}
            j_1 - j_2 + i_1 - i_2\\
            \sqrt{3}(j_1 + j_2 + i_1 + i_2) 
    \end{bmatrix}= 
    \begin{bmatrix}
        r_{mx}\\
        r_{my}
    \end{bmatrix}.
\end{multline}
The difference between site $i$ and site $j$, indexed by $\Delta = j-i = (d_1, d_2)$, for integration path $\vec{r}_i \rightarrow \vec{r}_j$, corresponds to a position change of
\begin{align}
    \vec{r}_\Delta(i,j) &= \vec{r}_j -\vec{r}_i = 
    \begin{bmatrix}
        r_{jx} - r_{ix}\\
        r_{jy} - r_{iy}
    \end{bmatrix} \nonumber\\
    &= \begin{bmatrix}
        \dfrac{1}{2}(d_1 - d_2)\\
        \dfrac{\sqrt{3}}{2}(d_1 + d_2)
    \end{bmatrix} = \begin{bmatrix}
        r_{dx}\\
        r_{dy}
    \end{bmatrix}.
\end{align}

The integral in~\cref{eq:peierls-phase} is performed by integrating over the shortest straight line path from $\vec{r}_i$ to $\vec{r}_j$; in discrete form, it is (setting $\beta=1-\alpha$ for this section only) 
\begin{align}
    \varphi_{ij} &= \frac{2\pi}{\Phi_0} \vec{A} ( \vec{r}_m) \cdot (\vec{r}_j - \vec{r}_i) \nonumber\\
    &= \frac{2\pi}{\Phi_0} \frac{2 N_\phi \Phi_0}{\sqrt{3}a^2  N_1 N_2} \begin{bmatrix}
        -\alpha r_{my}\\
        (1-\alpha) r_{mx}
    \end{bmatrix} \cdot \begin{bmatrix}
        r_{dx}\\
        r_{dy}
    \end{bmatrix} \nonumber\\
    &= \frac{2\pi \cdot 2 \cdot N_\phi}{\sqrt{3}N_1 N_2}\left[ -\alpha r_{my} r_{dx} + \beta r_{mx} r_{dy}\right] \nonumber\\
    &= \boxed{p \left[ -\alpha r_{my} r_{dx} + \beta r_{mx} r_{dy}\right]}\label{eq:phase-interior}
\end{align}
where the prefactor $p = \dfrac{2\pi}{\Phi_0} B$ represents
\begin{equation}
    p = \frac{2\pi \cdot 2 \cdot N_\phi}{\sqrt{3}N_1 N_2}. \label{eq:prefactor}
\end{equation}

Note that \cref{eq:phase-interior} is the phase accumulated in the $c_i^\dagger c_j$ term, i.e. $j \rightarrow i$ electron hopping, while the integral is performed along path $i\rightarrow j$. 

\cref{eq:phase-interior} only works for internal points, when the path $j\rightarrow i$ does not cross the finite cluster boundary.
We deal with the boundary-crossing case by implementing modified periodic boundary conditions consistent with magnetic translation symmetry~\cite{Assaad2002,Xiao2010}. This amounts to ensuring that the Hamiltonian translated by $N_1 \vec{a}_1$ and $N_2 \vec{a}_2$ is the same as the original Hamiltonian by making the identification for $j$ in a finite cluster as (so that $j \pm N_1 \vec{a}_1$, $j \pm N_2 \vec{a}_2$ are out of bounds):
\begin{align}
    c_j &= c_{j \pm N_1 \vec{a}_1} \exp\left[i \frac{2\pi}{\Phi_0}\vec{A}( \pm N_1 \vec{a}_1) \cdot \vec{r}_j\right],\\
    c_j^\dagger &= c^\dagger_{j\pm N_1 \vec{a}_1} \exp\left[i \frac{2\pi}{\Phi_0}\vec{A}( \pm N_1 \vec{a}_1) \cdot (-\vec{r}_j)\right],
\end{align}
and 
\begin{align}
    c_j &= c_{\pm N_2 \vec{a}_2} \exp\left[i \frac{2\pi}{\Phi_0}\vec{A}(\pm N_2 \vec{a}_2) \cdot \vec{r}_j\right],\\
    c_j^\dagger &= c^\dagger_{j\pm N_2 \vec{a}_2} \exp\left[i \frac{2\pi}{\Phi_0}\vec{A}(\pm N_2 \vec{a}_2) \cdot (-\vec{r}_j)\right].
\end{align}

For example, consider one term in the Hamiltonian in a $4\times 4$ cluster, 
\begin{equation}
    c_{i}^\dagger c_{j} \exp\left[i\frac{2\pi}{\Phi_0} \vec{A}(\vec{r}_m(i,j)) \cdot \vec{r}_\Delta (i,j)\right].
\end{equation}
Suppose $i = (3,2)$ and $j = (4,2)$. $j$ is out of bounds, so we must ``wrap'' it to $jj = (0,2)$. The index offset caused by this wrapping in the $\vec{a}_1$ direction is $j - jj = (N_1,0)$, so
\begin{multline}
    c_{jj} = c_j \exp\left[i \frac{2\pi}{\Phi_0}\vec{A}(N_1 \vec{a}_1) \cdot \vec{r}_{jj}\right] \quad \Rightarrow \\
    \quad c_j = c_{jj} \exp\left[i \frac{2\pi}{\Phi_0}\vec{A}(N_1 \vec{a}_1) \cdot (-\vec{r}_{jj})\right],
\end{multline}
where the extra phase term is 
\begin{align}
    u_1 &= p \cdot (-1) \cdot 
    \begin{bmatrix}
        -\alpha (N_1 a_1)_y\\
        \beta (N_1 a_1)_x
    \end{bmatrix}\cdot 
    \begin{bmatrix}
        r_{jjx}\\
        r_{jjy}
    \end{bmatrix} \label{eq:wrap-x-phase}
    \\
    &= p \cdot (-1) \cdot \left[ -\alpha \cdot N_1 \cdot \frac{\sqrt{3}}{2} \cdot r_{jj x} + \beta \cdot N_1 \cdot \frac{1}{2} \cdot r_{jj y}\right]. \nonumber
\end{align}
Similarly, suppose $i = (1,3)$ and $j = (1,4)$. $j$ is out of bounds, so we must ``wrap'' it to $jj = (1,0)$. The index offset caused by this wrapping in the $\vec{a}_2$ direction is $j - jj = (0,N_2)$, so
\begin{equation}
    c_j = c_{jj} \exp\left[i \frac{2\pi}{\Phi_0}\vec{A}(N_2 \vec{a}_2) \cdot (-\vec{r}_{jj})\right].
\end{equation}
The extra phase term is 
\begin{align}
    u_2 &= p \cdot (-1) \cdot 
    \begin{bmatrix}
        -\alpha (N_2 a_2)_y\\
        \beta (N_2 a_2)_x
    \end{bmatrix}\cdot 
    \begin{bmatrix}
        r_{jjx}\\
        r_{jjy}
    \end{bmatrix} \label{eq:wrap-y-phase}
    \\
    &=  p \cdot (-1) \cdot \left[ -\alpha \cdot N_2 \cdot \frac{\sqrt{3}}{2} \cdot r_{jj x} + \beta \cdot N_1 \cdot \frac{-1}{2} \cdot r_{jj y}\right]. \nonumber
\end{align}

Now we consider what happens if the path $j\rightarrow i$ crosses two boundaries. Suppose $i = (3,3)$ and $j = (5,4)$. Then $j$ needs to be wrapped twice to return to the  $jj = (1,0)$ within the original cluster. The index offset caused by this wrapping is $j - jj = (N_1,N_2)$.
There are two possible routes to take, and they produce two different resulting boundary phases. 

Route 1:
\begin{multline}
    c_{(5,4)} = 
    \exp\left[\frac{2\pi}{\Phi_0} \vec{A}(N_2 \vec{a}_2) \cdot (-\vec{r}_{(5,0)})\right]  \cdot\\
    \exp\left[\frac{2\pi}{\Phi_0}\vec{A}(N_1 \vec{a}_1) \cdot (-\vec{r}_{(1,0)})\right]c_{(1,0)},
\end{multline}
so the phase produced here is 
\begin{multline}
    v_1 = p \cdot (-1) \cdot \left\{
    \begin{bmatrix}
        -\alpha (N_2 \vec{a}_2)_y \\
        \beta (N_2 \vec{a}_2)_x
    \end{bmatrix}\cdot
    \begin{bmatrix}
        r_{jjx} + (N_1 \vec{a}_1)_x\\
        r_{jjy} + (N_1 \vec{a}_1)_y
    \end{bmatrix} + \right.\\
    \left .
    \begin{bmatrix}
        -\alpha (N_1 \vec{a}_1)_y \\
        \beta (N_1 \vec{a}_1)_x
    \end{bmatrix}\cdot
    \begin{bmatrix}
        r_{jjx} \\
        r_{jjy} 
    \end{bmatrix}
    \right\}. \label{eq:xy-route1}
\end{multline}

Route 2:
\begin{multline}
    c_{(5,4)} = 
    \exp\left[\frac{2\pi}{\Phi_0} \vec{A}(N_1 \vec{a}_1) \cdot (-\vec{r}_{(1,4)})\right] \cdot \\
    \exp\left[\frac{2\pi}{\Phi_0}\vec{A}(N_2 \vec{a}_2) \cdot (-\vec{r}_{(1,0)})\right]c_{(1,0)}
\end{multline}
so the phase produced here is 
\begin{multline}
    v_2 = p \cdot (-1) \cdot \left\{
    \begin{bmatrix}
        -\alpha (N_1 \vec{a}_1)_y \\
        \beta (N_1 \vec{a}_1)_x
    \end{bmatrix}\cdot
    \begin{bmatrix}
        r_{jjx} + (N_2 \vec{a}_2)_x\\
        r_{jjy} + (N_2 \vec{a}_2)_y
    \end{bmatrix} \right .\\
    \left .+ 
    \begin{bmatrix}
        -\alpha (N_2 \vec{a}_2)_y \\
        \beta (N_2 \vec{a}_2)_x
    \end{bmatrix}\cdot
    \begin{bmatrix}
        r_{jjx} \\
        r_{jjy} 
    \end{bmatrix}
    \right\}. \label{eq:xy-route2}
\end{multline}

Comparing~\cref{eq:xy-route1} and~\cref{eq:xy-route2}, we find
\begin{align}
    v_1 = u + p \cdot (-1) \cdot 
    \begin{bmatrix}
        -\alpha (N_2 \vec{a}_2)_y \\
        \beta (N_2 \vec{a}_2)_x
    \end{bmatrix}\cdot
    \begin{bmatrix}
        (N_1 \vec{a}_1)_x\\
        (N_1 \vec{a}_1)_y
    \end{bmatrix} \label{eq:v1}\\
    = u + p \cdot (-1) \cdot \left[ \frac{\sqrt{3}}{4} N_1 N_2 (-\alpha - \beta)\right], \nonumber\\
    v_2 = u + p \cdot (-1) \cdot 
    \begin{bmatrix}
        -\alpha (N_1 \vec{a}_1)_y \\
        \beta (N_1 \vec{a}_1)_x
    \end{bmatrix}\cdot
    \begin{bmatrix}
        (N_2 \vec{a}_2)_x\\
        (N_2 \vec{a}_2)_y
    \end{bmatrix}\label{eq:v2}\\
    = u + p \cdot (-1) \cdot \left[ \frac{\sqrt{3}}{4} N_1 N_2 (\alpha + \beta)\right], \nonumber
\end{align}
where
\begin{align}
    u = p \cdot (-1) \cdot \left\{
    \begin{bmatrix}
        -\alpha (N_1 \vec{a}_1 + N_2 \vec{a}_2)_y \\
        \beta (N_1 \vec{a}_1 + N_2 \vec{a}_2)_x
    \end{bmatrix}\cdot
    \begin{bmatrix}
        r_{jjx}  \\
        r_{jjy} 
    \end{bmatrix}\right\}, \label{eq:wrap-general-phase}
\end{align}
which allows us to see that~\cref{eq:wrap-x-phase,eq:wrap-y-phase} are special cases of~\cref{eq:wrap-general-phase}.

Moreover, comparing~\cref{eq:v1,eq:v2} we find 
\begin{equation}
    v_1 - v_2 = p \frac{\sqrt{3}}{2} N_1 N_2  = \frac{2\pi \cdot 2 \cdot N_\phi}{\sqrt{3}N_1 N_2} \frac{\sqrt{3}}{2} N_1 N_2 = 2\pi N_\phi.
\end{equation}
Thus the two routes produce the same extra boundary phase factor after taking the complex exponential. In implementation, we just choose one term to add, either $v_1$ or $v_2$.

\section{Simulation Parameters} 
\label{sec:simulation_parameters}

Determinant quantum Monte Carlo (DQMC) data are obtained from simulations performed using $2\times 10^4$ to $5\times 10^4$ warm-up sweeps and $2\times 10^5$ to $2 \times 10^6$ measurement sweeps through the auxiliary field. We run 20-400 independently seeded Markov chains for each individual set of parameters. For all parameter values, the imaginary time discretization interval $\Delta\tau \leq 0.1/t$, and the number of imaginary time slices $L = \beta/\Delta \tau \geq 20$. 
The imaginary time discretization interval in DQMC simulations satisfies the conventional heuristic $U\Delta \tau^2 \lesssim 1/W$, where $W$ is the bandwidth of the tight-binding model, for all $U/t \leq 10$ used in this work. The smallest cluster size is $6\times 6$ and the largest cluster size is $12\times 12$. We show that our DQMC results have minimal finite-size effects in~\cref{sec:finite_size}.

In all DQMC simulations, multiple equal-time measurements are taken in each full measurement sweep through the auxiliary field. Each Markov chain with $M$ measurement sweeps collects $M L /5$ equal-time measurements.
The standard error of the mean, where shown, are estimated via jackknife resampling of independent Markov chains.

\section{DQMC Fermion sign}
\label{sec:fermion_sign}
\begin{figure}[htpb]
    \includegraphics[width=\linewidth]{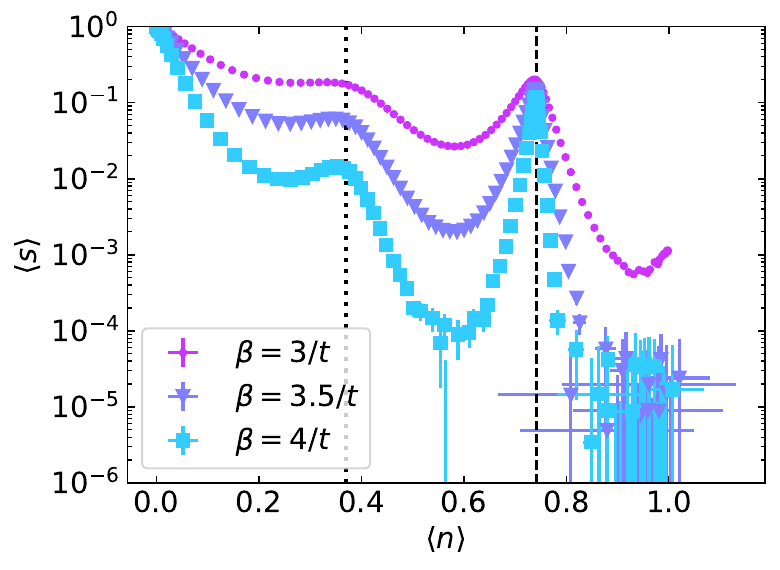} 
    \caption{Temperature dependence of fermion sign at field strength $\Phi/\Phi_0 = 30/81$ on $9\times 9$ cluster. Hubbard $U/t=10$. Dotted line marks $\nu=1$ and dashed line marks $\nu=2$.}
    \label{fig:fermion-sign}
\end{figure}
Historically, there have been limited DQMC studies of the Hubbard model on the triangular lattice (and other non-bipartite lattices), because these systems in general have bad fermion sign problem~\cite{Iglovikov2015}. The fermion sign worsens with larger cluster size, larger Hubbard $U$, and lower temperatures. In~\cref{fig:fermion-sign}, we show the average fermion sign $\expval{s}$ for one set of representative parameters. We observe improvements of fermion sign at more incompressible regions, consistent with previous studies~\cite{Ding2022,Mai2023}. The sign problem limits the temperatures we can access via DQMC to $\beta \leq 5/t$ at $U=10t$ and $\beta \leq 8/t$ at $U=4t$. 
As detailed in~\cref{sec:simulation_parameters}, we run a large number of sweeps and bins to overcome the sign problem, which allows us to see indicators of ground state properties even at relatively high temperatures accessible via DQMC.

\section{DQMC finite-size analysis}
\label{sec:finite_size}
\begin{figure}[htpb]
    \includegraphics[width=\linewidth]{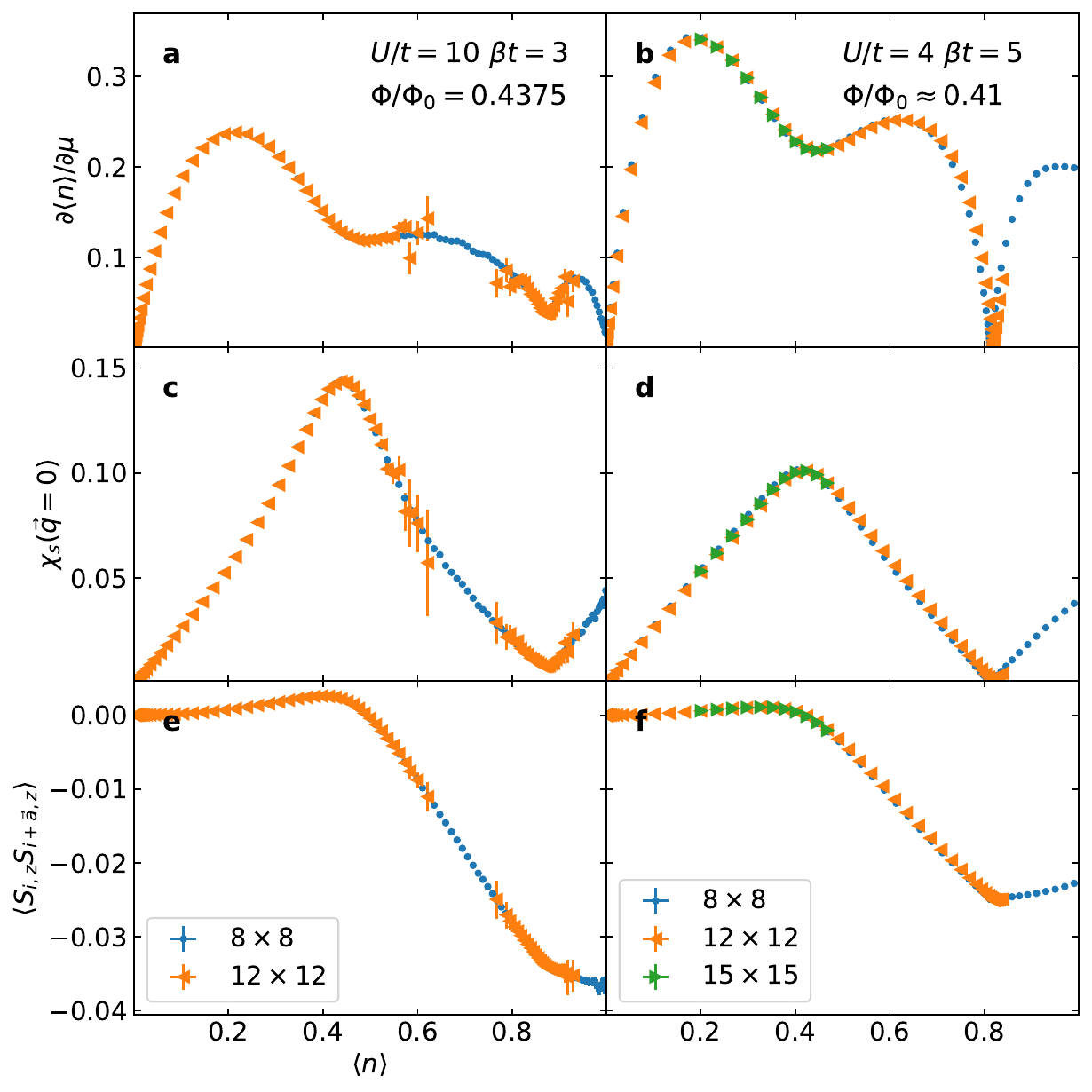} 
    \caption{Comparison of DQMC results obtained on $8\times 8$, $12 \times 12$, and $15\times 15$ clusters for \textbf{a}-\textbf{b} charge compressibility, \textbf{c}-\textbf{d} uniform spin susceptibility, and \textbf{e}-\textbf{f} nearest neighbor spin correlation. Each column shows one set of representative parameters, so that \textbf{a}, \textbf{c}, \textbf{e} corresponds to fixed field strength $\Phi/\Phi_0 = 28/64 = 63/144$, Hubbard interaction $U/t=10$ and inverse temperature $\beta=3/t$. \textbf{b}, \textbf{d}, \textbf{f} corresponds to approximately fixed field strength $\Phi/\Phi_0 \approx 0.41 \approx 26/64 \approx 59/144 \approx 92/225$, Hubbard interaction $U/t=4$ and inverse temperature $\beta=5/t$. The field strength in the latter case can only be approximately fixed, because the magnetic field strength takes quantized values in finite clusters.}
    \label{fig:finite-size}
\end{figure}

\begin{figure}[htpb]
    \fbox{\includegraphics[width=0.9\linewidth]{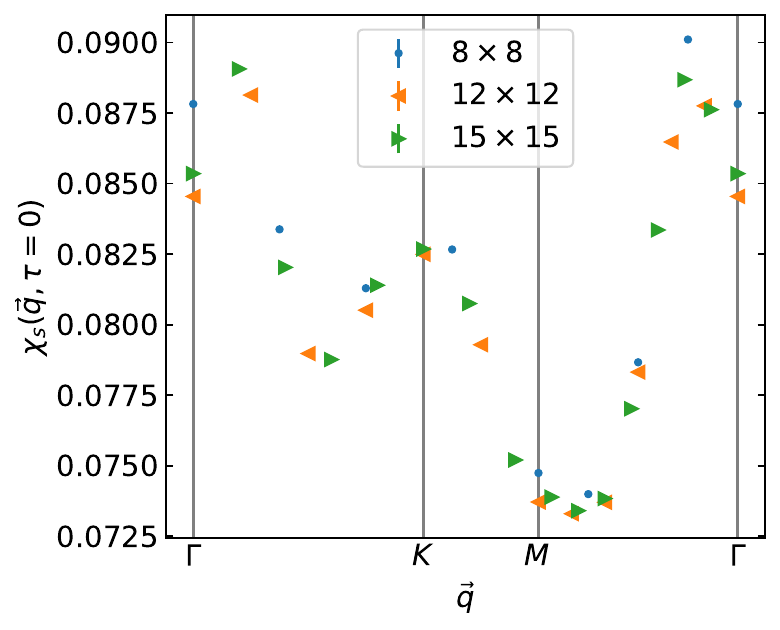}}
    \caption{Comparison of DQMC results obtained on $8\times 8$, $12 \times 12$, and $15\times 15$ clusters for equal time spin correlation $\chi_s(\vec{q},\tau=0)$ along the high symmetry path $\Gamma - K - M - \Gamma$. Fixed inverse temperature $\beta = 5/t$, Hubbard interaction $U/t = 4$, and particle density $\langle n\rangle = 0.33$. Approximately fixed field strength $\Phi/\Phi_0 \approx 0.41 \approx 26/64 \approx 59/144 \approx 92/225$. The field strength can only be approximately fixed, because the magnetic field takes quantized values in finite clusters.}
    \label{fig:qmax-fs}
\end{figure}

\begin{figure}[htpb]
    \fbox{\includegraphics[width=\linewidth]{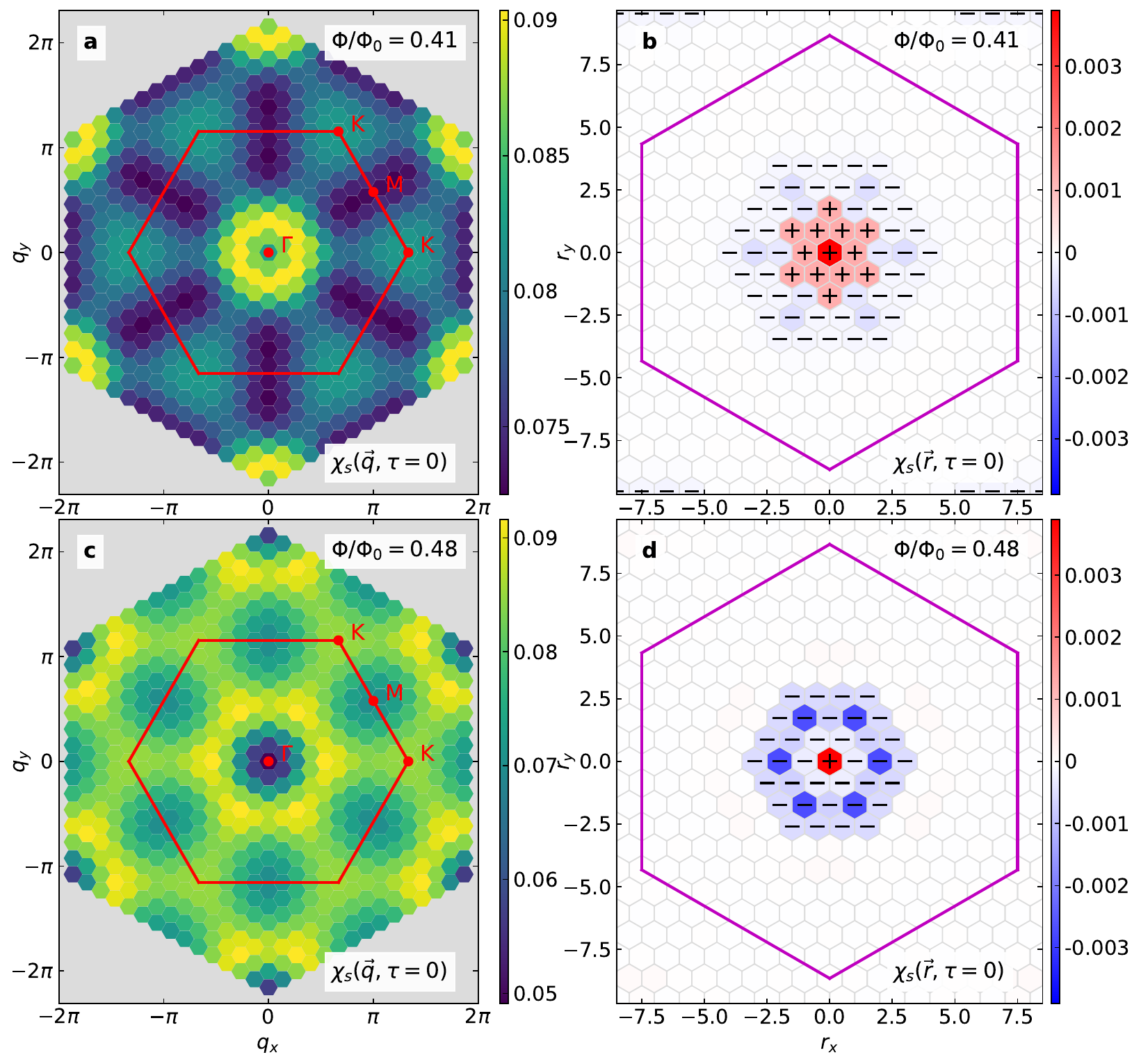}}
    \caption{Same as~\cref{fig:suscept-sample}\textbf{e}-\textbf{h} in main text, except using DQMC data obtained on a $15\times 15$ cluster. All plots have fixed inverse temperature $\beta = 6/t$, Hubbard interaction $U/t = 4$, and particle density $\langle n\rangle = 0.34$.}
    \label{fig:suscept-15x15}
\end{figure}

As discussed in~\cite{Mai2023}, the finite-size effect in DQMC simulations of the HH model depends on magnetic field strength $\Phi/\Phi_0$ and interaction strength $U/t$. In particular, the finite-size effect is in general smaller for ``more incommensurate'' values of $\Phi/\Phi_0$. This means that it is advantageous, from the perspective of minimizing finite-size effect on a fixed-size cluster, to consider magnetic field strength $\Phi/\Phi_0 = p/q$ with a large denominator $q$, where $p$ and $q$ are co-prime. Intuitively, we may understand this effect by noting that a large $q$ breaks the crystal Brillouin zone into magnetic Brillouin zones which are $q$ times smaller. This removes artificial degeneracies caused by the limited momentum resolution in a small cluster (termed ``shell effect'' in~\cite{Iglovikov2015}). This has been exploited in~\cite{Assaad2002,Schattner2016}, where one magnetic flux quantum was introduced to reduce finite-size effects for zero-field models.~\cref{fig:finite-size} shows the minimal cluster size dependence of charge compressibility and uniform spin susceptibility at two typical parameter combinations. \cref{fig:qmax-fs} shows similarly minimal finite size dependence for the $\vec{q}$-resolved equal-time spin susceptibilities. \cref{fig:suscept-15x15}, when viewed side-by-side with \cref{fig:suscept-sample}, demonstrates that salient spin correlation features of the high-field metal are also independent of system size. Many other similar checks allow us to be confident that our conclusions are valid in the thermodynamic limit.

\section{Non-interacting results}

\begin{figure}[htpb]
    \includegraphics[width=\linewidth]{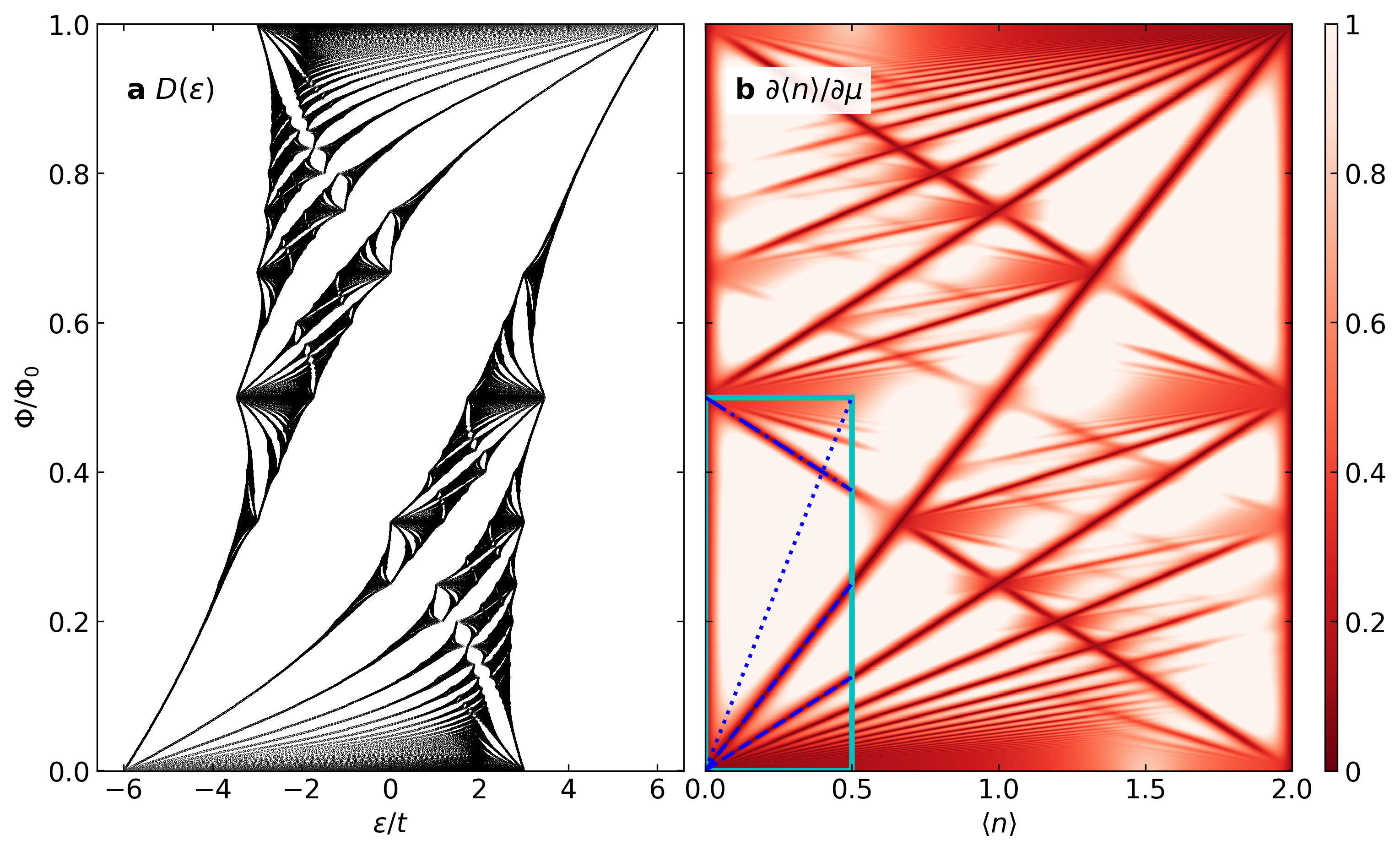} 
    \caption{Noninteracting Hofstadter model band structure. \textbf{a} density of states, AKA Hofstadter butterfly, and \textbf{b} Wannier diagram with 2-fold spin degeneracy, at inverse temperature $\beta=20/t$. In \textbf{b}, the cyan box demarcate the region of parameter space we focus on in this study; dotted blue line denotes $\nu=1$, dashed lines denote $\nu=2$ and $\nu=4$, while dot-dashed line denotes $n=-4 (\Phi/\Phi_0) + 2$, consistent with styles used in the main text. }
    \label{fig:nonint-dos-wannier}
\end{figure}

\begin{figure}[htpb]
    \includegraphics[width=\linewidth]{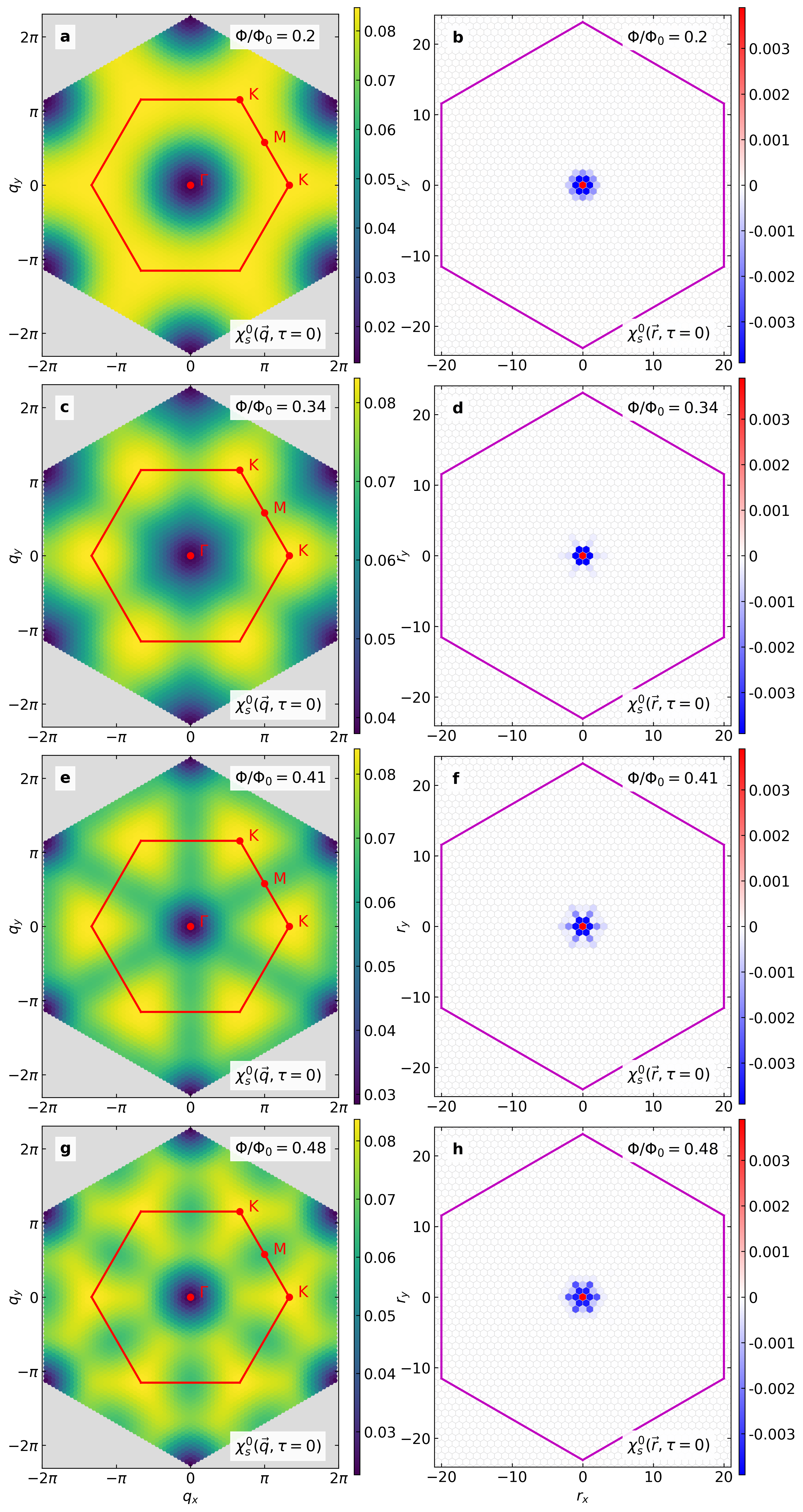} 
    \caption{Non-interacting counterpart of~\cref{fig:suscept-sample} in main text. All plots have fixed inverse temperature $\beta = 6/t$, and particle density $\langle n\rangle = 0.34$.}
    \label{fig:nonint-suscept}
\end{figure}

In this section, we show the band stucture and Wannier diagram (\cref{fig:nonint-dos-wannier}), and Lindhard susceptibilities (\cref{fig:nonint-suscept}) calculated in the non-interacting Hofstadter model, for easy comparison with results presented in the main text. All results are obtained by diagonalizing the Hofstadter model on a $40\times 40$ cluster.

The non-interacting band structure of the Hofstadter model, as shown in~\cref{fig:nonint-dos-wannier}, is $180^\circ$ rotation symmetric about $\Phi/\Phi_0=0.5$ and $\epsilon=0$ (or $\Phi/\Phi_0=0.5$ and $n=1$), reflection symmetric about $\Phi/\Phi_0 = \pm 1, \pm 3, ...$ and periodic along the $\Phi/\Phi_0$ direction with a period of 2.

The equal time spin susceptibilities shown in \cref{fig:suscept-sample,fig:nonint-suscept} are defined as
\begin{equation}
\chi_{s}(\vec{r},\tau=0) = \frac{1}{N} \sum_{i} e^{-i \vec{q}\cdot \vec{r}} \expval*{S_{i,z}S_{i+\vec{r},z}}, \label{eq:suscept-r}
\end{equation}
and 
\begin{align}
\chi_{s}(\vec{q},\tau=0) &= \frac{1}{N} \sum_{i,j} e^{-i \vec{q}\cdot (\vec{r}_i- \vec{r}_j)} \expval*{S_{i,z}S_{j,z}}, \label{eq:suscept-q}\\
&= \sum_{\vec{r}} e^{-i \vec{q}\cdot \vec{r}} \chi_{s}(\vec{r},\tau=0) \nonumber
\end{align}
where
\begin{equation}
    S_{i,z} = \frac{\hbar}{2}\left(c^\dagger_{i\uparrow}c_{i\uparrow} - c^\dagger_{i\downarrow}c_{i\downarrow}\right). \nonumber
\end{equation}
\newline

\section{Supplemental data}
\label{sec:supplemental-data}

\begin{figure}[htpb]
    \includegraphics[width=\linewidth]{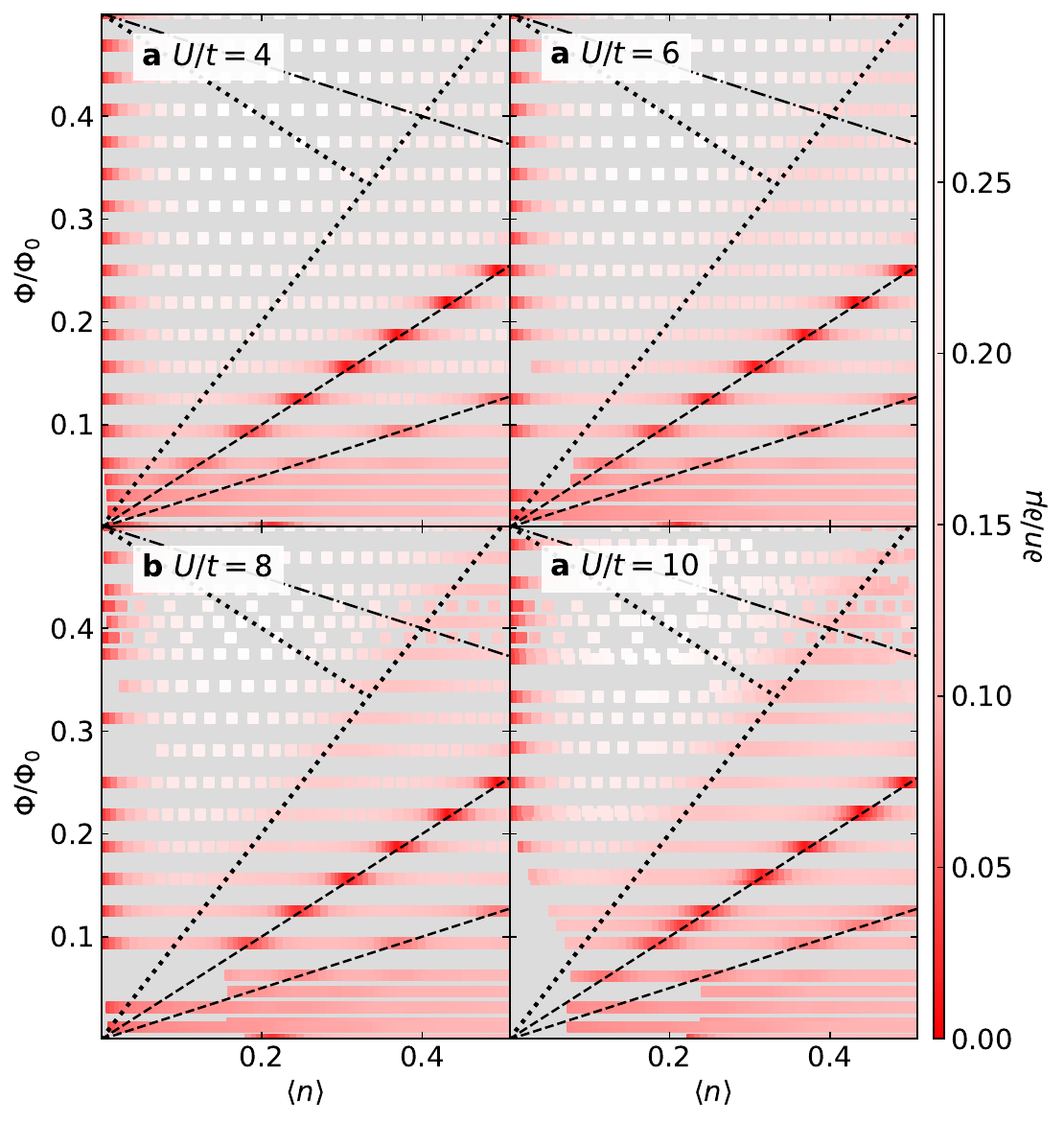} 
    \caption{Wannier diagram for Hubbard $U/t=4$ to $U/t=10$. Color: charge compressibity $\partial \langle n \rangle / \partial \mu$. Fixed inverse temperature $\beta=4/t$. Dotted lines denote $\nu=1$ and $n=-2 (\Phi/\Phi_0) + 1$, dashed lines denote $\nu=2$ and $\nu=4$, and dot-dashed lines denote $n=-4 (\Phi/\Phi_0) + 2$. Grey regions are where we don't have simulation data. All subplots share the same colorbar. DQMC data are obtained on clusters of size $8\times 8$ and $9\times 9$.}
    \label{fig:wannier}
\end{figure}

\begin{figure}[htpb]
    \includegraphics[width=\linewidth]{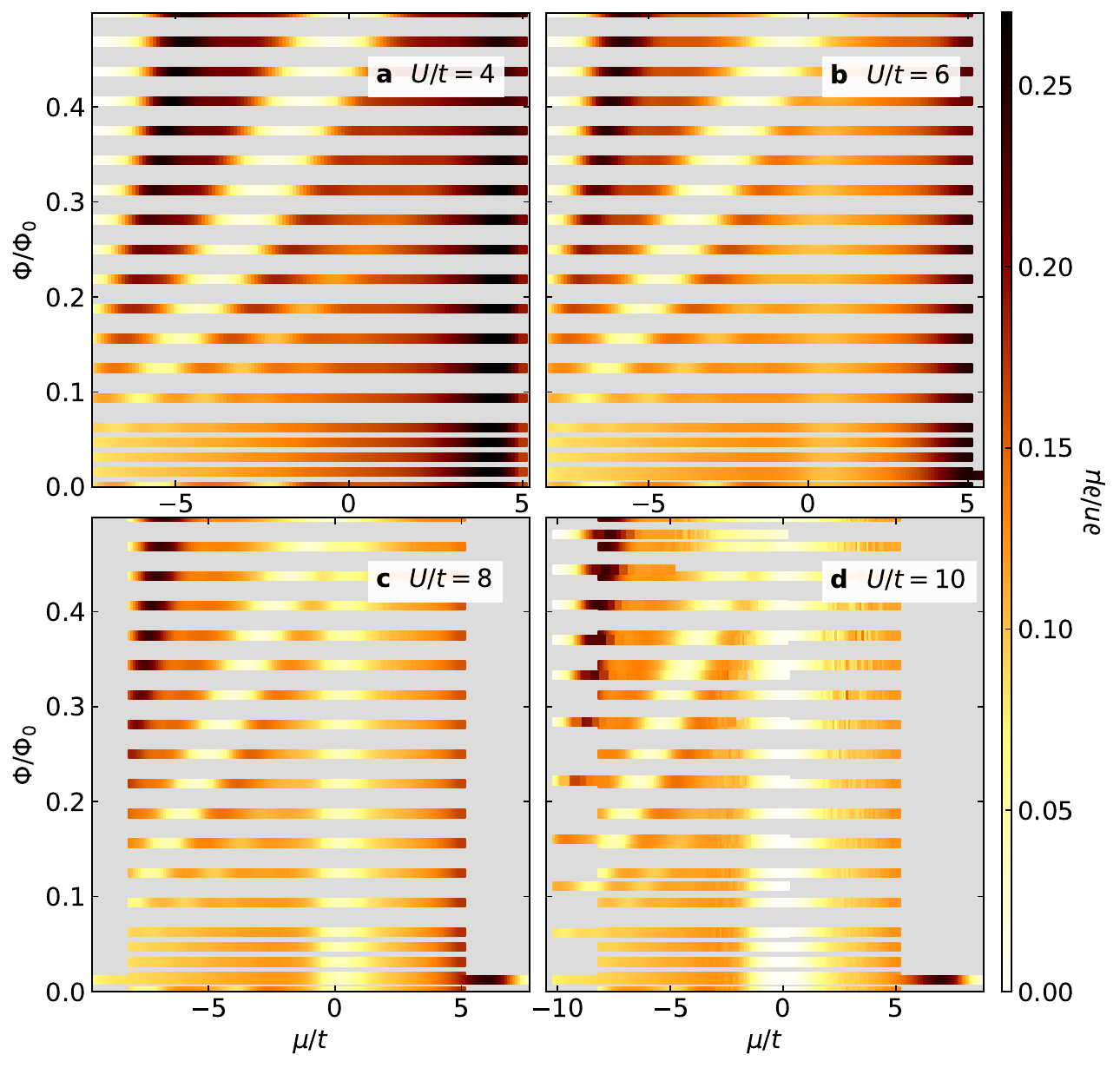}
    \caption{Correlated Hofstadter butterfly for Hubbard $U/t=4$ to $U/t=10$. Color: charge compressibity $\partial \langle n \rangle / \partial \mu$. Fixed inverse temperature $\beta=3/t$. Grey regions are where we don't have simulation data. All subplots share the same colorbar. DQMC data are obtained on clusters of size $8\times 8$ and $9\times 9$.}
    \label{fig:corr-butterfly}
\end{figure}

\begin{figure}[htpb]
    \includegraphics[width=\linewidth]{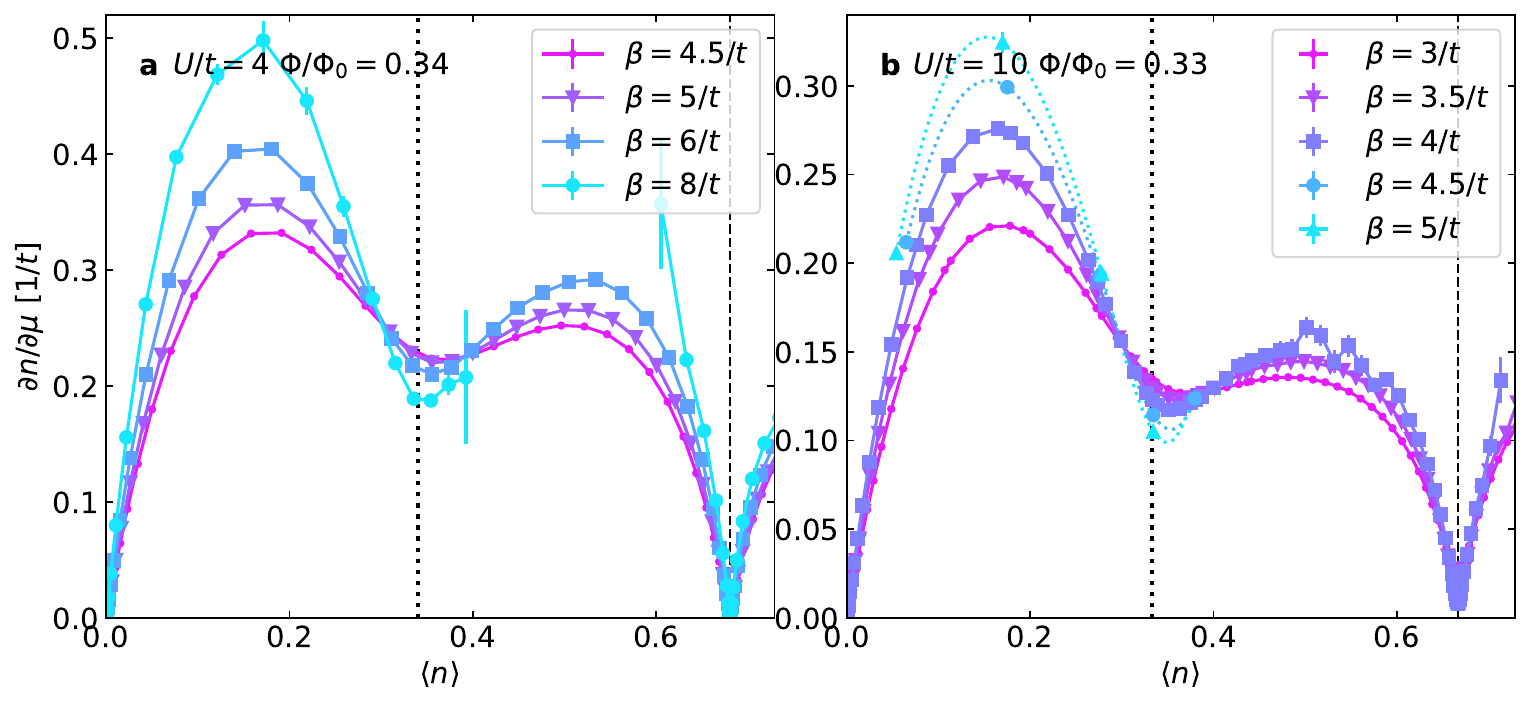} 
    \caption{Temperature dependence of charge compressibility $\partial n/\partial \mu$ at fixed magnetic field strength $\Phi/\Phi_0 \approx 0.33$ for \textbf{a} $U/t=4$ and \textbf{b} $U/t=10$. Dotted black lines denote $\nu=1$, and dashed black lines denote $\nu=2$. DQMC data are obtained on \textbf{a} $12\times 12$ cluster and \textbf{b} $8\times8$ and $9\times 9$ clusters, respectively. In \textbf{b}, dotted lines going through data points for $\beta=4.5/t$ and $\beta=5/t$ are guides to the eye obtained via cubic spline fitting.}
    \label{fig:compress-T-dep}
\end{figure}

Charge compressibility $\partial \langle n \rangle /\partial \mu$ is measured in DQMC using the formula
\begin{equation}
\frac{\partial \langle n \rangle}{\partial \mu} = \frac{\beta}{N} \sum_{ij}\left[\langle n_i n_j\rangle  - \langle n_i \rangle \langle n_j \rangle \right].
\end{equation}
Plots of charge compressibility at fixed temperature and for $U/t = 4$-$10$ are shown as
Wannier diagrams (\cref{fig:wannier}) and ``correlated Hofstadter butterflies'' (\cref{fig:corr-butterfly}). 
The temperature dependence of charge compressibility at fixed magnetic field strength is shown in \cref{fig:compress-T-dep}.

\cref{fig:compress-T-dep} shows the temperature dependence of charge compressibility $\chi$ at fixed magnetic field strength $\Phi/\Phi_0 \approx0.33$. The $\nu=2$ state (dashed line) is highly incompressible and shows limited temperature dependence. On the other hand, the charge compressibility at $\nu=1$ decreases with temperature, but not quickly enough to obtain a good Arrhenius fit. We can also see that down to the lowest accessible temperatures $\beta=8$ for $U/t=4$ and $\beta=5$ for $U/t=10$, the system is still not fully gapped. The DQMC data suggests that if there exists a temperature-independent charge gap scale, we have not reached temperatures low enough in DQMC simulations to observe it clearly. \cref{fig:compress-T-dep} thus allows us to estimate upper bounds on the $\nu=1$ charge gap: $\Delta_{\nu=1}(U/t=4) < t/8$ and $\Delta_{\nu=1}(U/t=10) < t/5$.

\begin{figure}[htpb]
    \includegraphics[width=\linewidth]{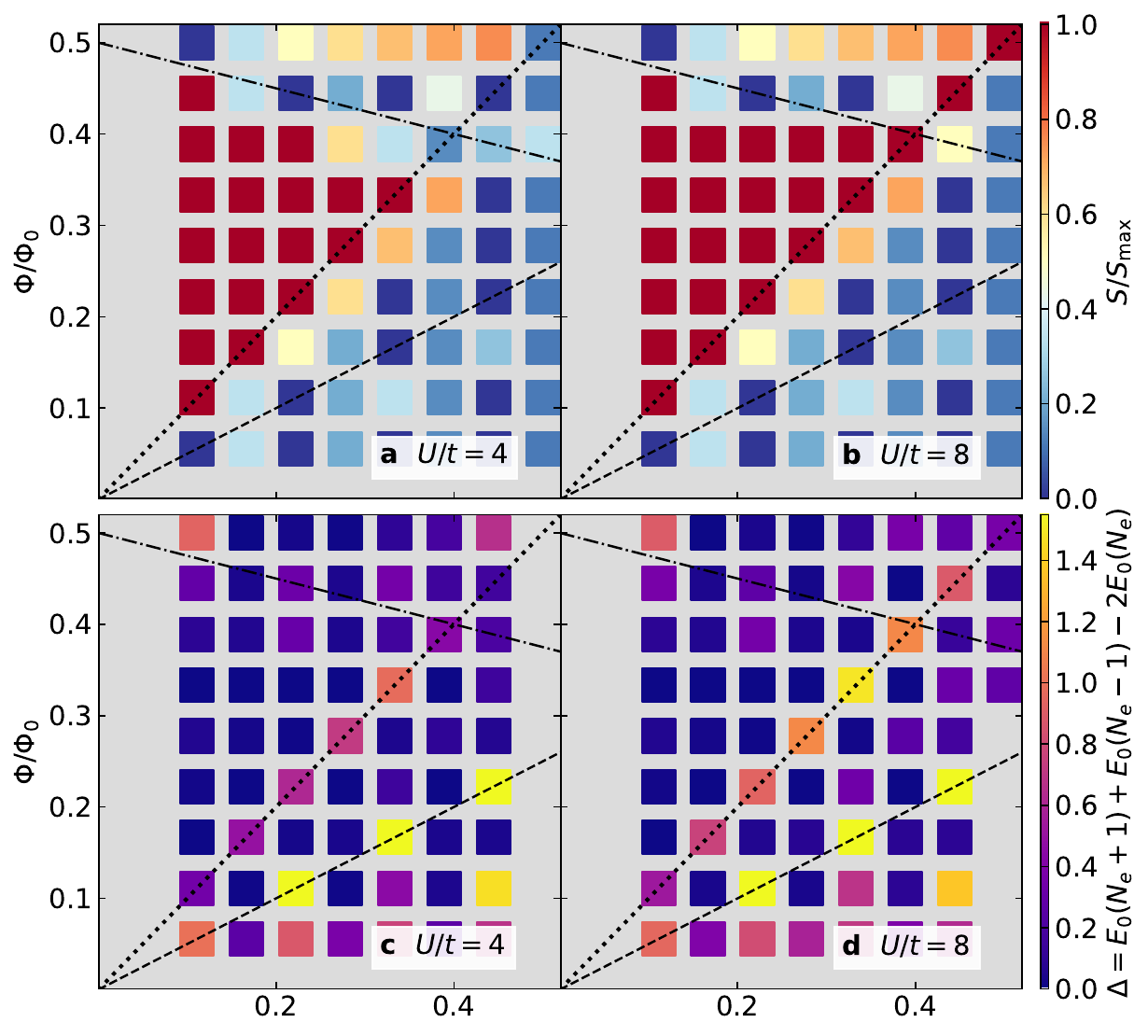}
    \caption{DMRG ground state spin degeneracy (top row) and charge gap (bottom row). \textbf{a}-\textbf{b} Spin degeneracies are plotted as $S/S_{\mathrm{max}}$, where $S_{\mathrm{max}} = N_e/2$. States differing in energy by less than $10^{-4} t$ 
    are treated as degenerate. Red color indicates that the ground state is maximally spin polarized in its particle sector, while blue color indicates that the ground state is a singlet configuration. \textbf{c}-\textbf{d} Charge addition-removal gap, $\Delta(N_e)$ as defined in \cref{eq:gs-delta}. Due to the rapid growth of entanglement with system size in the torus geometry, DMRG data are obtained on $3\times6$ cluster. Maximum bond dimensions is 2000 and truncation error is of order $10^{-6}$ or below. The colorbar in \textbf{c}-\textbf{d} are saturated in the maximum direction in order to improve the visibility of the $\nu=1$ gaps.}
    \label{fig:GS-DMRG-torus}
\end{figure}

We also examined incompressible states along $\nu=1$ and $\nu=2$ via DMRG simulations on a small cluster by calculating the particle addition-removal gap
\begin{equation}
    \Delta(N_e) = E_0(N_e+1) + E_0(N_e-1) - 2E_0(N_e) \label{eq:gs-delta},
\end{equation}
where $E_0(N_e)$ denotes the ground state energy in the $N_e$-electron sector. The charge addition-removal gap is identical to inverse charge compressibility $\partial\mu/\partial n$ as $T\rightarrow 0$ in the thermodynamic limit. $\Delta(N_e)$ obtained by DMRG are shown in~\cref{fig:GS-DMRG-torus}\textbf{c} for $U/t=4$ and~\cref{fig:GS-DMRG-torus}\textbf{d} for $U/t=8$. The charge gap along $\nu=2$ is robust against $U$, which matches the DQMC data and shows that this Hofstadter band gap is not closed by the Hubbard interaction~\cite{Ding2022}. Along $\nu=1$, 
a smaller gap which we associate with $\nu=1$ QHFM emerges. It is larger for higher Hubbard $U$, with a maximum gap value of $\Delta=1.5t$ for $U/t=8$ on the $3\times 6$ cluster. Along $\nu=1$, the charge gap is greatly reduced above $\Phi/\Phi_0 = 1/3$ for $U/t=4$, while it persists to higher field strengths, near $\Phi/\Phi_0 \sim 0.5$, for $U/t=8$, which matches the extent of ferromagnetic ground states observed in \cref{fig:GS-DMRG}.

\cref{fig:GS-DMRG-torus}\textbf{a} and \cref{fig:GS-DMRG-torus}\textbf{b} show ground state spin degeneracies for $U/t=4$ and $U/t=8$ on a $3\times 6$ cluster with modified periodic boundary conditions. These results are qualitatively consistent with results presented in the in the main text (\cref{fig:GS-DMRG}), which were obtained on a $3\times 25$ system with cylindrical boundary conditions.

\begin{figure*}[htpb]
    \centering
    \includegraphics[width=0.85\linewidth]{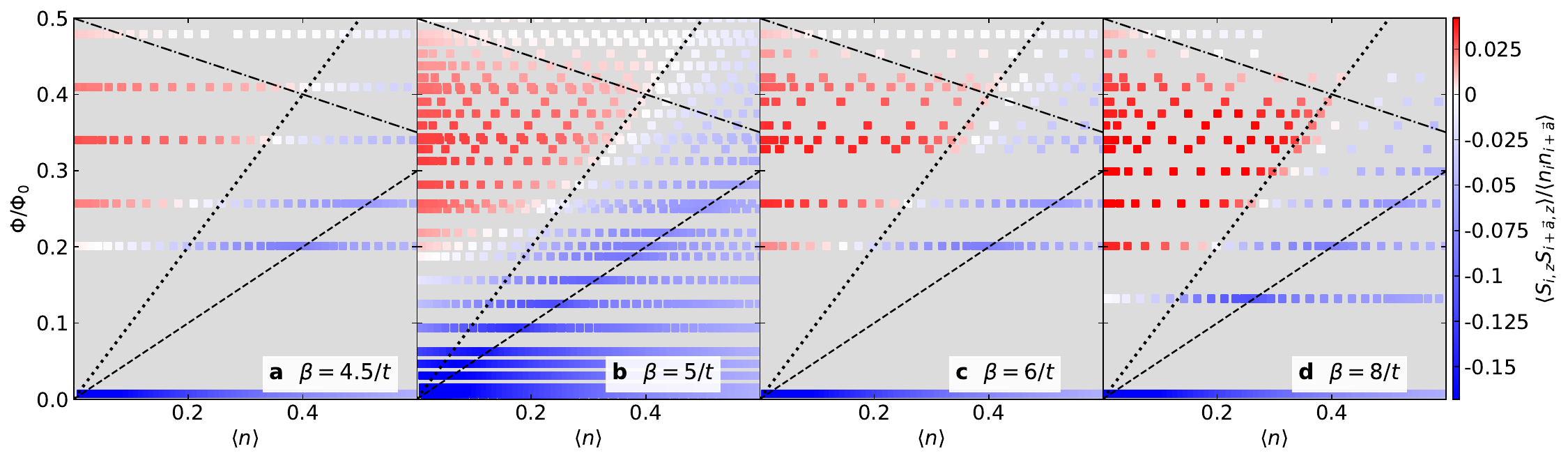}
    \caption{Extended temperature dependence of equal-time nearest-neighbor spin correlation normalized by nearest-neighbor charge correlation, $\langle S_{i,z}  S_{i+\vec{a},z}\rangle/\langle n_i  n_{i+\vec{a}}\rangle$, obtained by DQMC. Fixed Hubbard $U/t=4$. Dotted lines denote $\nu=1$, dashed lines denote $\nu=2$, and dot-dashed lines denote $n=-4 (\Phi/\Phi_0) + 2$. The colorbar is saturated in the minimum direction. DQMC data are obtained on clusters of size $8\times 8$ and $12\times 12$. }
    \label{fig:extend-T}
\end{figure*}

Extended temperature dependence of nearest neighbor spin correlation, obtained via DQMC, is shown in \cref{fig:extend-T}.

\begin{figure*}[htpb]
    \centering
    \includegraphics[width=\linewidth]{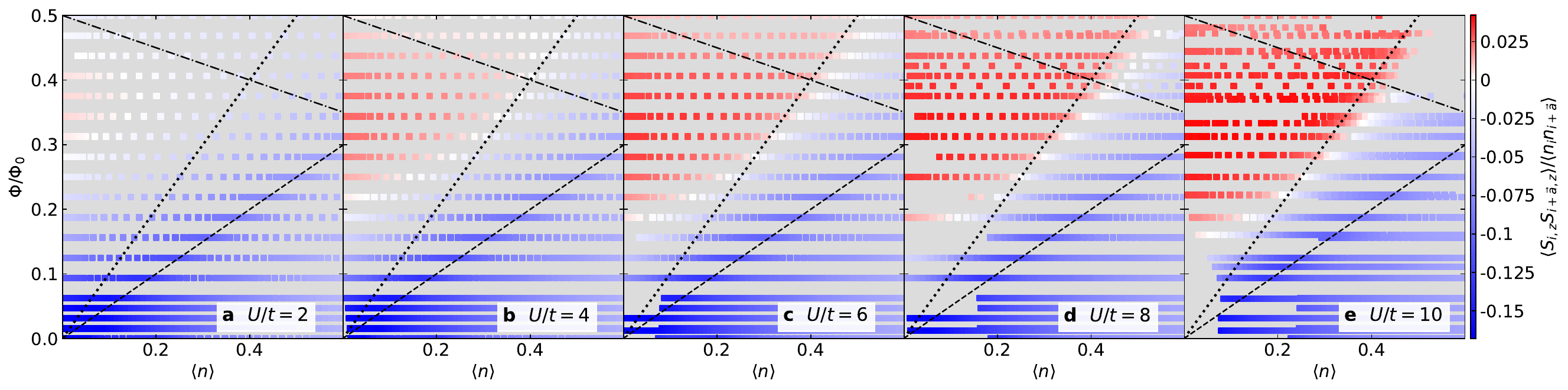}
    \caption{Extended Hubbard $U$ dependence of equal-time nearest-neighbor spin correlation normalized by nearest-neighbor charge correlation, $\langle S_{i,z}  S_{i+\vec{a},z}\rangle/\langle n_i  n_{i+\vec{a}}\rangle$, obtained by DQMC. Fixed inverse temperature $\beta=4/t$. Dotted lines denote $\nu=1$, dashed lines denote $\nu=2$, and dot-dashed lines denote $n=-4 (\Phi/\Phi_0) + 2$. The colorbar is saturated in the minimum direction. DQMC data are obtained on clusters of size $6\times 6$, $8\times 8$, and $9\times 9$.}
    \label{fig:extend-U}
\end{figure*}

Extended Hubbard interaction dependence of nearest neighbor spin correlation, obtained via DQMC, is shown in \cref{fig:extend-U}.

\begin{figure*}[htpb]
    \centering
    \includegraphics[width=\linewidth]{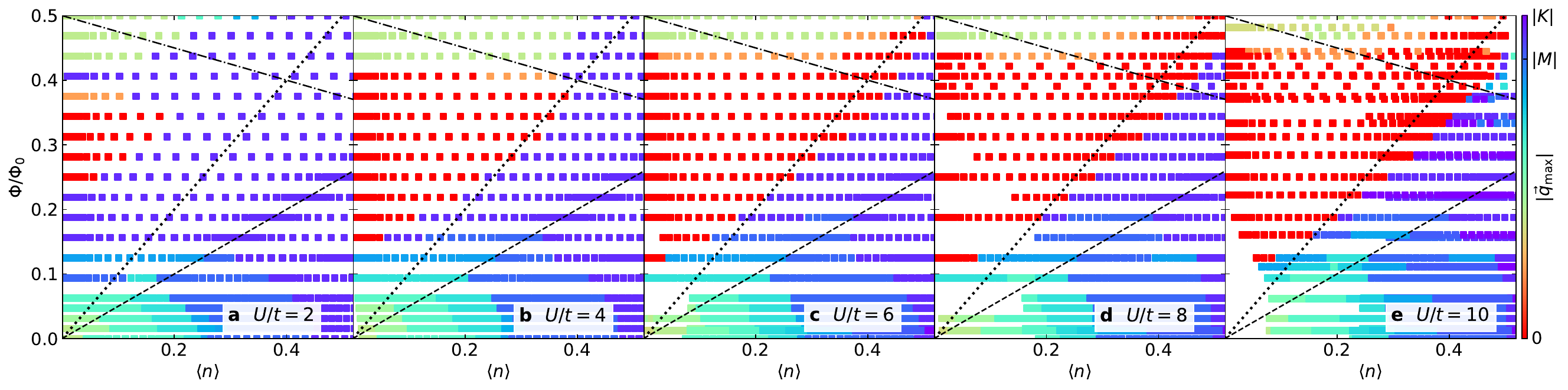}
    \caption{Extended Hubbard $U$ dependence of $|q_{\mathrm{max}}|$, obtained by DQMC. Fixed inverse temperature $\beta=4/t$. Dotted lines denote $\nu=1$, dashed lines denote $\nu=2$, and dot-dashed lines denote $n=-4 (\Phi/\Phi_0) + 2$. DQMC data are obtained on clusters of size $8\times 8$ and $9\times 9$.}
    \label{fig:extend-qmax-U}
\end{figure*}

Extended Hubbard interaction dependence of $|q_{\mathrm{max}}|$, obtained via DQMC, is shown in \cref{fig:extend-qmax-U}.

\end{document}